\input epsf
%
%
%
\def\unredoffs{} 

%
%
%
%
\newbox\leftpage \newdimen\fullhsize \newdimen\hstitle \newdimen\hsbody
\tolerance=1000\hfuzz=2pt
\catcode`\@=11 
%
\magnification=1200\unredoffs\baselineskip=16pt plus 2pt minus 1pt
\hsbody=\hsize \hstitle=\hsize 
%
%
%
\newcount\yearltd\yearltd=\year\advance\yearltd by -1900

%
%

\def\draftmode{\message{ DRAFTMODE }\def\draftdate{{\rm preliminary draft:
\number\month/\number\day/\number\yearltd\ \ \hourmin}}%
\headline={\hfil\draftdate}\writelabels\baselineskip=20pt plus 2pt minus 2pt
 {\count255=\time\divide\count255 by 60 \xdef\hourmin{\number\count255}
  \multiply\count255 by-60\advance\count255 by\time
  \xdef\hourmin{\hourmin:\ifnum\count255<10 0\fi\the\count255}}}
\def\nolabels{\def\wrlabeL##1{}\def\eqlabeL##1{}\def\reflabeL##1{}}
\def\writelabels{\def\wrlabeL##1{\leavevmode\vadjust{\rlap{\smash%
{\line{{\escapechar=` \hfill\rlap{\sevenrm\hskip.03in\string##1}}}}}}}%
\def\eqlabeL##1{{\escapechar-1\rlap{\sevenrm\hskip.05in\string##1}}}%
\def\reflabeL##1{\noexpand\llap{\noexpand\sevenrm\string\string\string##1}}}
\nolabels
%
\global\newcount\secno \global\secno=0
\global\newcount\meqno \global\meqno=1
\def\newsec#1{\global\advance\secno by1\message{(\the\secno. #1)}
\global\subsecno=0\eqnres@t\noindent{\bf\the\secno. #1}
\writetoca{{\secsym} {#1}}\par\nobreak\medskip\nobreak}
\def\eqnres@t{\xdef\secsym{\the\secno.}\global\meqno=1\bigbreak\bigskip}
\def\sequentialequations{\def\eqnres@t{\bigbreak}}\xdef\secsym{}
\global\newcount\subsecno \global\subsecno=0
\def\subsec#1{\global\advance\subsecno by1\message{(\secsym\the\subsecno. #1)}
\ifnum\lastpenalty>9000\else\bigbreak\fi
\noindent{\it\secsym\the\subsecno. #1}\writetoca{\string\quad
{\secsym\the\subsecno.} {#1}}\par\nobreak\medskip\nobreak}
\def\appendix#1#2{\global\meqno=1\global\subsecno=0\xdef\secsym{\hbox{#1.}}
\bigbreak\bigskip\noindent{\bf Appendix #1. #2}\message{(#1. #2)}
\writetoca{Appendix {#1.} {#2}}\par\nobreak\medskip\nobreak}
%
%
\def\eqnn#1{\xdef #1{(\secsym\the\meqno)}\writedef{#1\leftbracket#1}%
\global\advance\meqno by1\wrlabeL#1}
\def\eqna#1{\xdef #1##1{\hbox{$(\secsym\the\meqno##1)$}}
\writedef{#1\numbersign1\leftbracket#1{\numbersign1}}%
\global\advance\meqno by1\wrlabeL{#1$\{\}$}}
\def\eqn#1#2{\xdef #1{(\secsym\the\meqno)}\writedef{#1\leftbracket#1}%
\global\advance\meqno by1$$#2\eqno#1\eqlabeL#1$$}
%
\newskip\footskip\footskip6pt plus 1pt minus 1pt 
\def\footnotefont{\ninepoint}\def\f@t#1{\footnotefont #1\@foot}
\def\f@@t{\baselineskip\footskip\bgroup\footnotefont\aftergroup\@foot\let\next}
\setbox\strutbox=\hbox{\vrule height9.5pt depth4.5pt width0pt}
\global\newcount\ftno \global\ftno=0
\def\foot{\global\advance\ftno by1\footnote{$^{\the\ftno}$}}
%
\newwrite\ftfile
\def\footend{\def\foot{\global\advance\ftno by1\chardef\wfile=\ftfile
$^{\the\ftno}$\ifnum\ftno=1\immediate\openout\ftfile=foots.tmp\fi%
\immediate\write\ftfile{\noexpand\smallskip%
\noexpand\item{f\the\ftno:\ }\pctsign}\findarg}%
\def\footatend{\vfill\eject\immediate\closeout\ftfile{\parindent=20pt
\centerline{\bf Footnotes}\nobreak\bigskip\input foots.tmp }}}
\def\footatend{}
%
%
\global\newcount\refno \global\refno=1
\newwrite\rfile
\def\ref{[\the\refno]\nref}
\def\nref#1{\xdef#1{[\the\refno]}\writedef{#1\leftbracket#1}%
\ifnum\refno=1\immediate\openout\rfile=refs.tmp\fi
\global\advance\refno by1\chardef\wfile=\rfile\immediate
\write\rfile{\noexpand\item{#1\ }\reflabeL{#1\hskip.31in}\pctsign}\findarg}
\def\findarg#1#{\begingroup\obeylines\newlinechar=`\^^M\pass@rg}
{\obeylines\gdef\pass@rg#1{\writ@line\relax #1^^M\hbox{}^^M}%
\gdef\writ@line#1^^M{\expandafter\toks0\expandafter{\striprel@x #1}%
\edef\next{\the\toks0}\ifx\next\em@rk\let\next=\endgroup\else\ifx\next\empty%
\else\immediate\write\wfile{\the\toks0}\fi\let\next=\writ@line\fi\next\relax}}
\def\striprel@x#1{} \def\em@rk{\hbox{}}
\def\lref{\begingroup\obeylines\lr@f}
\def\lr@f#1#2{\gdef#1{\ref#1{#2}}\endgroup\unskip}
\def\semi{;\hfil\break}
\def\addref#1{\immediate\write\rfile{\noexpand\item{}#1}} 
\def\startrefs#1{\immediate\openout\rfile=refs.tmp\refno=#1}
\def\xref{\expandafter\xr@f}\def\xr@f[#1]{#1}
\def\refs#1{\count255=1[\r@fs #1{\hbox{}}]}
\def\r@fs#1{\ifx\und@fined#1\message{reflabel \string#1 is undefined.}%
\nref#1{need to supply reference \string#1.}\fi%
\vphantom{\hphantom{#1}}\edef\next{#1}\ifx\next\em@rk\def\next{}%
\else\ifx\next#1\ifodd\count255\relax\xref#1\count255=0\fi%
\else#1\count255=1\fi\let\next=\r@fs\fi\next}
%

%
\newwrite\ffile\global\newcount\figno \global\figno=1
\def\fig{fig.~\the\figno\nfig}
\def\nfig#1{\xdef#1{fig.~\the\figno}%
\writedef{#1\leftbracket fig.\noexpand~\the\figno}%
\ifnum\figno=1\immediate\openout\ffile=figs.tmp\fi\chardef\wfile=\ffile%
\immediate\write\ffile{\noexpand\medskip\noexpand\item{Fig.\ \the\figno. }
\reflabeL{#1\hskip.55in}\pctsign}\global\advance\figno by1\findarg}
\def\xfig{\expandafter\xf@g}\def\xf@g fig.\penalty\@M\ {}
\def\figs#1{figs.~\f@gs #1{\hbox{}}}
\def\f@gs#1{\edef\next{#1}\ifx\next\em@rk\def\next{}\else
\ifx\next#1\xfig #1\else#1\fi\let\next=\f@gs\fi\next}
\newwrite\lfile
{\escapechar-1\xdef\pctsign{\string\%}\xdef\leftbracket{\string\{}
\xdef\rightbracket{\string\}}\xdef\numbersign{\string\#}}

\def\writestop{\def\writestoppt{\immediate\write\lfile{\string\pageno%
\the\pageno\string\startrefs\leftbracket\the\refno\rightbracket%
\string\def\string\secsym\leftbracket\secsym\rightbracket%
\string\secno\the\secno\string\meqno\the\meqno}\immediate\closeout\lfile}}
\def\writestoppt{}\def\writedef#1{}
\def\seclab#1{\xdef #1{\the\secno}\writedef{#1\leftbracket#1}\wrlabeL{#1=#1}}
\def\subseclab#1{\xdef #1{\secsym\the\subsecno}%
\writedef{#1\leftbracket#1}\wrlabeL{#1=#1}}
\newwrite\tfile \def\writetoca#1{}
\def\leaderfill{\leaders\hbox to 1em{\hss.\hss}\hfill}
\def\writetoc{\immediate\openout\tfile=toc.tmp
   \def\writetoca##1{{\edef\next{\write\tfile{\noindent ##1
   \string\leaderfill {\noexpand\number\pageno} \par}}\next}}}

%
\catcode`\@=12 
%
\edef\tfontsize{\ifx\answ\bigans scaled\magstep3\else scaled\magstep4\fi}
\font\titlerm=cmr10 \tfontsize \font\titlerms=cmr7 \tfontsize
\font\titlermss=cmr5 \tfontsize \font\titlei=cmmi10 \tfontsize
\font\titleis=cmmi7 \tfontsize \font\titleiss=cmmi5 \tfontsize
\font\titlesy=cmsy10 \tfontsize \font\titlesys=cmsy7 \tfontsize
\font\titlesyss=cmsy5 \tfontsize \font\titleit=cmti10 \tfontsize
\skewchar\titlei='177 \skewchar\titleis='177 \skewchar\titleiss='177
\skewchar\titlesy='60 \skewchar\titlesys='60 \skewchar\titlesyss='60
\def\titlefont{\def\rm{\fam0\titlerm}
\textfont0=\titlerm \scriptfont0=\titlerms \scriptscriptfont0=\titlermss
\textfont1=\titlei \scriptfont1=\titleis \scriptscriptfont1=\titleiss
\textfont2=\titlesy \scriptfont2=\titlesys \scriptscriptfont2=\titlesyss
\textfont\itfam=\titleit \def\it{\fam\itfam\titleit}\rm}
 \ifx\answ\bigans\else scaled\magstep1\fi
\ifx\answ\bigans\else

 \font\absi=cmmi10 scaled\magstep1
\font\absis=cmmi7 scaled\magstep1 \font\absiss=cmmi5 scaled\magstep1
\font\abssy=cmsy10 scaled\magstep1 \font\abssys=cmsy7 scaled\magstep1
\font\abssyss=cmsy5 scaled\magstep1 
\skewchar\absi='177 \skewchar\absis='177 \skewchar\absiss='177
\skewchar\abssy='60 \skewchar\abssys='60 \skewchar\abssyss='60
\fi
\font\ninerm=cmr9 \font\sixrm=cmr6 \font\ninei=cmmi9 \font\sixi=cmmi6
\font\ninesy=cmsy9 \font\sixsy=cmsy6 \font\ninebf=cmbx9
\font\nineit=cmti9 \font\ninesl=cmsl9 \skewchar\ninei='177
\skewchar\sixi='177 \skewchar\ninesy='60 \skewchar\sixsy='60
\def\ninepoint{\def\rm{\fam0\ninerm}
\textfont0=\ninerm \scriptfont0=\sixrm \scriptscriptfont0=\fiverm
\textfont1=\ninei \scriptfont1=\sixi \scriptscriptfont1=\fivei
\textfont2=\ninesy \scriptfont2=\sixsy \scriptscriptfont2=\fivesy
\textfont\itfam=\ninei \def\it{\fam\itfam\nineit}\def\sl{\fam\slfam\ninesl}%
\textfont\bffam=\ninebf \def\bf{\fam\bffam\ninebf}\rm}
%
%
\def\noblackbox{\overfullrule=0pt}
\hyphenation{anom-aly anom-alies coun-ter-term coun-ter-terms}
\def\inv{^{\raise.15ex\hbox{${\scriptscriptstyle -}$}\kern-.05em 1}}

\def\Dsl{\,\raise.15ex\hbox{/}\mkern-13.5mu D} 
\def\dsl{\raise.15ex\hbox{/}\kern-.57em\partial}

\def\lspace{\ifx\answ\bigans{}\else\qquad\fi}
\def\lbspace{\ifx\answ\bigans{}\else\hskip-.2in\fi} 
\def\boxeqn#1{\vcenter{\vbox{\hrule\hbox{\vrule\kern3pt\vbox{\kern3pt
        \hbox{${\displaystyle #1}$}\kern3pt}\kern3pt\vrule}\hrule}}}
\def\mbox#1#2{\vcenter{\hrule \hbox{\vrule height#2in
                \kern#1in \vrule} \hrule}}  
%
   
 \def\CH{{\cal H}}

\def\darr#1{\raise1.5ex\hbox{$\leftrightarrow$}\mkern-16.5mu #1}

\def\half{{\textstyle{1\over2}}} 
\def\roughly#1{\raise.3ex\hbox{$#1$\kern-.75em\lower1ex\hbox{$\sim$}}}
\hyphenation{Mar-ti-nel-li}

\def\1{\;1\!\!\!\! 1\;}

\def\etal{{\it et al.}}
\def\rhs{right hand side}

\def\frac#1#2{{{#1}\over {#2}}}
\def\half{\hbox{${1\over 2}$}}

\def\smallfrac#1#2{\hbox{${{#1}\over {#2}}$}}

\def\GeV{{\rm GeV}}\def\TeV{{\rm TeV}}

\def\MS{\hbox{$\overline{\rm MS}$}}

\def\QMS{Q$_0$\MS}

\catcode`@=11 
\def\slash#1{\mathord{\mathpalette\c@ncel#1}}
 \def\c@ncel#1#2{\ooalign{$\hfil#1\mkern1mu/\hfil$\crcr$#1#2$}}
\def\lsim{\mathrel{\mathpalette\@versim<}}
\def\gsim{\mathrel{\mathpalette\@versim>}}
 \def\@versim#1#2{\lower0.2ex\vbox{\baselineskip\z@skip\lineskip\z@skip
       \lineskiplimit\z@\ialign{$\m@th#1\hfil##$\crcr#2\crcr\sim\crcr}}}
\catcode`@=12 

\def\PR{{\it Phys.~Rev.~}}

\def\NP{{\it Nucl.~Phys.~}}

\def\PL{{\it Phys.~Lett.~}}

\def\SJNP{{\it Sov.~Jour.~Nucl.~Phys.~}}
\def\SPJETP{{\it Sov.~Phys.~J.E.T.P.~}}

\def\JHEP{{\it Jour.~High~En.~Phys.~}}
\def\EPJ{{\it Euro.~Phys.~Jour.}}
\def\vol#1{{\bf #1}}\def\vyp#1#2#3{\vol{#1} (#2) #3}

\def\as{\alpha_s}

\def\ash{\widehat\alpha_s}

\noblackbox
\pageno=0\nopagenumbers\tolerance=10000\hfuzz=5pt
\baselineskip 12pt
\line{\hfill CERN-PH-TH/2007-212}
\line{\hfill Edinburgh 2007-11}
\vskip 24pt
\centerline{\titlefont Resummation of Hadroproduction}\vskip 10pt
\centerline{\titlefont Cross-sections at High Energy}
\vskip 36pt\centerline{Richard D.~Ball}
\vskip 24pt
\centerline{\it School of Physics, University of Edinburgh}
\centerline{\it JCMB, KB, Mayfield Rd, Edinburgh EH9 3JZ, Scotland}
\vskip 6pt
\centerline{\it and}
\vskip 6pt
\centerline{\it CERN, Department of Physics, Theory Division}
\centerline{\it CH-1211 Gen\`eve 23, Switzerland}
\vskip 60pt
\centerline{\bf Abstract}
{\narrower\baselineskip 10pt
\medskip\noindent We reconsider the high energy resummation of 
photoproduction, electroproduction and hadroproduction cross-sections, 
in the light of recent progress in the resummation of perturbative 
parton evolution to NLO in logarithms of $Q^2$ and $x$. We show 
in particular that the when the coupling runs the dramatic enhancements
seen at fixed coupling, due to infrared singularities in the 
partonic cross-sections, are substantially reduced, 
to the extent that they are largely accounted for by the usual
NLO and NNLO perturbative corrections. This leads to a 
novel explanation of the large $K$-factors commonly found 
in perturbative calculations of hadroproduction cross-sections.
We give numerical estimates of high energy resummation effects 
for inclusive $B$-production, inclusive jets, Drell-Yan and vector boson 
production, along with their rapidity distributions. We 
find that resummation modifies the $B$-production cross-section 
at the LHC by at most 15\%, but that the enhancement of 
gluonic $W$-production may be as large as 50\% at large rapidities.}
\vfill
\line{CERN-PH-TH/2007-212\hfill }
\line{October 2007\hfill}
\eject \footline={\hss\tenrm\folio\hss}

\lref\glapd{
V.N.~Gribov and L.N.~Lipatov,
\SJNP\vyp{15}{1972}{438}\semi  
L.N.~Lipatov, \SJNP\vyp{20}{1975}{95}\semi    
G.~Altarelli and G.~Parisi,
\NP\vyp{B126}{1977}{298}\semi  
see also
Y.L.~Dokshitzer,
{\it Sov.~Phys.~JETP~}\vyp{46}{1977}{691}.} 
\lref\nlo{G.~Curci, W.~Furma\'nski and R.~Petronzio,
\NP\vyp{B175}{1980}{27}\semi 
E.G.~Floratos, C.~Kounnas and R.~Lacaze,
\NP\vyp{B192}{1981}{417}.} 
\lref\nnlo{S.A.~Larin, T.~van~Ritbergen, J.A.M.~Vermaseren,
\NP\vyp{B427}{1994}{41}\semi  
S.A.~Larin \etal, \NP\vyp{B492}{1997}{338}.} 
\lref\bfkl{L.N.~Lipatov,
\SJNP\vyp{23}{1976}{338}\semi 
 V.S.~Fadin, E.A.~Kuraev and L.N.~Lipatov,
\PL\vyp{60B}{1975}{50}; 
 {\it Sov. Phys. JETP~}\vyp{44}{1976}{443}; 
\vyp{45}{1977}{199}\semi 
 Y.Y.~Balitski and L.N.Lipatov,
\SJNP\vyp{28}{1978}{822}.} 
\lref\CH{
S.~Catani and F.~Hautmann,
\PL\vyp{B315}{1993}{157}; 
\NP\vyp{B427}{1994}{475}.} 
\lref\fl{V.S.~Fadin and L.N.~Lipatov,
\PL\vyp{B429}{1998}{127}\semi  
V.S.~Fadin et al, \PL\vyp{B359}{1995}{181}; 
\PL\vyp{B387}{1996}{593}; 
\NP\vyp{B406}{1993}{259}; 
\PR\vyp{D50}{1994}{5893}; 
\PL\vyp{B389}{1996}{737};  
\NP\vyp{B477}{1996}{767};  
\PL\vyp{B415}{1997}{97};  
\PL\vyp{B422}{1998}{287}.} 
\lref\cc{G.~Camici and M.~Ciafaloni,
\PL\vyp{B412}{1997}{396}; 
\PL\vyp{B430}{1998}{349}.} 
\lref\dd{V.~del~Duca, \PR\vyp{D54}{1996}{989}; 
\PR\vyp{D54}{1996}{4474}\semi 
V.~del~Duca and C.R.~Schmidt,
\PR\vyp{D57}{1998}{4069}\semi 
Z.~Bern, V.~del~Duca and C.R.~Schmidt,
\PL\vyp{B445}{1998}{168}.}
\lref\ross{
D.~A.~Ross,
Phys.\ Lett.\ B {\bf 431}, 161 (1998) 
}
\lref\jar{T.~Jaroszewicz,
\PL\vyp{B116}{1982}{291}.}
\lref\ktfac{S.~Catani, F.~Fiorani and G.~Marchesini,
\NP\vyp{B336}{1990}{18}\semi 
S.~Catani et al.,
\NP\vyp{B361}{1991}{645}.}
\lref\bfsum{R.~D.~Ball and S.~Forte,
\PL\vyp{B351}{1995}{313}; 
\PL\vyp{B359}{1995}{362}\semi
R.K.~Ellis, F.~Hautmann and B.R.~Webber,
\PL\vyp{B348}{1995}{582}.}
\lref\bfalf{R.~D.~Ball and S.~Forte,
\PL\vyp{B358}{1995}{365}.}
\lref\bfafp{R.~D.~Ball and S.~Forte,
\PL\vyp{B405}{1997}{317}.}
\lref\DGPTWZ{A.~De~R\'ujula {\it et al.},
\PR\vyp{D10}{1974}{1649}.}
\lref\bfdas{R.~D.~Ball and S.~Forte,
\PL\vyp{B335}{1994}{77}; 
\vyp{B336}{1994}{77}\semi 
{\it Acta~Phys.~Polon.~}\vyp{B26}{1995}{2097}.}
\lref\esw{
See {\it e.g.}  R.~K.~Ellis, W.~J.~Stirling and B.~R.~Webber,
``QCD and Collider Physics'' (C.U.P., Cambridge 1996).}
\lref\bfklfits{R.~D.~Ball and S.~Forte,
{\tt hep-ph/9607291}\semi 
I.~Bojak and M.~Ernst, \PL\vyp{B397}{1997}{296};
\NP\vyp{B508}{1997}{731}\semi
J.~Bl\"umlein  and A.~Vogt,
\PR\vyp{D58}{1998}{014020}.}
\lref\flprobs{R.~D.~Ball  and S.~Forte,
{\tt hep-ph/9805315}.
}
\lref\salam{G.~Salam, \JHEP\vyp{9807}{1998}{19}\semi
M.~Ciafaloni, D.~Colferai and G.~P.~Salam,
  \JHEP\vyp{9910}{1999}{17}.}
\lref\bfsxap{R.~D.~Ball and S.~Forte,
\PL\vyp{B465}{1999}{271}.}
\lref\abfsxres{G. Altarelli, R.~D. Ball and S. Forte,
\NP\vyp{B575}{2000}{313}.}  
\lref\abfsxphen{G. Altarelli, R.~D.~Ball and S. Forte,
\NP\vyp{B599}{2001}{383};  
see also {\tt hep-ph/0104246}.}  
\lref\Liprun{L.N.~Lipatov,
\SPJETP\vyp{63}{1986}{5}.}
\lref\ColKwie{J.~C.~Collins and J.~Kwiecinski, 
\NP\vyp{B316}{1989}{307}.}
\lref\ciafqz {M.~Ciafaloni,
  \PL\vyp{B356}{1995}{74}.}  
\lref\ccfac{  G.~Camici and M.~Ciafaloni,
  \NP\vyp{B496}{1997}{305};
  [Erratum-ibid.\vyp{B607}{2001}{431}].}
\lref\ccsrengrp{  M.~Ciafaloni, D.~Colferai and G.~P.~Salam,
  \PR\vyp{D60}{1999}{114036}.} 
\lref\ccsfact{
M.~Ciafaloni, D.~Colferai and G.~P.~Salam,
\JHEP\vyp{0007}{2000}{054}.}
\lref\ciafrc{M.~Ciafaloni, M.~Taiuti and A.~H.~Mueller,
\NP\vyp{B616}{2001}{349}\semi 
M.~Ciafaloni et al., \PR\vyp{D66}{2002}{054014}.}
\lref\ciafsplit{M.~Ciafaloni et al,
  \PR\vyp{D68}{2003}{114003};
\PL\vyp{B635}{2006}{320}.}
\lref\ccfact{M.~Ciafaloni and D.~Colferai,
\JHEP\vyp{0509}{2005}{069}.}
\lref\abfairy{
G.~Altarelli, R.~D.~Ball and S.~Forte,
\NP\vyp{B621}{2002}{359}; 
\NP\vyp{B674}{2003}{459}.}
\lref\abfrundual{  R.~D.~Ball and S.~Forte,
  \NP\vyp{B742}{2006}{158}.} 
\lref\abfrunfin{G.~Altarelli, R.~D.~Ball and S.~Forte,
  \NP\vyp{B742}{2006}{1}.}  
\lref\rdbrke{ R.~D.~Ball and R.~K.~Ellis,
  \JHEP\vyp{0105}{2001}{053}.}  
\lref\NDE{  P.~Nason, S.~Dawson and R.~K.~Ellis,
   ``The Total Cross-Section for the Production of Heavy Quarks in Hadronic
  \NP\vyp{B303}{1988}{607}.} 
\lref\MNR{  M.~L.~Mangano, P.~Nason and G.~Ridolfi,
  Nucl.\ Phys.\  B {\bf 373}, 295 (1992).}
\lref\BNMSS{ W.~Beenakker et al,
  \NP\vyp{B351}{1991}{507}.}   
\lref\FrixMan{  S.~Frixione and M.~L.~Mangano,
  \NP\vyp{B483}{1997}{321}.}   
\lref\BCMN{  R.~Bonciani, S.~Catani, M.~L.~Mangano and P.~Nason,
  \NP\vyp{B529}{1998}{424}.}   
\lref\ellross{  R.~K.~Ellis and D.~A.~Ross,
  \NP\vyp{B345}{1990}{79}.}   
\lref\cchglu{  S.~Catani, M.~Ciafaloni and F.~Hautmann,
  \PL\vyp{B242}{1990}{97};   
  \NP\vyp{B366}{1991}{135}.}   
\lref\colell{  J.~C.~Collins and R.~K.~Ellis,
  \NP\vyp{B360}{1991}{3}.}   
\lref\ellisnason{  R.~K.~Ellis and P.~Nason,
  \NP\vyp{B312}{1989}{551}.}   
\lref\mbffnnlo{S.~Marzani, R.~D.~Ball, P.~Falgari and S.~Forte,
  arXiv:0704.2404 [hep-ph].}  
\lref\CFMNR{M.~Cacciari et al,
\JHEP\vyp{0407}{2004}{033}.} 
\lref\inst{J.C.~Collins and J.~Kwiecinski,
\NP\vyp{B316}{1989}{307}\semi
Y.V.~Kovchegov and A.H.~Mueller, \PL\vyp{B439}{1998}{428}\semi
N.~Arnesto, J.~Bartels and M.A.~Braun, \PL\vyp{B442}{1998}{459}.}
\lref\thorne{R.S.~Thorne, \PL\vyp{B474}{2000}{372}, 
\PR\vyp{D64}{2001}{074005}.}
\lref\DYNNLO{R.~Hamberg,W.L.~van~Neerven and T.~Matsuura, 
\NP\vyp{B359}{1991}{343} [Erratum: \NP\vyp{B644}{2002}{403}].}
\lref\schwinger{J.S.~Schwinger, \PR\vyp{82}{1951}{664}.}
\lref\cchphot{S.~Catani, M.~Ciafaloni and F.~Hautmann,
  DESY HERA Workshop 1991, 690-711.}  
\lref\esw{R.K.~Ellis, W.J.~Stirling and B.R.~Webber,
{\it QCD and Collider Physics}, C.U.P. 1996.}
\lref\abfnf{G.~Altarelli, R.D.~Ball and S.~Forte, paper in preparation.}
\lref\nnpdf{L.~Del~Debbio {\it et al}, [NNPDF Collaboration],
  \JHEP\vyp{0703}{2007}{039}\semi 
  J.~Rojo {\it et al.}  [NNPDF Collaboration],
  arXiv:0706.2130 [hep-ph].}
\lref\higgs{F.~Hautmann,
  \PL\vyp{B535}{2002}{159}\semi
A.~V.~Lipatov and N.~P.~Zotov,
  \EPJ\vyp{C44}{2005}{559}.}
\lref\thornefits{
  C.~D.~White and R.~S.~Thorne,
  Phys.\ Rev.\  D {\bf 75} (2007) 034005.}

\newsec{Introduction}
\noindent 

At the LHC we hope to 
separate with confidence a tiny fraction of interesting events from an 
overwhelming background of collisions involving gluons carrying only 
a small fraction of the momentum in the beams. The success of this 
enterprise depends crucially on our ability to control 
high energy logarithms in perturbative QCD at leading twist when we 
calculate inclusive cross-sections. Currently however no reliable 
calculations of high energy resummation corrections to any hadronic 
process have been made. The purpose of this paper is to remove the 
one remaining obstacle to performing such calculations.

In recent years there has been considerable progress by several groups
in understanding the resummation of parton evolution, so that we now
know how to simultaneously resum all collinear and small-$x$ logarithms 
at NLO. This programme depended on several key ingredients: 
$k_T$-factorisation \refs{\ktfac,\CH}, 
NLLx corrections \refs{\fl,\cc,\dd}, the recognition of the need to 
simultaneously resum collinear, anti-collinear and high energy logarithms 
\refs{\flprobs,\salam}, the
use of high energy duality to achieve this \refs{\bfafp,\abfsxres}, and the 
understanding of running coupling effects 
\refs{\ciafrc,\abfairy,\abfrunfin,\ciafsplit}.
It is now possible to perform
precise and reliable calculations of small-$x$ resummation corrections 
to parton distribution functions.

Hadroproduction processes have received much less attention 
\refs{\colell,\cchglu,\rdbrke,\higgs}. The reason for this is 
partly their additional kinematic complexity, but also because 
of a difficult conceptual problem standing in the way of reliable 
results. This problem relates to the infrared singularity which 
appears when two gluons collide at high energy, due to the 
possibility of all the energy
going into a timelike gluon which may then go almost on-shell \cchglu. 
Though it has been understood for some time that this singularity might
produce substantial enhancements of hadronic cross-sections, it has
been difficult to make reliable predictions, particularly 
when the coupling runs \rdbrke. This is the problem we resolve 
in this paper. We will find that the singularity is less dangerous
than naive arguments suggest, that the enhancements it produces at 
high energy are modest, and in fact may be well approximated by the NLO and
NNLO perturbative results.  

The structure of the rest of the paper is as follows: in chapter 2 we 
summarise the main ideas used in the resummation of high energy logarithms 
in order to set the scene and fix notation, 
and then explain the difficulties encountered in applying these ideas to 
the resummation of hadronic cross-sections due to infrared singularities. 
In chapter 3 we consider the simpler scenario of 
photoproduction and electroproduction processes, and in 
particular how there infrared singularities 
may be dealt with using an exponentiation trick. We apply this trick to the 
inclusive photoproduction of $b\bar{b}$-pairs, providing quantitative 
estimates of resummation effects. We then in chapter 4 move 
on to the more interesting case of hadroproduction, construct the 
gluon-gluon luminosity, describe the singularity structure of the 
partonic cross-sections, and show how the same trick used in 
photoproduction works here also. We provide generic estimates for 
resummation effects in various hadroproduction processes at the Tevatron, LHC 
and a notional VLHC, and consider in detail the particular case of 
hadroproduction of $b\bar{b}$ pairs. We also consider the 
stability of the resummed perturbative expansion. In chapter 5 we 
consider how we may compute rapidity distributions in this framework, 
and offer estimates 
of resummation corrections for the rapidity distributions of $b\bar{b}$ 
pairs and $W$ bosons at LHC. Finally in chapter 6 we summarise our results, 
and suggest several directions for future work.

\newsec{High Energy Singularities}

\subsec{High Energy Factorization}

We consider electron-hadron, photon-hadron and 
hadron-hadron processes in which 
$Q$ is the hard (transverse) scale (for example the photon virtuality, 
a heavy quark mass, or the invariant mass of some particular particles 
in the final state), $S$ the square of the centre-of-mass 
energy, and $\rho\equiv Q^2/S$. The dimensionless  
cross-section $\Sigma\equiv Q^2\sigma$ 
is a function of $\rho$ and $Q$, the scale of $Q$ being set by 
$\Lambda_{QCD}$. 
For hard processes $Q\gg\Lambda_{QCD}$, while for 
high energy processes $S \gg Q^2$, so $\rho\ll 1$. 

In an electroproduction or  
photoproduction process, if $x$ and $\bf{k}$ are the longitudinal and 
transverse momenta of the struck parton, the square of the centre-of-mass 
energy of the hard process is $s = xS$, and the dimensionless 
hard cross-section $\Sigma_{\gamma j}\equiv Q^2\sigma_{\gamma j}$ is a 
function of $Q^2/s=\rho/x$, ${\bf{k}}/Q$ and $\mu/Q$, where $\mu$ 
is the factorization and renormalization scale (here set equal), and j 
labels the struck parton. The (unintegrated) parton distribution function
${\cal F}_j$ depends only on $x$, ${\bf{k}}^2$ and $\mu^2$. 
Factorization (or more specifically ``$k_T$-factorization'' 
\refs{\ktfac,\CH,\colell,\cchglu}) is 
then the statement that the photon-hadron cross-section
may be written as

\eqn\photofac{\Sigma_{\gamma {\rm h}}(\rho,Q) 
= \sum_{j=g,q,\bar{q}}\int_\rho^1 \! {dx\over x}\int \! 
{d^2{\bf{k}}\over\pi{\bf{k}}^2} 
\,\Sigma_{\gamma j}\big({\rho\over x},{{\bf{k}}\over Q},\as(\mu^2)\big)
{\cal F}_j\big(x,{\bf{k}}^2,\mu^2\big),
}
up to terms which vanish as inverse powers of the hard scale $Q$.

\topinsert
\vbox{
\vskip-8.5truecm
\epsfxsize=21truecm
\centerline{\hskip1.5truecm\epsfbox{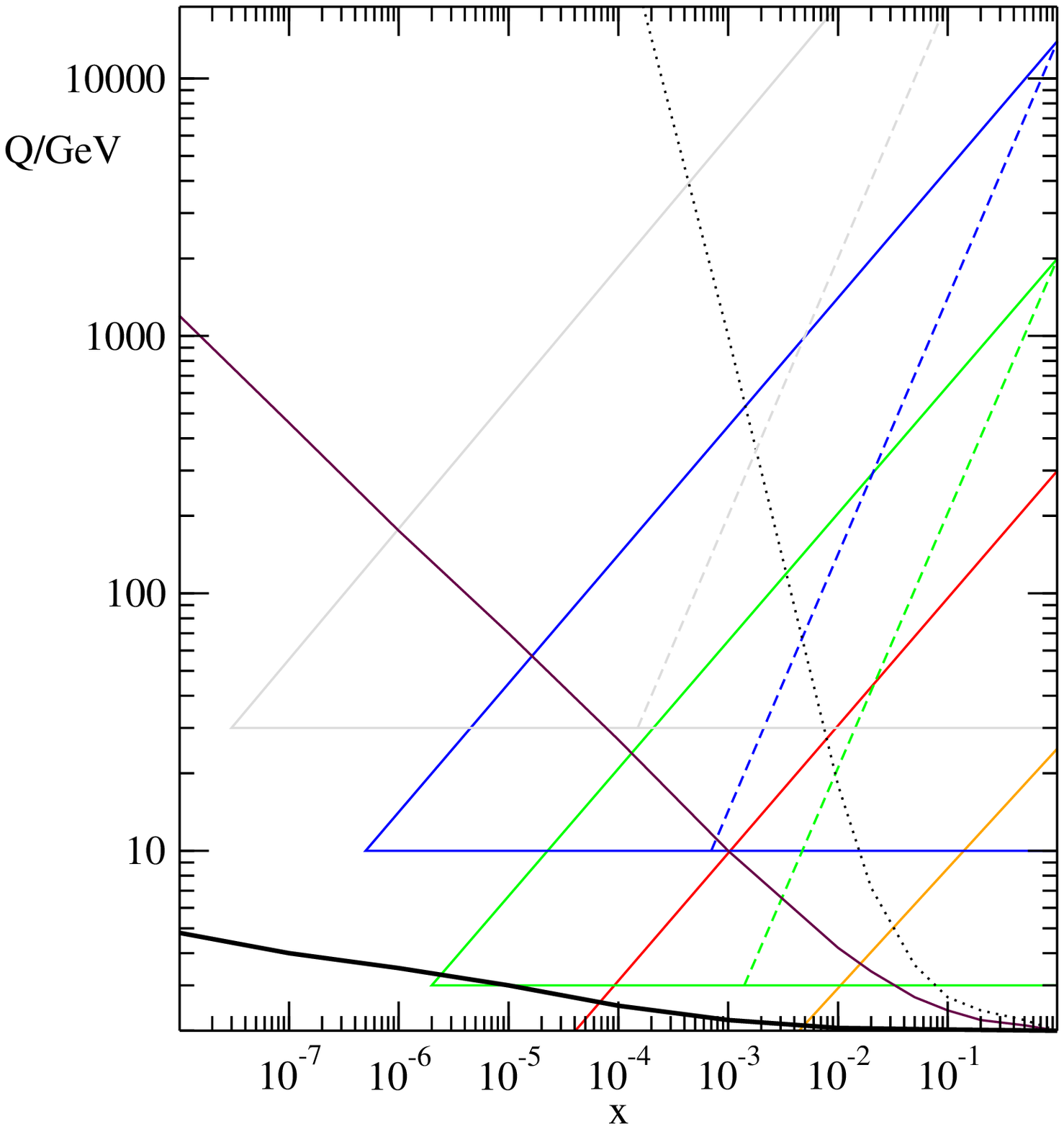}}
\vskip-1.5truecm
\bigskip
\hbox{
\vbox{\footnotefont\baselineskip6pt\narrower\noindent Figure 1: the 
kinematic plane for electro-hadron and hadron-hadron collisions:
$x$ is the fraction of longitudinal momentum in the parton, and 
$Q$ is the hard scale in $\GeV$ (which for photoproduction and 
hadroproduction is the 
invariant mass of the particles produced in the hard process). Shown are
the regions accessible in fixed target experiments (orange), at HERA (red),
at the Tevatron (green), at LHC (blue) and at a notional VLHC with 
$S=200\TeV$ (grey). The dashed lines for the hadron colliders show 
the central rapidity region, where $x_1=x_2=x$: when the rapidity is nonzero,
the values of $x_1$ and $x_2$ may be read off for a given $Q$ by choosing
points symmetrically placed about this line. The solid black diagonal line
shows the points at which $\dot L_z=L_z'$, and thus the 
logarithms of $Q^2$ are as important as the logarithms of 
$x$ (see eqn.(6.1) below): the dotted black line is determined by 
 $\dot L_z=0.1 L_z'$, while the heavy black line is determined by 
$\dot L_z=10L_z'$.
 }}\vskip-0.5truecm}
\endinsert 

For a purely hadronic process, the centre-of-mass 
energy of the hard process is $s = x_1x_2 S$, where $x_1$ and $x_2$ are 
the longitudinal momentum fractions of the colliding partons $j_1$ and $j_2$ 
in hadrons $h_1$ and $h_2$. The dimensionless hard cross-section 
$\Sigma_{j_1j_2}\equiv Q^2\sigma_{j_1j_2}$ is then a  
function of $Q^2/s =\rho/x_1x_2$, ${\bf{k_1}}/Q$, ${\bf{k_2}}/Q$ 
and $\mu/Q$, so factorization is the statement that the hadron-hadron 
cross-section may be written as

\eqn\hadrofac{\eqalign{\Sigma_{\rm hh}(\rho,Q) 
&= \sum_{j_1,j_2=g,q,\bar{q}}
\int_\rho^1 \! {dx_1\over x_1}\int_\rho^1 \! {dx_2\over x_2}
\int \! {d^2{\bf{k_1}}\over\pi{\bf{k}}_1^2}
\int \! {d^2{\bf{k_2}}\over\pi{\bf{k}}_2^2}\cr 
&\qquad\qquad\Sigma_{j_1j_2}\big({\rho\over x_1x_2},
{{\bf{k_1}}\over Q},{{\bf{k_2}}\over Q},\as(\mu^2)\big)
{\cal F}_{j_1}\big(x_1,{\bf{k}}_1^2,\mu^2\big)
{\cal F}_{j_2}\big(x_2,{\bf{k}}_2^2,\mu^2\big),
}}
again up to terms which vanish as inverse powers of the hard scale $Q$.
The dependence on $\mu$ will be suppressed in what follows: in practice 
we will take $\mu=Q$.

\subsec{Gluon dominance at high energy}

The range of $x$ and $Q$ relevant at lepton-hadron and hadron-hadron 
colliders (specifically HERA with $\sqrt{S}=320\GeV$, 
the Tevatron with $\sqrt{S}=1.8\TeV$, the LHC with $\sqrt{S}=14\TeV$ 
and a notional VLHC with $\sqrt{S}=200\TeV$) is shown in fig.1. It is
obvious from the figure that over most of the kinematic reach of these 
machines $x$ is small, and thus small $x$ logarithms 
are potentially large:
only for processes at the highest scales (for which the cross-sections are 
correspondingly small) can the small $x$ region be altogether excluded.

High energy processes thus usually involve collisions of small $x$ 
partons, and 
at small $x$ and high $Q^2$ it is now well known that 
all singlet parton distributions 
show a steep rise, that of the gluon distribution $G(x,Q^2)$ being 
driven by the nonabelian splitting $g\to gg$, and that of the singlet quark 
distribution $Q(x,Q^2)$ by $g\to q\bar{q}$ \refs{\DGPTWZ,\bfdas}. 
Because the latter process is $O(\as)$, at small $x$ and high $Q^2$ 
the singlet quark distribution is always 
smaller by a power of $\as(Q^2)$ than the gluon, i.e.  
$Q(x,Q^2)\simeq \as(Q^2) G(x,Q^2)$: at small $x$ most partons are 
gluons.

\topinsert
\vskip0.5truecm
\vbox{\hbox{\centerline{
\hskip1truecm
\epsfxsize=15truecm
\epsfbox{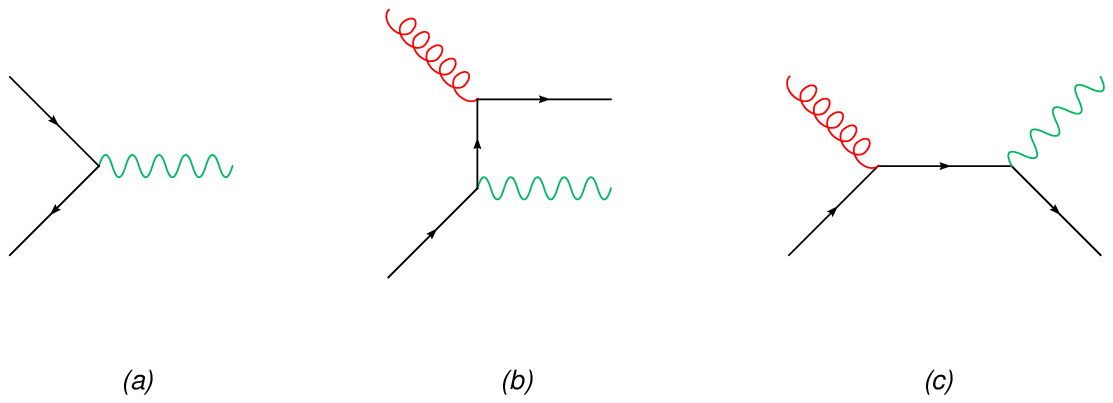}
}}
\vskip0.5truecm
\hbox{
\vbox{\footnotefont\baselineskip6pt\narrower\noindent Figure 2: the quark 
induced Drell-Yan process, or vector boson production 
(a) the $O(\alpha)$ $q\bar{q}$ annihilation 
process and (b),(c) the $O(\alpha\as)$ $qg$ (or $\bar{q}g$) 
fusion process, with initial and final state radiation respectively.
 }}\hskip1truecm}
\vbox{\hbox{\centerline{
\hskip1truecm
\epsfxsize=15truecm
\epsfbox{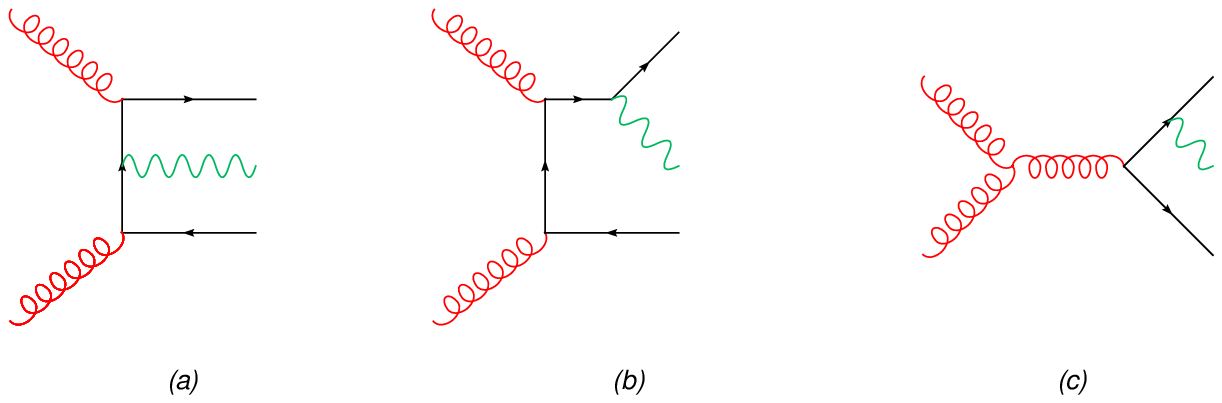}
}}
\vskip0.5truecm
\hbox{
\vbox{\footnotefont\baselineskip6pt\narrower\noindent Figure 3: the 
gluon induced 
Drell-Yan process, or vector boson production: the 
three main classes of contribution (a) with 
initial state radiation, (b) with both initial and final state radiation 
and (c) with final state radiation only. All these processes are formally 
$O(\alpha\as^2)$.
 }}
}
\endinsert 

This means that when we compute high energy partonic cross-sections 
we can no longer rely on the simple counting of powers of $\as$ as we do 
when $Q$ and $S$ are comparable. Consider for 
example a Drell-Yan type of process: formally the LO 
process is $q\bar{q}$ annihilation, fig.2(a), and is $O(\alpha)$, while 
the $qg$ (and $\bar{q}g$) scattering processes fig.2(b) and fig.2(c) 
are both NLO, i.e. $O(\alpha\as)$, and the $gg$ processes fig.3 
are all $O(\alpha\as^2)$ and thus NNLO. However at high energy, 
if we take into account 
the relative suppression of the quark relative to the gluon, all of 
these contributions are in practice $O(\alpha\as^2)$, and thus should 
be considered as leading order. Subleading contributions to the 
$q\bar{q}$ and $qg$ contributions in fig.2 are known \DYNNLO: the subleading 
contributions to the $gg$ processes fig.3 are not known at present, 
since formally they would be NNNLO.

In this paper we shall only consider processes with gluons in the initial 
state. In practice this means that we simply drop the summation over 
partons in eqns.\photofac,\hadrofac. This simplification is sufficient 
to discuss most of the issues in high energy resummation, but
will of necessity mean that our numerical results are estimates rather 
than calculations of complete cross-sections. 

\subsec{Double Mellin Transforms}

The convolutions over $x$ and $\bf{k}$ in the factorizations 
\photofac\ and \hadrofac\ may be undone by taking Mellin transforms with 
respect to $\rho$ and $Q$. Explicitly if we define 

\eqn\xsecMel{\Sigma_{\gamma {\rm h}}(N,M)= \int_0^1 \! {d\rho\over \rho}\rho^{N}
\int_0^\infty \! {dQ^2\over Q^2}\Big({Q^2\over \Lambda^2}\Big)^{-M}
\Sigma_{\gamma {\rm h}}(\rho,Q^2),}
while for the hard cross-section
\eqn\ghxsecMel{{\cal C}(N,M)= \int_0^1 \! {d\rho\over \rho}\rho^{N}
\int {d^2{\bf{k}}\over\pi {\bf{k}}^2}\Big({{\bf{k}}^2\over Q^2}\Big)^{M}
\Sigma_{\gamma g}\big({\rho},{{\bf{k}}\over Q}),}
then if the double Mellin transform of the unintegrated gluon distribution 
${\cal{G}}\equiv {\cal{F}}_g$ is 
\eqn\GMel{{\cal{G}}(N,M) = \int_0^1 \! {dx\over x}x^{N}
\int_0^\infty \! {d{{k^2}}\over{{k^2}}}
\Big({{{k^2}}\over \Lambda^2}\Big)^{-M} 
\, {\cal{G}}(x,{{k^2}}),}
the factorization \photofac\ becomes simply algebraic: the purely 
gluonic contribution to the transformed cross-section is
\eqn\photofacMel{\Sigma_{\gamma {\rm h}}(N,M) = {\cal C}(N,M){\cal{G}}(N,M).}

It is easier to make contact with phenomenology if instead of working with the
unintegrated gluon distribution we define the integrated distribution
\eqn\intglu{G(x,Q^2) = \int_0^{Q^2}{d{k^2}\over{k^2}}
{\cal G}(x,k^2).}
This is the distribution that would be proportional to the physical 
cross-section if the hard process were pointlike, i.e. the hard cross-section 
was simply proportional to $\Theta(Q^2-k^2)$.
Then in Mellin space 
\eqn\intvsunint{G(N,M) = M^{-1}{\cal G}(N,M),} 
and eqn.\photofacMel\ 
becomes
\eqn\photofacMelint{\Sigma_{\gamma {\rm h}}(N,M) = C(N,M)G(N,M).}
where $C(N,M)\equiv M {\cal C}(N,M)$. 

To recover the physical cross-section we 
invert the two Mellin transforms \xsecMel:
\eqn\photmelinv{\Sigma_{\gamma {\rm h}}(\rho,Q) = 
\int_{-i\infty}^{i\infty} {dN\over 2\pi i} e^{\xi N}
\int_{-i\infty}^{i\infty} {dM\over 2\pi i} e^{t M}C(N,M)G(N,M),}
where for clarity in future discussions we have defined 
\eqn\xitdef{ \xi \equiv \log 1/\rho,\qquad t\equiv\log Q^2/\Lambda^2.}
Here $N$ and $M$ are both complex variables: the contours  
in the integrations over $N$ and $M$ keep just 
to the right of the singularities near $N=0$ and $M=0$. 
The contour in $N$
is always closed on the left (since $\xi > 0$): 
the contour in $M$ is closed on the left in 
the ultraviolet ($t>0$), but on the right in the infrared ($t<0$). 

For hadronic processes we proceed similarly: defining 
\eqn\hhxsecMel{{\cal H}(N,M_1,M_2)= \int_0^1 \! {d\rho\over \rho}\rho^{N}
\int {d^2{\bf{k_1}}\over\pi {\bf{k}}_1^2}
\Big({{\bf{k}}_1^2\over Q^2}\Big)^{M_1}
\int {d^2{\bf{k_2}}\over\pi {\bf{k}}_2^2}
\Big({{\bf{k}}_2^2\over Q^2}\Big)^{M_2}
\Sigma_{g g}\big({\rho},{{\bf{k_1}}\over Q},{{\bf{k_2}}\over Q}),}
the factorization formula \hadrofac\ becomes
\eqn\hadrofacMelint{\Sigma_{\rm hh}(N,M_1,M_2) 
= H(N,M_1,M_2)G(N,M_1)G(N,M_2).}
where $H(N,M_1,M_2)= M_1 M_2 {\cal H}(N,M_1,M_2)$. The hadronic 
cross-section is thus
\eqn\hadmelinv{\Sigma_{\rm hh}(\rho,Q) = 
\int_{-i\infty}^{i\infty} \!{dN\over 2\pi i}\, e^{\xi N}
\int_{-i\infty}^{i\infty} \!{dM_1\over 2\pi i}
{dM_2\over 2\pi i}\,
 e^{t (M_1+M_2)}H(N,M_1,M_2)G(N,M_1)G(N,M_2),}
and we have to deal with functions of the three complex variables $N$, 
$M_1$ and $M_2$ integrated along three contours. Clearly $H(N,M_1,M_2)=
H(N,M_2,M_1)$.

\subsec{Duality at fixed coupling}

Perturbative expansions of the hard cross sections are contaminated by
logarithms of $Q$ and $\rho$, corresponding in Mellin space 
to inverse powers of $M$ and $N$ respectively. This may be seen directly 
by considering Laurent expansions around $M=N=0$ of the integrands 
of \photmelinv\ and \hadmelinv: every extra inverse power of $M$ or $N$ 
must be compensated by a positive power from expansion of the exponential, 
yielding an extra factor of $t$ or $\xi$. These logarithms are at most 
single logarithms, in the sense that in perturbation theory there is at 
most one extra logarithm of each type whenever there is an extra power of 
$\as$: a typical contribution to the integrand is $\as^lM^{-m}N^{-n}$ 
where $m,n\leq l$. To obtain meaningful results in perturbation theory 
these logarithms must be resummed (at LO, where $m+n=2l$, NLO, 
where $m+n=2l+1$, etc.) and factored into the gluon distribution $G(M,N)$.
In particular without resummation the GLAP splitting function is unstable 
in the small $x$ region, while the BFKL kernel is unstable in 
both the collinear and anticollinear regions, which means that for 
reasonable values of $\as$ it is also unstable in the small $x$ region 
\flprobs.

Resummation of the transverse and longitudinal logarithms 
$t=\log Q^2/\Lambda^2$ and $\xi = \log 1/\rho$ proceeds by solution of the
GLAP and BFKL equations respectively:
\eqn\glap{ \frac{dG}{dt} = \int_\rho^1 \! {dx\over x} 
\,P\big({\rho\over x},\as(Q^2)\big)
G\big(x,Q^2\big),}
where $P$ is the gluon splitting function, and 
\eqn\bfkl{\frac{d{\cal G}}{d\xi}=\int_0^\infty \! 
{dk^2\over k^2} 
\,{\cal K}\big({k^2\over Q^2},\as(k^2)\big)
{\cal G}\big(x,k^2\big),}
where ${\cal K}$ is the (angular averaged) BFKL kernel. 
Taking double Mellin transforms both equations 
simplify to algebraic equations, since the convolutions reduce to products 
and the derivatives to factors of $M$ and $N$ respectively:
\eqn\evolmel{\eqalign{MG(N,M) =& \, G_0(N)+\gamma(N;\ash)\,G(N,M),\cr
NG(N,M) = & \, \bar G_0(M)+M^{-1}\chi(M,\ash)M\,G(N,M),}}
where $\gamma$ and $\chi$ are the Mellin transforms of the respective kernels
\eqn\gamchi{\gamma(N;\as)\equiv  \int_0^1 \! {dx\over x}\, x^{N} 
\,P(x;\as),\quad
\chi(M;\ash)\equiv 
\int_0^\infty \! {dk^2\over k^2}\,\Big({k^2
\over Q^2}\Big)^{M} \,{\cal K}\big({k^2\over Q^2},\as(k^2)\big),}
and $G_0(N)$, $\bar{G}_0(M)$ are (nonperturbative) boundary conditions. The 
transform of the BFKL equation has been written in terms of the integrated 
distribution using eqn.\intvsunint.

The main complication in the derivation of eqns.\evolmel\ from 
eqns\glap,\bfkl\ is the running of the coupling: in Mellin space 
$\as(t)$ becomes an operator
\eqn\alfop{\ash\equiv \as\big(-\smallfrac{\partial}{\partial M}\big).}
This leads to various difficulties, discussed at length in 
refs\refs{\abfairy,\abfrunfin,\Liprun,\abfrundual,\mbffnnlo}. In the 
remainder of this section we sidestep this issue, and consider 
the coupling to be fixed.

At fixed coupling, $\ash\to\as$, and the evolution equations have the 
simple algebraic solution
\eqn\evolsol{G(N,M) = \frac{1}{M-\gamma(N;\as)}G_0(N)
 = \frac{1}{N-\chi(M;\as)}{\bar G}_0(M).}
Since the leading twist perturbative singularities can always be 
factorised from the singularities in the nonperturbative boundary conditions, 
the poles in the perturbative factors must coincide: at the pole
\eqn\pole{M=\gamma(N;\as),\qquad N=\chi(M;\as).}
The functions $\gamma$ and $\chi$ must thus satisfy the consistency conditions
\eqn\duality{M = \gamma(\chi(M;\as);\as),\qquad N = \chi(\gamma(N;\as);\as),}
i.e. as functions of their main arguments 
$\gamma = \chi^{-1}$, $\chi=\gamma^{-1}$.
These are the duality relations \refs{\bfafp,\colell,\jar,\bfsum}: 
$\chi$ determines the high energy ($N=0$) 
singularities of $\gamma$, just as $\gamma$ determines the collinear ($M=0$) 
singularities of $\chi$. For example if $\gamma \sim \as/N $, 
$\chi \sim \as/M$ and vice versa.
Conservation of longitudinal momentum implies that $\gamma(1,\as)=0$ to 
all orders in perturbation theory: duality then tells us that 
$\chi(0;\as)=1$ to all orders in perturbation theory, i.e. that the 
collinear ($M=0$) singularities in the expansion of 
$\chi(M;\as)$ in powers of $\as$ 
resum to unity \abfsxres. 

Using these ideas, and the symmetry under the exchange of $Q$ and $k$ of 
the BFKL kernel (which in Mellin space translates into a symmetry under 
$M\to 1-M$ \salam), it is possible to use the the known results for the 
GLAP anomalous dimension $\gamma(N;\as)$ at LO and NLO in powers 
of $\as$ to resum the collinear ($Q^2\gg k^2$, thus $M=0$) 
and anticollinear ($Q^2\ll k^2$, thus $M=1$) 
singularities in the BFKL kernel $\chi(M;\as)$ 
at LO and NLO. Indeed the resummation of these 
singularities is essential to obtain a meaningful expansion 
of the kernel for reasonable values of $\as$.
The resummed kernel then in turn through duality gives 
an anomalous dimension in which 
the high energy $N=0$ singularities are also resummed. This resummed 
anomalous dimension can then be used to evolve an initial (integrated) 
gluon distribution at small $x$ \refs{\abfsxres,\abfsxphen}.

The small-$x$ behaviour of the fixed coupling anomalous dimension (or 
rather its associated splitting function) is given by the behaviour 
around the minimum of $\chi(M,\as)$: fixed coupling duality implies that this
leads to a square root branch cut at $N=c(\as)\equiv\chi(\half;\as)$, at
which the anomalous dimension rises to one half:
\eqn\gamcut{\gamma(N;\as)\sim \half 
- \sqrt{\frac{N-c(\as)}{\half\kappa(\as)}},}
where $\kappa(\as)\equiv \chi''(\half,\as)$ is the curvature at the minimum.
This cut in turn gives rise to the famous $x^{-c(\as)}$ growth in the 
splitting function.

Fixed coupling duality may also be used to resum the high energy logarithms 
in hard cross-sections. Since all the 
collinear and high energy logarithms have now been absorbed into the 
integrated gluon distribution, the hard cross-sections are regular 
in both $N$ and $M$ close
to the origin, and may thus be Taylor expanded: for example
\eqn\xsectaylor{\eqalign{
C(N,M;\as) &= \sum_{l=1}^\infty\sum_{m=0}^\infty 
c^l_{m}(N)\as^lM^m,\cr
H(N,M_1,M_2) &= \sum_{l=2}^\infty
\sum_{m_1=0}^\infty\sum_{m_2=0}^\infty h^l_{m_1m_2}(N)
\as^lM_1^{m_1}M_2^{m_2},}}
where $c^l_m(N)$ and $h^l_{m_1m_2}(N)=h^l_{m_2m_1}(N)$ are also 
regular in the neighbourhood of $N=0$. The only singularity close to the origin
is thus that in $G(N,M)$, i.e. \evolsol, and this may be used to perform one 
of the  
photoproduction inverse Mellin transforms \photmelinv, or two of the 
hadroproduction inverse Mellin transforms \hadmelinv. 
The usual procedure is to perform
the integrals over $M$ in this way, leaving the single integral over $N$ to be 
done numerically: 
\eqn\factninvmel{\eqalign{
\Sigma_{\gamma {\rm h}}(\rho,Q) &= 
\int_{-i\infty}^{i\infty} {dN\over 2\pi i} e^{\xi N}
e^{t \gamma(N;\as)}C(N,\gamma(N;\as);\as)G_0(N),\cr
\Sigma_{\rm hh}(\rho,Q) &= 
\int_{-i\infty}^{i\infty} {dN\over 2\pi i}\, e^{\xi N}
 e^{2t\gamma(N;\as) }H(N,\gamma(N;\as),\gamma(N;\as);\as)
G_0(N)^2.}}

Note that since $\gamma(N;\as)$ contains resummed poles in $N$, (i.e. terms 
of the form $\as^lN^{-n}$, with $n\leq l$) this procedure
effectively resums the high energy singularities in the collinearly factorised 
hard cross-sections, obtained by expanding $C(N,\gamma(N;\as);\as)$ and 
in $H(N,\gamma(N;\as),\gamma(N;\as);\as)$ respectively in powers of $\as$
\refs{\CH,\cchglu,\bfsum}.
Conversely, comparing eqns.\factninvmel\ with the more conventional 
expressions 
\eqn\factninvmel{\eqalign{
\Sigma_{\gamma {\rm h}}(\rho,Q) &= 
\int_{-i\infty}^{i\infty} {dN\over 2\pi i} e^{\xi N}
e^{t \gamma(N;\as)}c(N;\as)G_0(N),\cr
\Sigma_{\rm hh}(\rho,Q) &= 
\int_{-i\infty}^{i\infty} {dN\over 2\pi i}\, e^{\xi N}
 e^{2t\gamma(N;\as) }f(N;\as) G_0(N)^2,}}
obtained in collinear factorization, we see that 
\eqn\fixedordersings{\eqalign{
c(N;\as) =& \sum_{l=0}^\infty \as^lc_l(N) = C(N,\gamma(N;\as);\as),\cr
f(N;\as) =& \sum_{l=0}^\infty \as^lf_l(N) = 
H(N,\gamma(N;\as),\gamma(N;\as);\as).}}
Thus by substituting the fixed order expansion 
$\gamma(N;\as) = \sum_{l=0}^\infty \as^l\gamma_l(N)$ into the expansions
\xsectaylor\ of the \rhs\ of \fixedordersings\ the high energy 
singularities of the fixed order hard cross-sections $c_l(N)$ and $f_l(N)$ 
may be economically computed. This makes matching to the fixed order 
calculations particularly straightforward. 

This is the procedure used in most previous studies 
\refs{\CH,\cchglu,\rdbrke,\higgs,\bfsum,\abfsxphen} of the effects of high 
energy resummation on high energy cross-sections. It leads to a strong 
growth in cross-sections, particularly hadronic cross-sections, due to 
infrared singularities at positive values of $M$ which, unlike the 
collinear singularities near $M=0$, have not been resummed into the 
perturbative evolution. It is the nature of these
infrared singularities, and the correct treatment of them, that is the main 
subject of this paper. First however we must consider how this simple 
picture is modified when the coupling runs.

\subsec{Duality at running coupling}

When the coupling runs the evolution equations \evolmel\ become 
differential equations in $M$, since the running coupling transforms 
to a differential operator \alfop. This operator commutes with $N$ but not 
with $M$: 
\eqn\comm{[\ash,M] = \beta(\ash),}
where $\beta(\as)=-\as^2\beta_0+\cdots$ is the QCD beta-function. 
This means that the BFKL 
operator $\chi(M;\ash)$ must be defined very carefully: different
orderings of $\ash$ and $M$ will give different results, because the 
arguments of the couplings will be different \abfrundual.

It was show some time ago both by saddle point arguments 
for the Mellin inversion \ccfac\ and by explicit solution of 
the $M$-space differential equation \bfsxap\ that when the coupling runs, 
the naive duality \duality\ is modified by terms 
proportional to $\beta$: for example at NLLx, writing 
$\chi(M,\as) = \as\chi_0(M)+\as^2\chi_1(M)+\cdots$,
\eqn\NLLxrun{\chi_1\to \chi_1+\beta_0 \frac{\chi_0''\chi_0}{2\chi_0'},}
where the primes denote derivatives with respect to $M$. 
In ref.\abfrundual\ it was shown that the fixed 
coupling duality \duality\ may be extended formally to relations 
between the operators $\gamma(N;\ash)$ and $\chi(M;\ash)$,
which may then be used to generate systematically
running coupling corrections using a purely algebraic algorithm, 
order by order in perturbation theory. 
This technique was used recently \mbffnnlo\ to compute an estimate of the 
leading twist BFKL kernel $\chi_2$ at NNLLx.

When the coupling runs it becomes important to specify carefully 
the factorization scheme, since a redefinition of the gluon 
distribution $G(N,M)\to Z(M)G(N,M)$ changes $\chi(M,\ash)$ by a commutator:
\eqn\schemechange{\chi(M,\ash)\to \chi(M,\ash)+Z^{-1}(M)[\chi(M;\ash),Z(M)].} 
It is thus possible to shuffle the running coupling terms between the 
evolution and the hard cross-section. In this paper we will use 
the \QMS\ scheme \refs{\ciafqz,\ccsfact,\ccfact}, a variant on the 
\MS\ scheme in which all the running coupling terms are absorbed 
into the evolution. The two schemes are equivalent in fixed order 
perturbation theory at LO, NLO and NNLO, but begin to differ at NNNLO.

Note that provided $Z(M)$ is regular at $M=0$ the second term on 
the \rhs\ of \schemechange\ is subleading compared to the first, 
since the commutator necessarily introduces an extra power of $\ash$ 
(see \comm, and note that $\beta(\as)=-\beta_0\as^2+O(\as^3)$). This means
that if we work at NLLx in the resummation of the singularities, it is 
only necessary to consider the hard cross-sections $C(N,M;\ash)$ and 
$H(N,M_1,M_2;\ash)$ at LLx: terms at NLLx in the cross-section are 
NNLLx in the evolution \refs{\CH,\bfsxap}. Similarly to this order 
we may ignore the running of $\alpha_s$ in the hard cross-section, setting
$\ash\to\as$ up to subleading terms.

\topinsert
\vbox{
\vskip-7truecm
\epsfxsize=22truecm
\centerline{\hskip2truecm\epsfbox{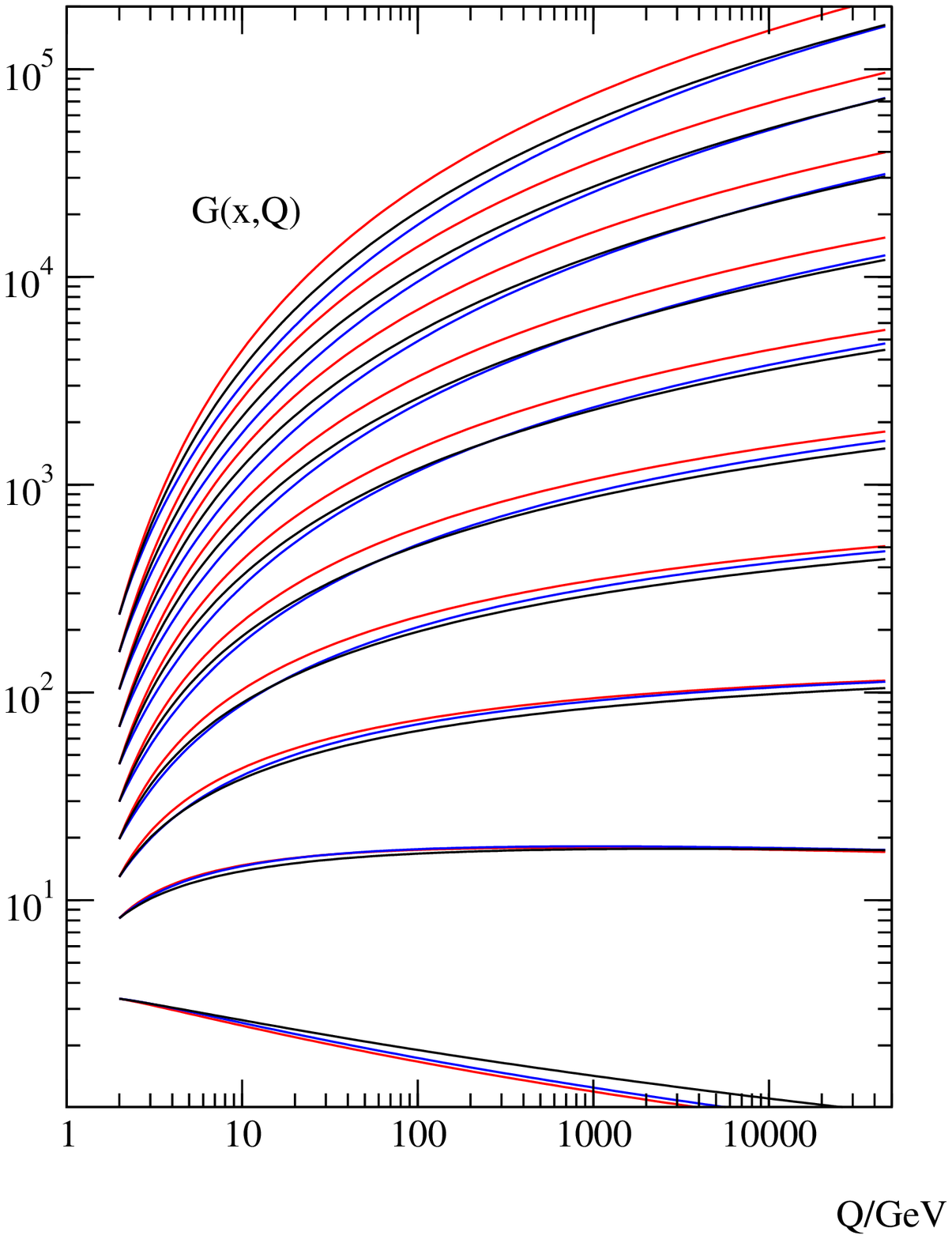}}
\vskip-1.0truecm
\bigskip
\hbox{
\vbox{\footnotefont\baselineskip6pt\narrower\noindent Figure 4: The gluon
distribution $G(x,Q)$ plotted against $Q$ in $\GeV$ for $x=10^{-10},10^{-9},
\ldots,0.01,0.1$ (from top to bottom). 
The blue curves are 
evolved with the NLO resummation
described in the text: the black and red curves  
are with LO and NLO GLAP evolution respectively.
 }}\hskip1truecm}
\endinsert 

Away from the region of very small $x$ the running coupling corrections to 
naive duality are small. However in the 
small $x$ limit, they become very large. This may be seen immediately from
\NLLxrun: at small $x$ and fixed coupling, the integrals over $M$ are 
dominated by the minimum of $\chi(M,\as)$, and thus by $M\sim\half$, 
ie the region in which $\chi_0'(M)\sim 0$ and the correction \NLLxrun\  
becomes large. It was thought at one time that these terms become so large
that they drive instabilities in the gluon distribution leading to 
negative cross-sections \refs{\ciafrc,\ccsfact,\inst,\thorne}. However these 
instabilities are due to diffusion into the infrared: if the singularities are
resummed, they may be factorised into the (nonperturbative) initial 
distribution, resulting in stable evolution \refs{\abfairy,\abfrunfin,
\ciafsplit}.

This further resummation of the running coupling singularities at $M=\half$ is
accomplished by expanding around the minimum of $\chi$, solving for $G(N,M)$ 
and then performing the inverse Mellin with respect to $M$ exactly, rather 
than using a saddle point expansion. The simplest version of this argument 
gives rise to an Airy function \Liprun: a full calculation \abfrunfin\ 
summing up the leading singularities in $\beta_0$ gives
the running coupling resummed anomalous dimension
\eqn\batandim{\gamma_B(N;\as)=
\half-\beta_0\bar\as 
+\frac{1}{A}\frac{K_{2B}^\prime((\beta_0\as
A)^{-1})}{K_{2B}((\beta_0\as A)^{-1})},
}
where $A$ and $B$ are simple functions of $\as$ and $N$ computed from the 
value of $\chi(M,\as)$ and its curvature at the minimum, and $K_\nu(x)$ 
is the Bateman function. The small $x$ behaviour resulting from \batandim\ 
is qualitatively different from that obtained with the fixed coupling  
result \gamcut, since the cut is replaced by a simple pole
located at $N=c_B(\as)$, with $c_B(\as)$ given by the rightmost zero 
of the Bateman function. Since $c_B(\as)$ is rather less than $c(\as)$,
the $x^{-c(\as)}$ growth of the splitting function at 
small $x$ is softened by the running coupling effects to 
$x^{-c_B(\as)}$ \abfrunfin. 
However since the new singularity is now a pole and not a cut 
$M$ can now grow indefinitely rather than saturating at $M=\half$: 
in effect the region between $M=0$ 
and $M=\half$ is stretched to infinity so that the effective $\chi(M;\as)$ 
(the naive dual of $\gamma(N;\as)$) is analytic for all $M>0$.

By combining the running coupling resummation with a 
collinear and anticollinear resummation of $\chi(M;\as)$ using running 
duality, a completely resummed anomalous dimension 
may be computed, and used to resum both 
high energy and collinear singularities in the gluon distribution 
at LO and NLO  \refs{\abfrunfin,\ciafsplit}. 
Such a NLO resummed gluon distribution 
is plotted in fig.4 for $x$ down to $10^{-10}$ and $Q$ up to $50~\TeV$. 
The initial distribution was chosen to be 
proportional to $x^{-0.18}(1-x)^5$ at $2~\GeV$, with the first moment 
(i.e. the integrated longitudinal momentum) normalised to unity. We
take $\as(m_Z)=0.118$, $\as(Q)$ evaluated using standard two loop running
with thresholds at $m_b$ and $m_t$.
For comparison we also plot the same distribution evolved using the 
LO and NLO GLAP anomalous dimensions. All evolution is performed with 
$n_f$ set to zero in the anomalous dimensions, 
consistent with our suppression of all quark induced processes. 
It can be seen from the plot that the effect of 
the resummation is modest, even over such a wide kinematic range, and 
is generally such as to smoothly soften the growth at small $x$ 
and large $Q^2$.

\subsec{Soft Singularities in Hard Cross-sections}

All that remains to be done is to combine this resummed 
gluon distribution with a hard cross-section. However when 
the coupling runs it is no longer so easy to justify 
the use of the pole approximation \evolsol\ to perform the 
integrals over $M$ in
\photmelinv\ and \hadmelinv\ as we did at fixed coupling, eqn.\factninvmel, 
since for the gluon distribution itself we know 
that at small $x$ the pole approximation no longer gives the correct 
asymptotic behaviour. Moreover, since when the coupling runs the 
resummed anomalous dimension $\gamma(N,\as)$ rises indefinitely as 
$N$ decreases, the position $M=\gamma(n,\as)$ of the pole in the evolved 
gluon will also 
grow large, and thus eventually the contour (which is always to the 
right of this pole) will become entangled with the various 
infrared singularities in the hard cross-sections. 

\topinsert
\vbox{
\hbox{\centerline{
\hskip1truecm
\epsfxsize=3truecm
\epsfbox{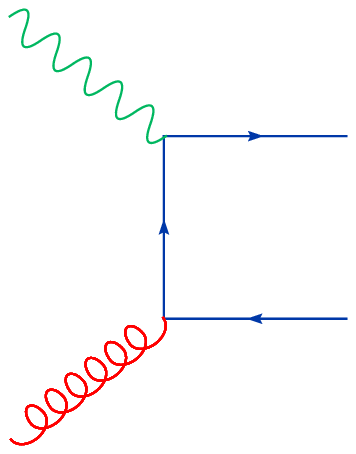}
}}
\vskip1.0truecm
\hbox{
\vbox{\footnotefont\baselineskip6pt\narrower\noindent Figure 5: 
Photoproduction of heavy quarks. The photon is on-shell, while the gluon is 
off-shell by an amount $k^2$, in order to calculate the impact factor 
$C(N,M)$. In electroproduction the photon would also be taken off-shell.
 }}\hskip1truecm}
\endinsert 

Consider first the photoproduction or electroproduction 
hard cross-section $C(N,M)$, which may be computed at leading order 
from the graph shown in fig.5. The hard scale $Q$ is then the mass of the 
quark (for photoproduction of heavy quarks), or the photon virtuality 
(for electroproduction). The incoming gluon has virtuality $k^2$. 
At very high energies $S\gg Q^2$, relevant values of $N$ should be small, 
so can in the first instance be neglected. 
Then considered as a function of $M$, $C(0,M)$ 
is regular near $M=0$ (since the collinear singularities, with 
$Q^2\gg k^2$, have already been absorbed in the gluon distribution),
but has poles at $M=-1,-2,\ldots$ and $M=1,2,3,\ldots$.
The poles at negative values of $M$ correspond to higher twist singularities:
they lead to power corrections in inverse powers of $Q^2$, which are 
not relevant here. The poles at positive values of $M$ correspond instead to 
process dependent infrared (anticollinear) 
singularities with $Q^2\ll k^2$. It is these
singularities that enhance the cross-section at high energy.

When the coupling is fixed, the anomalous dimension for the gluon evolution 
saturates at 
the cut, and thus at high energy the dominant contribution to the $M$ integral 
is from the region $M=\half$. The resummation $K$-factor may be then be 
estimated at high energy to be 
$\smallfrac{C(0,1/2)}{C(0,0)}=\smallfrac{27\pi^2}{56}\sim 4.8$, a 
significant enhancement \cchglu.
However when the coupling runs, there is no such saturation, and the pole 
approximation to the integration over $M$ breaks down at high energy, 
since the pole at $M=\gamma(N,\as)$ in the evolved gluon can then 
approach the pole 
at $M=1$ in the hard cross-section, and the contour gets pinched between them.
We must then attempt to perform the
integration exactly, and afterwards factor out the gluon distribution: 
resummation of the hard cross-section is thus rather more difficult than it 
was at fixed coupling.

\topinsert
\vskip0.5truecm
\vbox{\hbox{\centerline{
\hskip1truecm
\epsfxsize=10truecm
\epsfbox{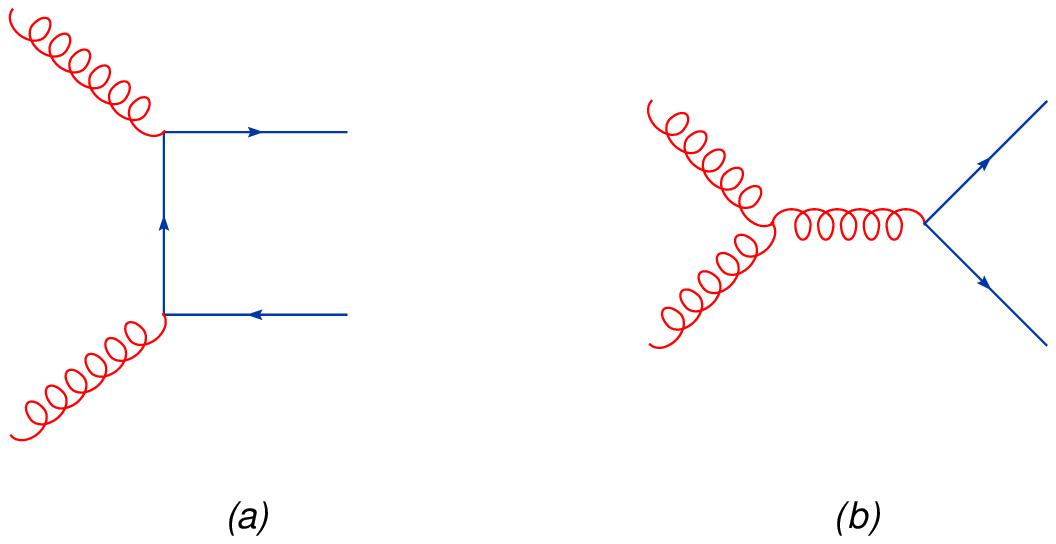}
}}
\vskip0.5truecm
\hbox{
\vbox{\footnotefont\baselineskip6pt\narrower\noindent Figure 6: 
Hadroproduction of heavy quarks: (a) the ``abelian'' diagram, with 
initial state radiation and (b) the ``nonabelian'' diagram with final 
state radiation. Both incoming gluons are taken
off-shell by amounts $k_1^2$ and $k_2^2$ in order to 
calculate the impact factor $H(0,M_1,M_2)$. The same graphs give 
hadroproduction of quark jets.
}}\vskip0.0truecm}
\vskip0.5truecm
\vbox{\hbox{\centerline{
\hskip1truecm
\epsfxsize=10truecm
\epsfbox{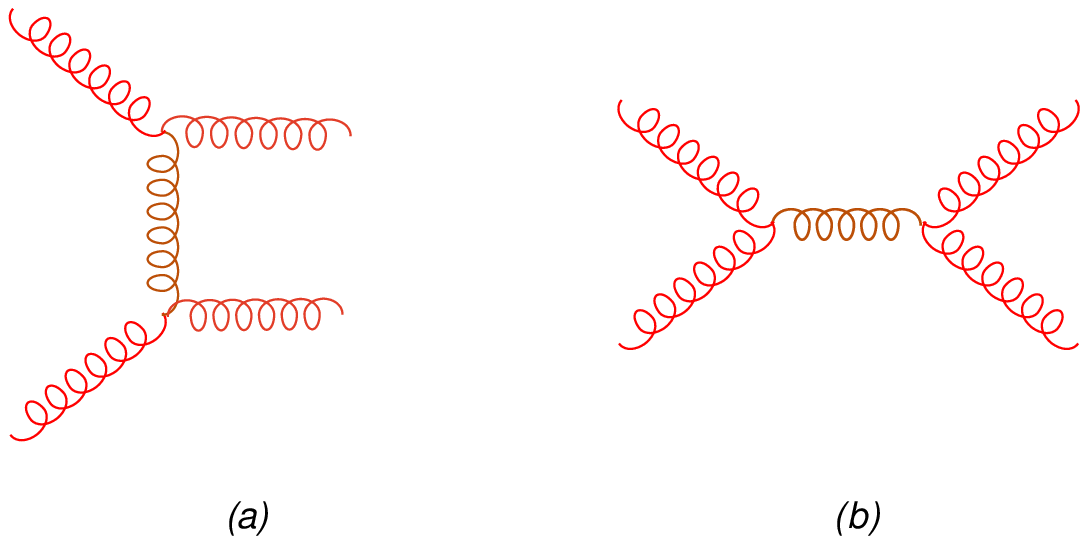}
}}
\vskip0.5truecm
\hbox{
\vbox{\footnotefont\baselineskip6pt\narrower\noindent Figure 7: 
Hadroproduction of gluonic jets: (a) the ``abelian'' diagram with initial 
state radiation and (b) the ``nonabelian'' diagram with final state 
radiation. Both incoming gluons are taken
off-shell. The two final state gluons are on-shell, with a specified $p_T$.
The four gluon vertex diagram is not shown, since it is nonsingular.
}}\vskip0.0truecm}
\endinsert 

This difficulty 
becomes much worse when we consider hadroproduction. Once again 
we focus on the singularities of the impact factor 
$H(0,M_1,M_2)$ in the double-Mellin plane of 
$M_1$ and $M_2$. This may be obtained from the graphs in fig.6 
(hadroproduction of heavy quarks, so $Q$ is the quark mass), or fig.6 and 
fig.7 (inclusive jets, so $Q$ is the transverse momentum of the jets).
Again the region around the origin is regular, and there are higher twist
poles at $M_1,M_2 = -1,-2,\ldots$ and infrared singularities at 
$M_1,M_2=1,2,\ldots$. However now we also have lines of infrared singularities 
at  $M_1+M_2=1,2,\ldots$: for example the heavy quark hadroproduction 
impact factor
\eqn\triplepole{H(M_1,M_2)\sim \as^2 \frac{\pi}{16}\frac{1}{(1-M_1-M_2)^3},}
when $M_1+M_2 \sim 1$ \refs{\cchglu,\rdbrke,\ccfac}. 
It is easy to see why this occurs by noting the form 
of the Mellin transform eqn.\hhxsecMel: an infrared singularity when 
$Q^2\ll k_1^2,k_2^2$ will become a singularity at $M_1+M_2=1,2,\ldots$. 

The degree of this singularity will depend on the nature of 
the infrared singularities of the individual graphs making 
up the process. The graphs with initial state radiation,
fig.6a and fig.7a, both have a single (anti)collinear singularity 
similar to that in photoproduction fig.5. However the singularity 
structure of the nonabelian graphs with final state 
radiation fig.6b and fig.7b
is more complicated. Firstly there is a soft singularity when the 
internal (timelike) gluon goes on-shell: the denominator of the propagator is 
$s=x_1x_2S-({\bf k_1}+{\bf k_2})^2$, and the singularity arises when 
$x_1x_2S\simeq ({\bf k_1}+{\bf k_2})^2 \ll {\bf k}_1^2\simeq {\bf k}_2^2$ 
\cchglu. On top of this there is the usual collinear (heavy quark 
production) or soft (jet production) singularity
for the emission of the final state partons, so altogether we find the 
triple pole eqn.\triplepole. Note that this kind of singularity is thus 
generic to most gluon-gluon hard cross-sections. In fact it is 
the dominant singularity, providing most of the cross-section 
at very high energy.

To evaluate the cross-section, we must perform the double Mellin inversion 
eqn.\hadmelinv. When the coupling is fixed, $M_1$ and $M_2$ both 
try to saturate at $M_1=M_2=\half$, and thus touch the singularity. 
This produces a strong enhancement of the cross-section \cchglu, 
stronger even than the BFKL growth. If $M_1$ and $M_2$ can grow beyond one 
half, as one expects at running coupling, the contours are pinched, and 
the cross-section seems to grow as fast as $\rho^{-2\as}$ \rdbrke. 
However this dramatic growth is also very unstable, 
in particular to $N$ dependent corrections, since these split 
the triple pole into a separate 
double pole (at $M_1+M_2=1+N$) and a single pole (at $M_1+M_2=1$).
This instability is a sure sign that the resummation is not under control. 
Indeed, when the coupling runs, the whole line of singularity 
become accessible (since both $M_1$ and $M_2$ may become large), and 
furthermore becomes entangled with
the collinear regions (since at $M_1\sim 0$, $M_2\sim 1$ and vice versa). 

So somehow we need a more reliable way of computing the integrals over $M_1$ 
and $M_2$. Since the singularity \triplepole\ does not 
factorise into a function of $M_1$ times a function of $M_2$, we 
actually need to perform both integrals simultaneously, and 
then factorise out the gluon distributions afterwards to obtain the 
resummation of the hard cross-section.

In the rest of this paper we will show how a simple trick may be used to solve
this problem, first for photoproduction and electroproduction 
(Section.~3) and then for hadroproduction (Section.~4). We consider 
processes with a single incoming gluon first because 
they are simpler: the more important hadroproduction processes will then be 
dealt with by a straightforward extension of essentially the same idea.

\newsec{Photoproduction and Electroproduction}

\subsec{Evaluating the Cross-section}

Consider first a photoproduction or electroproduction 
cross-section at high energy, and thus 
small $\rho$, but still far from the high energy limit. Then we might expect 
$M$ to remain ``small'' in some sense, and consider approximating the 
hard cross-section $C(N,M;\as)$ by the first few terms in its Taylor expansion 
\xsectaylor. The cross-section \photmelinv\ may then be written
\eqn\photnmelinv{\Sigma_{\gamma {\rm h}}(\rho,Q) = 
\int_{-i\infty}^{i\infty} {dN\over 2\pi i} 
e^{\xi N}\Sigma_{\gamma {\rm h}}(N,t),}
where
\eqn\photmelinvtaylor{
\Sigma_{\gamma {\rm h}}(N,t)=\as(t)\sum_{m=0}^\infty c^1_m(N)
\int_{-i\infty}^{i\infty} {dM\over 2\pi i} e^{t M}M^m G(N,M),}
where we have kept only the leading term in the expansion in $\as$, and 
(optimistically) changed the order of integration and 
summation over $m$, in order to do the integrals term by term. However for the 
first term this is an integral we already know: 
\eqn\Gevol{G(N,t) = \int_{-i\infty}^{i\infty} {dM\over 2\pi i} e^{t M} G(N,M),}
and consequently for all $m=0,1,2,\ldots$
\eqn\powtrick{\int_{-i\infty}^{i\infty} \frac{dM}{2\pi i}e^{tM}
\, M^m G(N,M) = \frac{\partial^m}{\partial t^m}G(N,t).}
However the general solution of the evolution equation \glap\ is simply
\eqn\gluevol{G(N,t)=\exp\Big(\int_0^t dt' \gamma(N,\as(t')) \Big) G_0(N),}
so all the partial derivatives in \powtrick\ may be evaluated in terms of the 
anomalous dimension and its derivatives:
\eqn\tderivs{\eqalign{\frac{\partial}{\partial t}G(N,t)&=
\gamma(N,t)G(N,t),\cr
\frac{\partial^2}{\partial t^2}G(N,t)&=(\gamma^2+ \dot{\gamma})G(N,t),\cr
\frac{\partial^3}{\partial t^3}G(N,t)&=
(\gamma^3+3\gamma \dot{\gamma}+\ddot{\gamma})G(N,t),\ldots}}
where the dot denotes partial derivatives with respect to $t$.

\topinsert
\vbox{
\vskip-0.5truecm
\hbox{\centerline{
\hskip1truecm
\epsfxsize=10truecm
\epsfbox{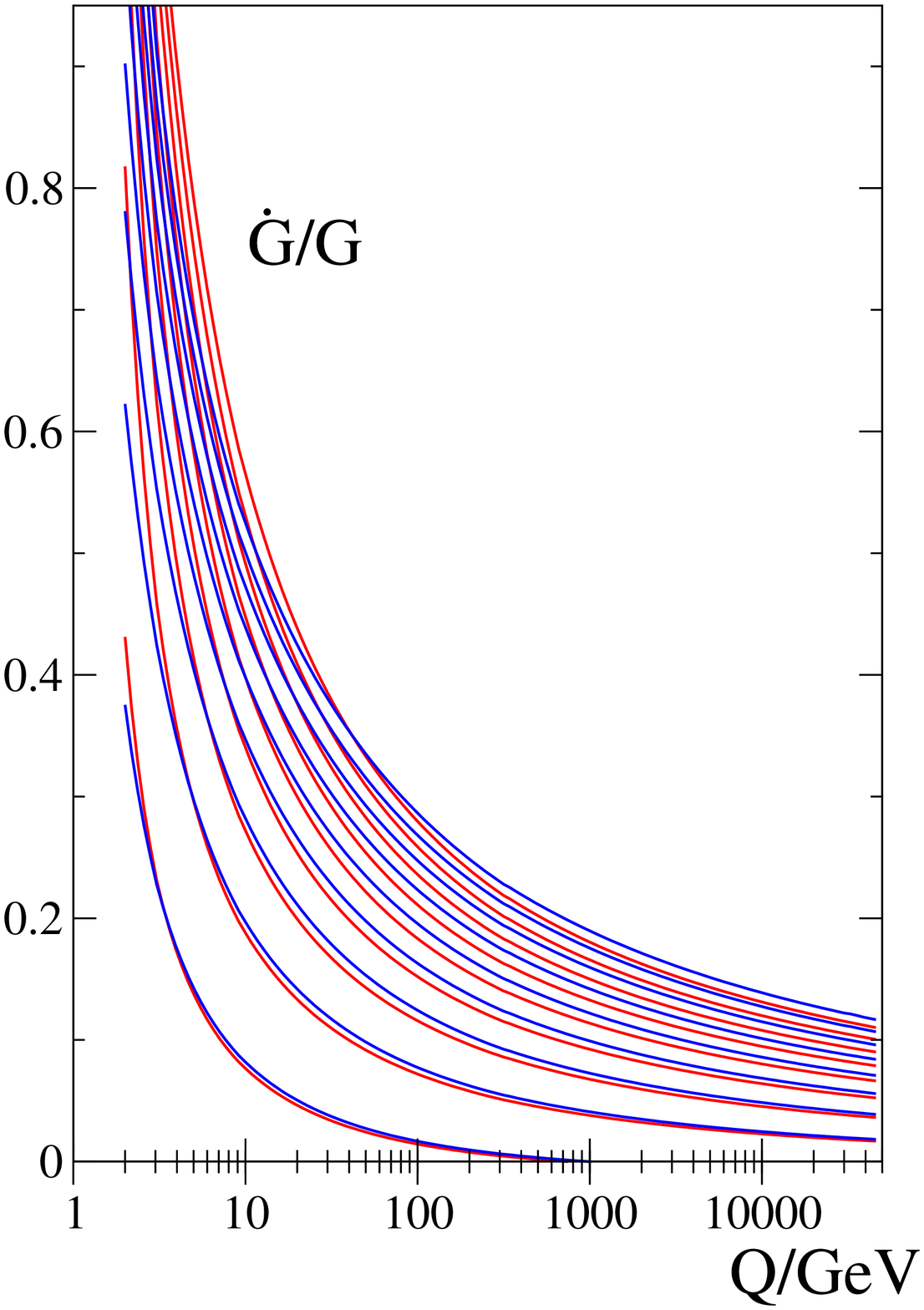}
\hskip-1truecm
\epsfxsize=10truecm
\epsfbox{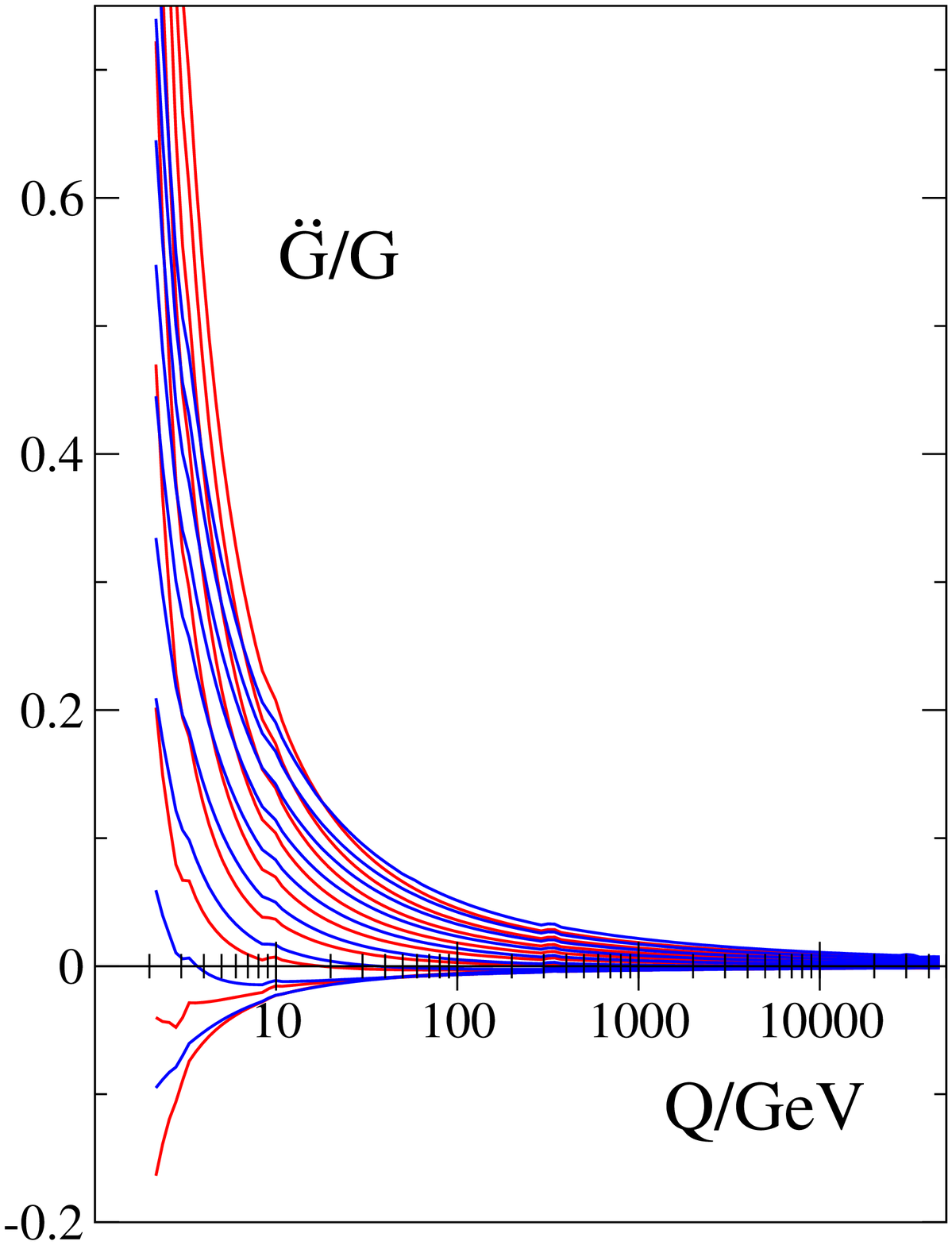}
}}
\hbox{
\vbox{\footnotefont\baselineskip6pt\narrower\noindent Figure 8: The first 
(left) and second (right) derivatives with respect to $t$ of the gluon 
distributions $G(x,Q)$ plotted in fig.4, normalised to $G$, plotted 
against $Q$ in $\GeV$, for $x=10^{-10},10^{-9},\ldots 0.01$ (from top 
to bottom). Again the blue curves are with NLO resummed evolution, the 
red with NLO GLAP evolution.
 }}\hskip1truecm}
\endinsert

Note that $\dot{\gamma}$ is formally subleading compared to $\gamma$, since 
\eqn\partialt{\frac{\partial}{\partial t} 
= \beta(\as)\frac{\partial}{\partial\as} 
= -\beta_0\as^2 \frac{\partial}{\partial\as},}
so every time we differentiate with respect to $t$ we add a power of $\as$.
So up to subleading terms partial derivatives of $G(N,t)$ with respect to 
$t$ simply result in powers of $\gamma(N;\as(t))$, and combining 
eqn.\photmelinvtaylor\ with \powtrick\ we find
\eqn\photsmallm{
\Sigma_{\gamma {\rm h}}(N,t)=\as(t)\sum_{m=0}^\infty c^1_m(N)
\frac{\partial^m}{\partial t^m}G(N,t)=\as(t)C(N,\gamma(N,\as(t)))G(N,t).
}
So provided we are not close to a pole, we find the same result 
when the coupling runs as we found at fixed coupling case 
through the pole dominance argument \factninvmel. 

Formally this method works provided the series \photsmallm\ converges. 
To see whether this is likely in practice, 
we plot in fig.8 the first and second derivatives of the gluon 
distributions fig.4, normalised to $G$: indeed the derivatives are 
much less than one for all but the lowest values of $x$ and $Q$. 
Moreover the second derivatives are much smaller than the first, 
suggesting that the series converges rather rapidly. In addition derivatives 
for the NLO resummed gluons are rather smaller at low $x$ and $Q$ 
than those for the NLO 
GLAP gluons, essentially because they evolve rather more slowly.
It follows that at all but the smallest $x$ and $Q$ we may use Taylor 
expansions in $M$, and we will show repeated examples of this in what follows.
It is worth noting in parenthesis that Taylor expansion in $M$ is little 
different at high energy to the usual fixed order expansion of the hard 
cross-section \fixedordersings, since it is only 
through the $M$ dependence of $C(N,M;\as)$ that
small $N$ singularities can be introduced, and at high energy it is these 
that dominate the cross-section. 

\subsec{Resumming an Infrared Singularity}

Now consider what happens when $C(N,M;\as)$ has a pole near $M=1$, 
so that the Taylor expansion around $M=0$ has radius of convergence one. This 
will generally be the case: such a singularity corresponds to 
logarithms of $k^2/Q^2$ as $k^2\to 0$. We will then need to do integrals 
of the form
\eqn\photmelinvpole{
\Sigma^{n}_{\gamma {\rm h}}(N,t)
=\int_{-i\infty}^{i\infty}\! {dM\over 2\pi i}\, 
\frac{1}{(1-M)^n}e^{t M} G(N,M),}
for $n=1,2,\ldots$. We can perform these integrals exactly using a 
simple trick \schwinger: we write
\eqn\expsing{\frac{1}{(1-M)^n} = \frac{1}{n!}\int_0^\infty \!d\tau\, 
\tau^{n-1}e^{-\tau(1-M)},}
thereby exponentiating the dependence on $M$ and transferring possible 
singularities in the complex variable $M$ to singularities in the 
integration over the real variable $\tau$. Substituting \expsing\ into 
\photmelinvpole\ the $M$ integration is now indeed trivial:
\eqn\exptrick{\eqalign{
\Sigma^{n}_{\gamma {\rm h}}(N,t)&= \int_{-i\infty}^{i\infty} 
\!\frac{dM}{2\pi i}\,
\frac{1}{n!}\int_0^\infty \!d\tau\, \tau^{n-1}e^{-\tau+M(t+\tau)}G(N,M)\cr
&=\frac{1}{n!}\int_0^\infty \!d\tau\, \tau^{n-1}e^{-\tau}G(N,t+\tau)
,}}
where in the second line we exchanged the order of the integrations over $M$
and $\tau$ and performed the integration over $M$. 
Since when the coupling runs the growth of $G(N,t)$
with $t$ is rather gentle (no greater than a power of $t$), 
the $\tau$ integral converges 
for all $t$. At small $x$ the $M=1$ singularity thus always leads to a 
modest enhancement of $G(x,t)$, since the growth with $t$ is monotonic. 

We may factorise this result using eqn.\gluevol: this gives
\eqn\exptrickfact{\Sigma^{n}_{\gamma {\rm h}}(N,t) 
=\Bigg[\frac{1}{n!}\int_0^\infty \!d\tau\, \tau^{n-1}
\exp\Big(-\tau+\!\int_0^\tau \! dt' \gamma(N,\as(t+t')) \Big)\Bigg]
\,G(N,t).}
The expression in square brackets is thus the resummed hard cross-section 
$C(N;\as(t))$.
 
Note that if we Taylor expand $G(N,t+\tau)$ in \exptrick\ in powers of $\tau$
\eqn\exptrickexp{\eqalign{\frac{1}{n!}\int_0^\infty d\tau \tau^{n-1}
e^{-\tau}G(N,t+\tau)
&=\int_0^\infty d\tau\sum_{m=0}^\infty \frac{\tau^{m+n-1}}{n!m!} 
e^{-\tau}\frac{\partial^m}{\partial t^m}G(N,t)\cr
&=\sum_{m=0}^\infty \frac{(m+n)!}{n!m!}\frac{\partial^m}{\partial t^m}G(N,t),}}
where in the second line we exchanged the order of summation and integration 
in order to perform the integrals over $\tau$. The result is of course 
precisely what one gets by first Taylor expanding $1/(1-M)^n$ in 
eqn.\photmelinvpole\ about $M=0$ and then using eqn.\powtrick. However 
the resulting series is useful only if $\gamma(N,t)$ and its derivatives 
are sufficiently small, as discussed in the previous section, whereas 
the integral representation eqn.\exptrick\ always converges. When the series
diverges, the integral representation resums it. 

It is instructive to consider explicitly two analytic examples of how 
this works in practice. First consider what happens at fixed coupling: then
at leading order $G(N,t)= e^{\gamma(N;\as) t}G_0(N)$, so taking 
$n=1$ for simplicity
\eqn\exptrickfc{\Sigma^{1}_{\gamma g}(N,t)
=\int_0^\infty \!d\tau\, 
e^{-\tau}e^{\gamma(N;\as) (t+\tau)}G_0(N)
= \frac{1}{1-\gamma(N,\as)}G(N,t).}
This is precisely the result expected from the pole dominance argument
described in sec.2.4: $M$ in the hard cross-section is replaced by 
$\gamma(N;\as)$ according to eqn.$\pole$. So the pole at $M=1$ results in 
a new pole when $\gamma(N;\as)=1$, to the right of the rightmost pole 
in $\gamma(N;\as)$: for example if $\gamma(N;\as)=\as/N$, the new pole is at
$N=\as$. Thus on performing the $N$ integration \photnmelinv\ the 
cross-section $\Sigma(\rho,Q)$ will exhibit a powerlike enhancement at 
high energy, growing faster than $G(\rho,Q)$.
\eqnn\kexact\eqnn\kapprox
\topinsert
\vskip-11.5truecm
\vbox{
\epsfxsize=18truecm
\centerline{\epsfbox{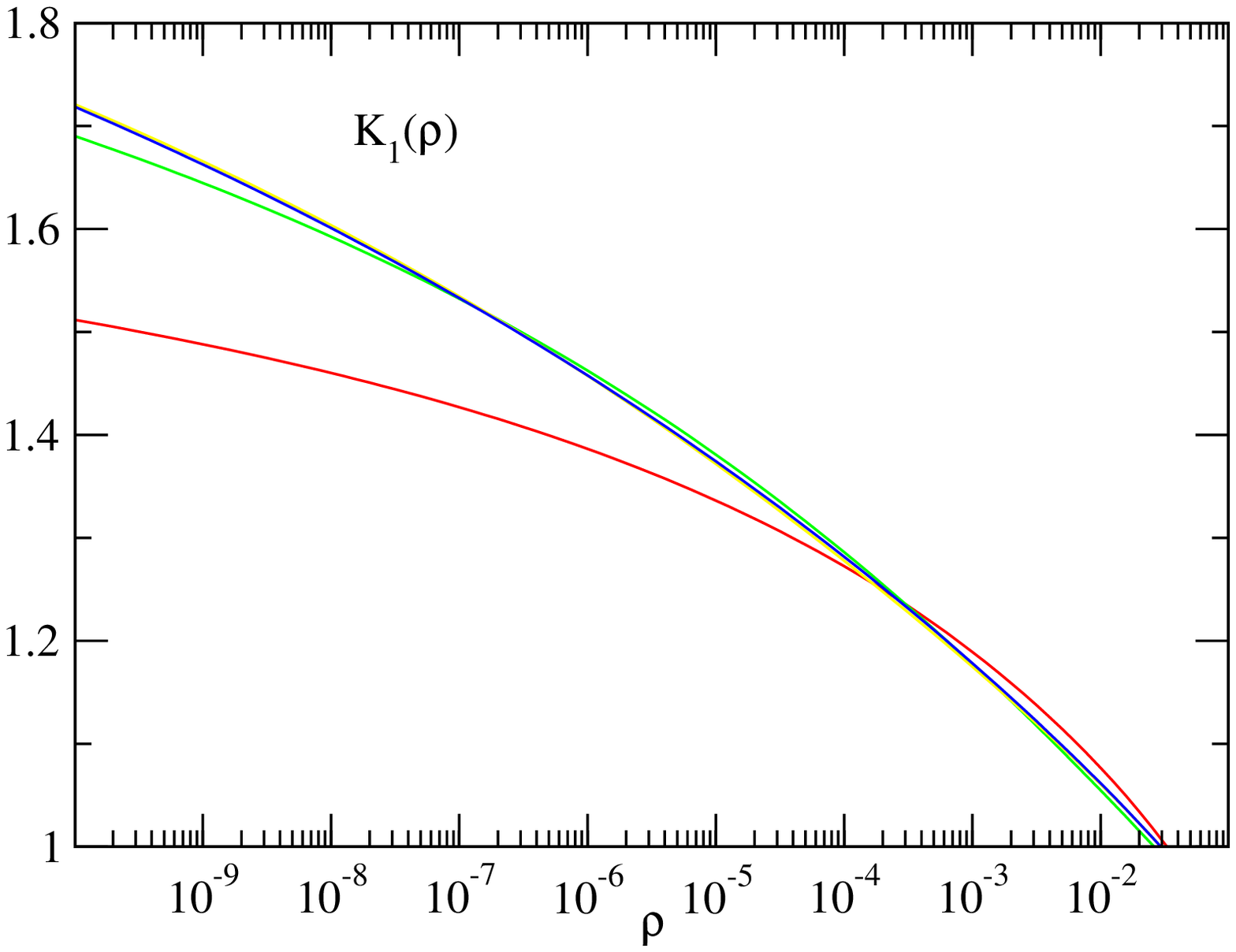}}
\vskip-2.0truecm
\bigskip
\hbox{
\vbox{\footnotefont\baselineskip6pt\narrower\noindent Figure 9: a generic 
photoproduction $K$-factor for a simple pole $1/(1-M)$. The blue (upper) line
corresponds to the exact expression eqn.\kexact, while the red, green 
and yellow lines (from bottom to top) are the NLO, NNLO 
and NNNLO approximation to it computed 
by including the second, third and fourth terms respectively of the 
series \kapprox. The hard scale is set at $10 \GeV$.}}
\vskip0.0truecm}
\endinsert 

Now consider what happens instead when the coupling runs: if we take for 
example $\gamma(N;\as(t))=\as(t)\gamma_0(N)$, and $\as(t)=1/\beta_0 t$, then
$G(N,t)= t^{\gamma_0(N)/\beta_0}G_0(N)$, and 
\eqn\exptrickrc{\Sigma^{1}_{\gamma {\rm h}}(N,t)=
\int_0^\infty d\tau 
e^{-\tau}(t+\tau)^{\gamma_0(N)/\beta_0}G_0(N) 
= t^{-\gamma_0(N)/\beta_0}e^t \Gamma(1+\beta_0^{-1}\gamma_0(N),t)G(N,t).}
where $\Gamma(a,t)\equiv \int_t^\infty ds s^{a-1}e^{-s}$ is the 
incomplete Gamma function. The only finite singularity in $\Gamma(a,t)$
is a branch cut from $t=0$ down the negative real axis: for $t$ real and 
positive, $\Gamma(a,t)$ is entire in $a$. Consequently the only singularities
of $\Sigma^{1}_{\gamma {\rm h}}(N,t)$ are those of $\gamma_0(N)$: in 
particular there is no new singularity when 
$\gamma(t)=\alpha_s(t)\gamma_0(N)\sim 1$. So when the coupling runs, 
$\Sigma(\rho,Q)$ rises asymptotically in the same way as $G(\rho,Q)$.

It is easy to see that this smoothing away of the $M=1$ pole (and 
indeed of any pole at $M=m_0$, provided only that $m_0$ is real and 
positive) is a generic feature when the coupling runs: the key 
ingredient is that $G(N,t)$ is a 
regular function of $\ln t$ rather than of $t$. As recalled in sec.2.5 the 
interval $M\in (0,\half)$ is then stretched to $(0,\infty)$, and so 
singularities at $M=1,2,\ldots$ are effectively pushed out to infinity. This 
means that when the coupling runs the 
expansion of the hard cross-section around $M=0$ provides 
a good approximation to the exact result after only a few terms: 
the running coupling resummation of the $M=\half$ singularity 
eqn.\NLLxrun\ in the evolution effectively deals with all the singularities
for $M>\half$, not only in the evolution but also in the hard cross-section,
ensuring that all such singularities are factorised into the the initial 
distribution.

In fact we can show that when the coupling runs, the expansion \exptrickexp\
is indeed asymptotic: returning to the specific example \exptrickrc, and 
using the asymptotic series
\eqn\incompgamexp{
\Gamma(a,t)\sim t^{a-1}e^{-t}[1+(a-1)t^{-1}+(a-1)(a-2)t^{-2}+\ldots],}
as $t\to\infty$, then as $\as(t)\to 0$ 
\eqn\exptrickrcexp{
\Sigma^{1}_{\gamma {\rm h}}(N,t)\sim 
[1+\gamma(t)+\gamma(t)^2(1-\smallfrac{\beta_0}{\gamma_0})
+\gamma(t)^3(1-\smallfrac{\beta_0}{\gamma_0})
(1-2\smallfrac{\beta_0}{\gamma_0})+\ldots]G(N,t),}
which is the same series term by term as \exptrickexp\ provided we 
evaluate the derivatives using eqns.\tderivs. So in the running 
coupling case the expansion in powers of derivatives is indeed an 
asymptotic series.

To see how well this asymptotic series works in practice, rather than 
in these simple examples, we define the $K$-factor as the ratio\foot{
Note that this $K$-factor is not 
the same as those used in phenomenological applications,
which also include a factor due to the difference between LO and higher
order partons.} 
\eqn\kayfacpol{K_1(\rho)=\frac{\Sigma^1_{\gamma {\rm h}}(\rho,Q)}
{\Sigma^0_{\gamma {\rm h}}(\rho,Q)},}
where $\Sigma^0_{\gamma {\rm h}}(\rho,Q)$ is simply the gluon distribution 
$G(\xi,t)$. With this 
definition the dependence on the gluon distribution largely 
cancels, so one sees the effect of the hard cross-section.
Then 
$$\eqalignno{K_1(\rho)
&=\frac{1}{G(\xi,t)}\int_{-i\infty}^{i\infty}\!\frac{dN}{2\pi i}\,e^{N\xi}
\int_0^\infty\!d\tau \,e^{-\tau}G(N,t+\tau)&\kexact\cr
&= 1+\frac{1}{G(\xi,t)}\frac{\partial}{\partial t}G(\xi,t)
+\frac{1}{G(\xi,t)}\frac{\partial^2}{\partial t^2}G(\xi,t)
+\frac{1}{G(\xi,t)}\frac{\partial^3}{\partial t^3}G(\xi,t)+\cdots,&\kapprox}$$
where in the first line we used \exptrick, and in the second \exptrickexp.
Using the resummed gluon distribution shown in fig.4 and its derivatives fig.8,
it is now a simple matter to compute the $K$-factor, using either \kexact\ 
or the series \kapprox. The result is shown in fig.9. 
It can be seen from the plot 
that even a simple pole at $M=1$ gives rise to a quite substantial $K$-factor,
and moreover that the NLO approximation to this $K$-factor, consisting of 
only the second term of \kapprox, is adequate only for $\rho\gsim 10^{-4}$. 
However the convergence thereafter is very rapid: the NNLO approximation
(the first three terms of \kapprox) is very good down to $\rho\sim 10^{-8}$, 
and the NNNLO approximation (four terms) is difficult to distinguish from 
the exact result. Thus although the series is only asymptotic,
in practice the first few terms give an excellent approximation to the 
full result.

\subsec{Photoproduction of Heavy Quarks}

We now consider the inclusive cross-section for the photoproduction of
a heavy quark pair. The hard scale in this case is the mass of the heavy 
quarks: $Q=2m_q$. The off-shell hard cross-section may be calculated from the 
diagram in fig.5: the result in \QMS\ scheme is \cchphot\foot{In 
refs.\refs{\cchglu,\rdbrke,\cchphot} this function is denoted by 
$j_\omega(\gamma)$.}
\eqn\jay{\eqalign{C(N,M) &=e_Q^2\alpha\alpha_s\pi 4^N 
\frac{14+20N+9N^2+N^3-M(10+7N+N^2)}{3-2M+2N}\cr
&\qquad\qquad\qquad\qquad\qquad\qquad\quad\quad\times
\frac{\Gamma(1-M+N)^3\Gamma(1+M)}{\Gamma(2-2M+2N)\Gamma(4+N)}.
}}
When $M=0$ this reduces to the usual \MS\ photoproduction cross-section 
relevant at large $\rho$, with poles 
at $N=-1,-\smallfrac{3}{2},-2,\ldots$, while 
at $N=0$ it reduces to the impact factor \cchglu\ 
\eqn\jayz{
C(0,M)=e_Q^2\alpha\alpha_s\frac{\pi}{3} 
\frac{7-5M}{3-2M}\frac{\Gamma(1-M)^3\Gamma(1+M)}{\Gamma(2-2M)},}
relevant for calculations at small $\rho$, i.e. at high energies $S\gg Q^2$,
with higher twist poles at $M=-1,-2,\ldots$ and infrared poles at 
$M=1,\smallfrac{3}{2},2,\ldots$. It is the latter that are relevant for 
the high energy limit: in particular near $M=1$ we have a double and a simple 
pole:
\eqn\jaypole{
C(0,M) \sim e_Q^2\alpha\alpha_s\frac{2\pi}{3}
\Big[\frac{2}{(1-M)^2}-\frac{1}{1-M}+O(1)\Big].}

Expanding eqn.\jayz\ about $M=0$ gives the Taylor expansion
\eqn\jayztaylor{\eqalign{
C(0,M)&=e_Q^2\alpha\alpha_s\smallfrac{7\pi}{9}\Big(1+\smallfrac{41}{21}M 
+\smallfrac{244}{63}M^2 + (\smallfrac{1460}{189}-2\zeta_3)M^3 
+O(M^4)\Big)\cr
&\simeq e_Q^2\alpha\alpha_s\,2.444(1+1.952M+3.87M^2+5.32M^3+\ldots).}}
This series has quite large coefficients, all of one sign, and growing: one 
thus expects it to converge rather slowly even for $|M|<1$. This is largely
due to the double pole at $M=1$: removing this by hand we find a series with 
rather smaller coefficients
\eqn\jaypoleapprox{\eqalign{
C(0,M) &\simeq e_Q^2\alpha\alpha_s\smallfrac{7\pi}{9}
\Big(\smallfrac{12}{7}\smallfrac{1}{(1-M)^2}-\smallfrac{5}{7}
-\smallfrac{31}{21}M -\smallfrac{80}{63}M^2 
+(\smallfrac{161}{189}-2\zeta_3)M^3+O(M^4)\Big)\cr
&\simeq e_Q^2\alpha\alpha_s\,2.444 (\smallfrac{1.714}{(1-M)^2}
-0.714-1.476M-1.27 M^2-1.54M^3+\ldots).}}
An even better series may be obtained by removing both the double and 
single poles by hand: 
\eqn\jaytwopoleapprox{\eqalign{
C(0,M) &\simeq e_Q^2\alpha\alpha_s\smallfrac{7\pi}{9}
\Big(\smallfrac{12}{7}\smallfrac{1}{(1-M)^2}
-\smallfrac{6}{7}\smallfrac{1}{(1-M)}
+\smallfrac{1}{7}
-\smallfrac{13}{21}M -\smallfrac{26}{63}M^2 
+(\smallfrac{326}{189}-2\zeta_3)M^3+O(M^4)\Big)\cr
&\simeq e_Q^2\alpha\alpha_s\,2.444(\smallfrac{1.714}{(1-M)^2}
-\smallfrac{0.857}{1-M}
+0.143-0.619M- 0.41M^2-0.68M^3+\ldots).}}

\topinsert
\vskip-11.5truecm
\vbox{
\epsfxsize=18truecm
\centerline{\epsfbox{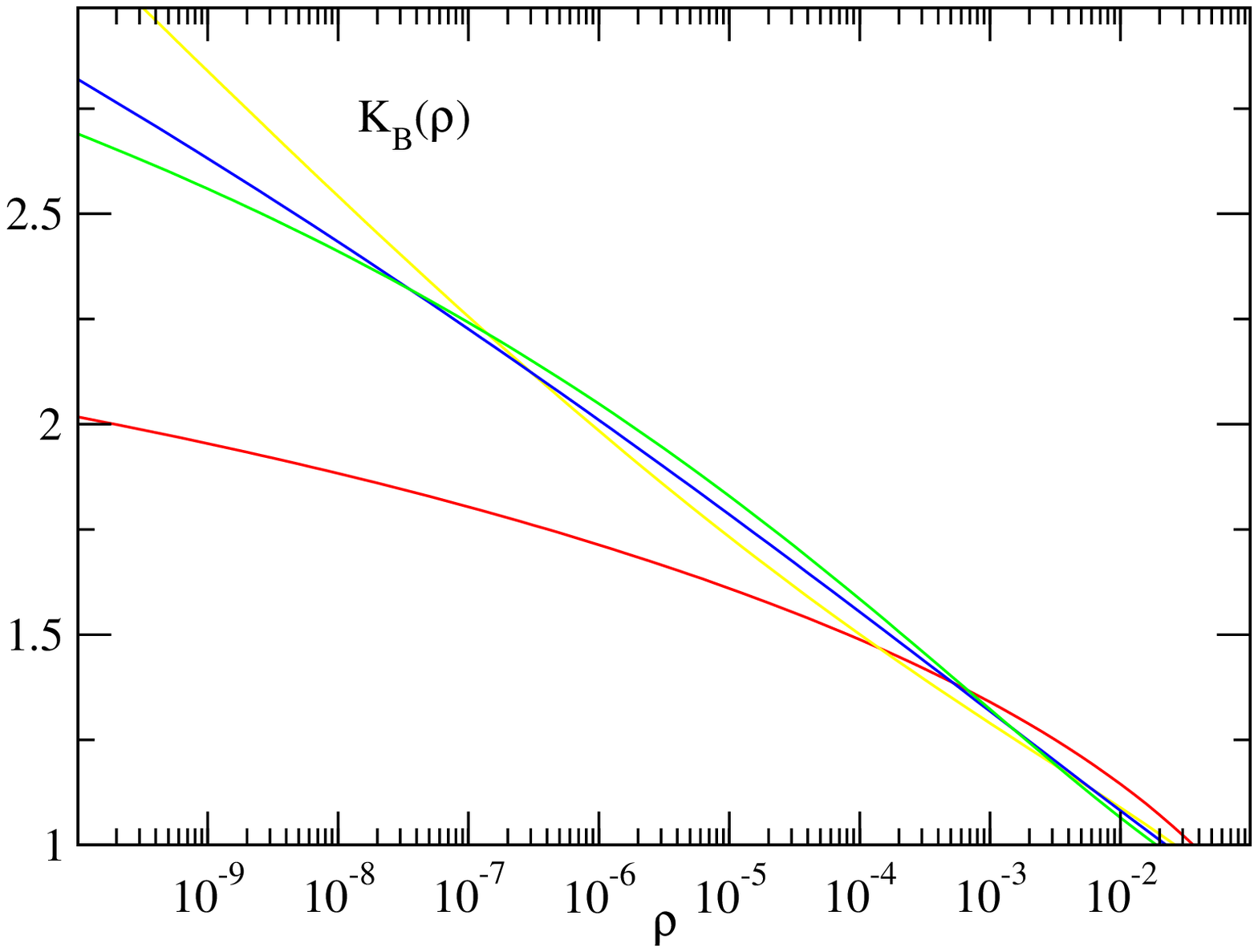}}
\vskip-1.5truecm
\bigskip
\hbox{
\vbox{\footnotefont\baselineskip6pt\narrower\noindent Figure 10: The 
$K$-factor for photoproduction of $b$ quarks. The blue curve (upper) is 
the resummed result, while the red, green 
and yellow curves (bottom to top) 
are the NLO, NNLO and NNNLO fixed order results computed 
as described in the text. The scale is $Q=10\GeV$.
 }}
\hskip0.5truecm}
\vskip-0.5truecm
\endinsert 

We can now use these expressions to compute the $K$-factor for inclusive $B$ 
photoproduction, 
\eqn\BKfac{K_B(\rho) = \frac{\Sigma^B_{\gamma {\rm h}}(\rho,m_B)}
{\Sigma^{O}_{\gamma {\rm h}}(\rho,m_B)},}
where $\Sigma^B_{\gamma {\rm h}}(\rho,m_B)$ is the cross-section computed using
resummation, and $\Sigma^{O}_{\gamma {\rm h}}(\rho,m_B)$ the cross-section 
computed using the same gluon distribution, but with the LO hard cross-section
(which here is simply $C(0,0)=\smallfrac{7\pi}{9}e_Q^2\alpha\alpha_s$, i.e. 
the first term in the expansion \jayztaylor). To calculate 
$\Sigma^B_{\gamma {\rm h}}$ we use 
the same techniques as in the previous section: poles at $M=1$ are dealt 
with using the exponentiation trick \exptrick, while powers are turned 
into derivatives according to \powtrick: the only new feature is that each term
gets multiplied by the various coefficients in \jayztaylor, \jaypoleapprox, or 
\jaytwopoleapprox. The results are plotted in fig.10: the blue curves are 
the resummed results computed using the three  approximations 
\jayztaylor, \jaypoleapprox, and \jaytwopoleapprox. The results are 
indistinguishable on the plot, and thus may be taken to be the exact 
result: as in the previous calculation, the Taylor 
series is an adequate approximation to the more accurate pole approximations
\jaypoleapprox\ and \jaytwopoleapprox\ for all $\rho\gsim 10^{-10}$.

Also in fig.10 we plot the $K$-factors for fixed order perturbation theory, 
at NLO, NNLO and NNNLO. Note that 
these are computed with the GLAP gluon distributions at the appropriate order
(LO, NLO and NNLO), not with the resummed distribution. To do these 
calculations we use the 
result eqn.\fixedordersings\ to evaluate the dominant contributions to 
the fixed order hard cross-section: these are $O(\smallfrac{\as}{N})$ at NLO
(these are the terms computed in \ellross, not the full result of 
ref.\ellisnason), $O(\smallfrac{\as^2}{N^2})$ and $O(\smallfrac{\as^2}{N})$ 
at NNLO, and $O(\smallfrac{\as^3}{N^3})$, $O(\smallfrac{\as^3}{N^2})$, and  
$O(\smallfrac{\as^2}{N})$ at NNNLO. They may thus be computed using the 
$O(M)$, $O(M^2)$ and $O(M^3)$ terms in the Taylor expansion \jayztaylor\ 
provided we use at least the LO, NLO and NNLO results respectively for 
the evolved gluon distributions. It may be seen from the plot that 
while the NLO calculation seriously underestimates the $K$-factor for 
$\rho\lsim 10^{-4}$, the NNLO calculation is fine down to 
$\rho\sim 10^{-8}$, while the NNNLO is worse, starting to overshoot 
below $10^{-7}$ or so. So it seems that while a NLO calculation of the 
hard cross-section is inadequate at high energy, a NNLO calculation can 
perform well, the fully resummed result being only really necessary at 
very small values of $\rho$.

\topinsert
\vskip-11.5truecm
\vbox{
\epsfxsize=18truecm
\centerline{\epsfbox{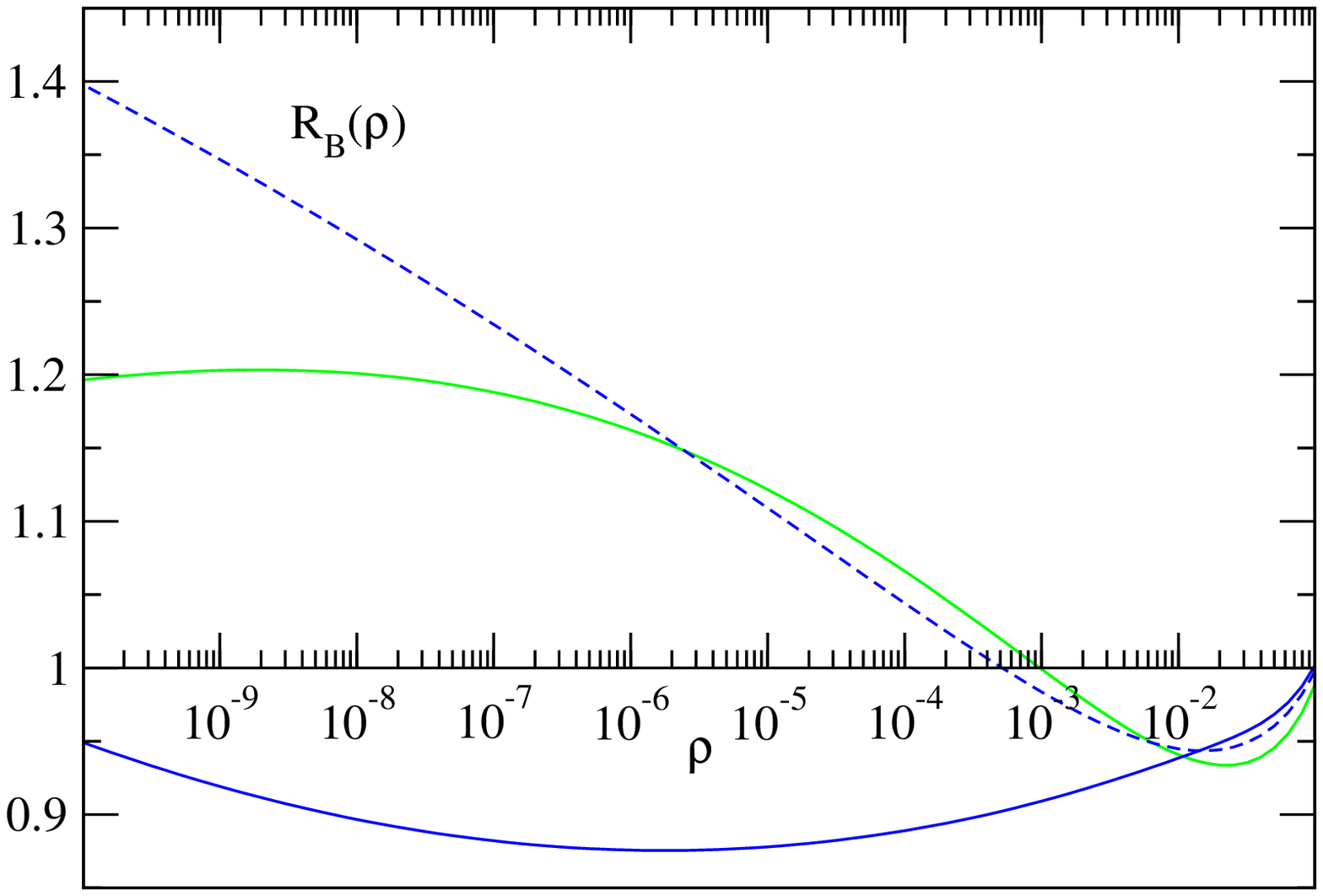}}
\vskip-2.0truecm
\bigskip
\hbox{
\vbox{\footnotefont\baselineskip6pt\narrower\noindent Figure 11: the 
resummation factor $R_B$ for photoproduction. The solid blue (lower) curve  
includes 
resummation in both evolution and hard cross-section, while the dashed 
blue curve only has resummation in the cross-section. The green (upper) 
curve is
the same computation comparing NNLO to NLO perturbation theory.
 }}\hskip1truecm}
\vskip-0.5truecm
\endinsert 

To explore the relative sizes of the various resummation contributions, 
in fig.11 we plot the ``resummation factor'' for photoproduction 
of $b\bar{b}$ pairs:
\eqn\rbdef{R_B(\rho) = \frac{\Sigma^B_{\gamma {\rm h}}(\rho,m_B)}
{\Sigma^{\rm NLO}_{\gamma {\rm h}}(\rho,m_B)},}
where $\Sigma^B_{\gamma {\rm h}}$ is computed as previously, while 
$\Sigma^{\rm NLO}_{\gamma {\rm h}}$ is computed using the NLO 
hard cross-section 
and NLO GLAP gluon distribution. The result is the solid blue curve: the
net effect of the resummation is to reduce the cross-section by between 
$5$ and $10\%$, rather uniformly over the full range of $\rho$. 
The dashed blue 
curve is the same calculation but with $\Sigma^B_{\gamma h}$ computed using 
the NLO GLAP gluon, rather than the resummed one, to assess the 
relative effects of the resummation of the cross-section and the 
resummation of the evolved gluon: clearly both effects are of similar 
importance, since when combined they almost cancel. The green curve is the 
ratio $\Sigma^{\rm NNLO}_{\gamma {\rm h}}/\Sigma^{\rm NLO}_{\gamma {\rm h}}$, 
with 
$\Sigma^{\rm NNLO}_{\gamma {\rm h}}$ computed with a consistent NNLO 
cross-section 
and gluon, for comparison with the resummed result: from this it is clear that
a full NNLO calculation does not give a good approximation to the resummed 
cross-section for $\rho\lsim 10^{-2}$, despite giving a good account of the 
hard cross-section. This is because the NNLO evolution is not very close 
to the resummed evolution, as it underestimates the suppression \abfrunfin.
Taken together, these three curves probably give a reasonable impression of 
the overall range of uncertainty in the $b\bar{b}$-photoproduction 
cross-section at high energy.

\newsec{Hadroproduction}

\subsec{The Gluon-Gluon Luminosity}

\topinsert
\vskip-7truecm
\vbox{
\epsfxsize=22truecm
\centerline{\hskip2truecm\epsfbox{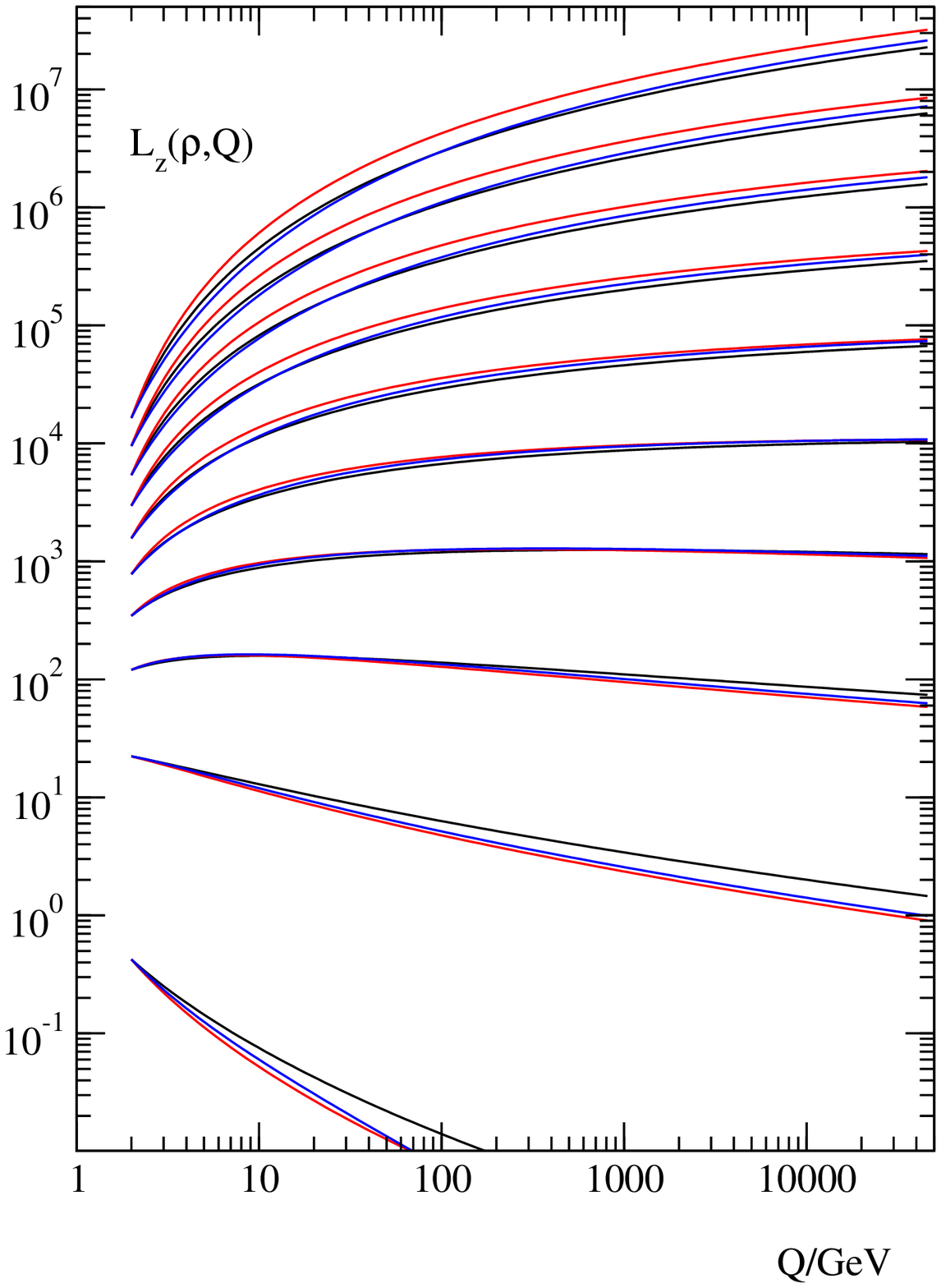}}
\vskip-1truecm
\bigskip
\hbox{
\vbox{\footnotefont\baselineskip6pt\narrower\noindent Figure 12: The 
gluon-gluon luminosity density $L_z(\rho,Q)$ plotted 
against $Q$ in $\GeV$ for $\rho =10^{-10},10^{-9},\ldots,
0.01,0.1$ (from top to bottom). 
The blue curves are evolved with the NLO resummation
described in the text: the black and red curves are with LO and NLO GLAP 
evolution.
 }}}
\vskip0.0truecm
\endinsert 

When we calculate hadronic cross-sections, using the factorization 
\hadrofac\ or equivalently \hadmelinv, it is convenient to first define 
the gluon-gluon 
luminosity density\foot{In the literature 
it is more usual to call this quantity
$d{\cal L}/d\tau$, where $\tau=x_1x_2$: after integration ${\cal L}$ is 
then the total gluon-gluon luminosity.}
\eqn\lumidef{\eqalign{L_z(z,Q_1,Q_2)&= 
\int_\rho^1 \!\frac{dx_1}{x_1}\int_\rho^1 \!\frac{dx_2}{x_2}\,\delta (z-x_1x_2)
G(x_1,Q_1)G(x_2,Q_2)\cr
&=\int_z^1 \!\frac{dy}{y}\, G(\frac{z}{y},Q_1)G(y,Q_2)\cr
&=\int_{-i\infty}^{i\infty} {dN\over 2\pi i}\, e^{\xi N}
\int_{-i\infty}^{i\infty} {dM_1\over 2\pi i}
{dM_2\over 2\pi i}\,e^{t_1 M_1+t_2 M_2}\, G(N,M_1)G(N,M_2),}}
where $\xi = \ln 1/z$, $t_1=\ln Q_1^2/\Lambda^2$, 
$t_2=\ln Q_2^2/\mu^2$. Thus we have in Mellin space simply 
\eqn\lumidefmel{L_z(N,M_1,M_2) = G(N,M_1)G(N,M_2),}
so eqn.\hadmelinv\ reads
\eqn\hadmelinvlumi{\eqalign{\Sigma_{\rm hh}(\rho,Q) &= 
\int_{-i\infty}^{i\infty} {dN\over 2\pi i}\, e^{\xi N}
\int_{-i\infty}^{i\infty} {dM_1\over 2\pi i}
{dM_2\over 2\pi i}\,
 e^{t (M_1+M_2)}H(N,M_1,M_2)L_z(N,M_1,M_2)\cr
&= \int_\rho^1 \! {dz\over z}
\int \! {d^2{\bf{k_1}}\over\pi\bf{k}_1^2}
\int \! {d^2{\bf{k_2}}\over\pi\bf{k}_2^2} 
\Sigma_{gg}\big({\rho\over z},
{{\bf{k_1}}\over Q},{{\bf{k_2}}\over Q}\big)L_z\big(z,{\bf{k}_1^2},
{\bf{k}_2^2}\big),}}
where now we have set $t_1=t_2=t=\ln Q^2/\Lambda^2$, $Q$ being the 
invariant mass in the final state. Substituting \lumidef\ into 
\hadmelinvlumi\ then gives back the gluon-gluon contribution to \hadrofac, 
as it should.

It is straightforward to compute $L_z(\rho,Q,Q)\equiv L_z(\rho,Q)$ 
from the gluon distribution 
$G(x,Q)$ shown in fig.4: the result is show in fig.12. Note that the luminosity
only starts to rise at large $Q$ when $\rho\lsim 10^{-5}$ or so, whereas 
for the gluon distribution the rise sets in already at $x \sim 10^{-3}$. Again 
the resummed luminosity grows rather more slowly under the resummed 
evolution than with NLO GLAP evolution.

Of course in a hadron collider it is not always possible to vary $\rho$ and 
$Q$ independently as one does in photoproduction: the inclusive 
cross-section for 
the hadroproduction of a final state of invariant mass $Q$ depends instead on
$L_z(Q)\equiv L_z(Q^2/S,Q,Q)$, where $S$ is the (fixed) centre-of-mass energy 
of the machine. This is plotted in fig.13 for three different colliders:
the Tevatron, the LHC,  
and a notional VLHC with $\sqrt{S}=200\TeV$. Note that $Q^{-1}L_z(Q)$ gives 
a rough estimate of the inclusive cross-section.

\topinsert
\vskip-8truecm
\vbox{
\epsfxsize=18truecm
\centerline{\epsfbox{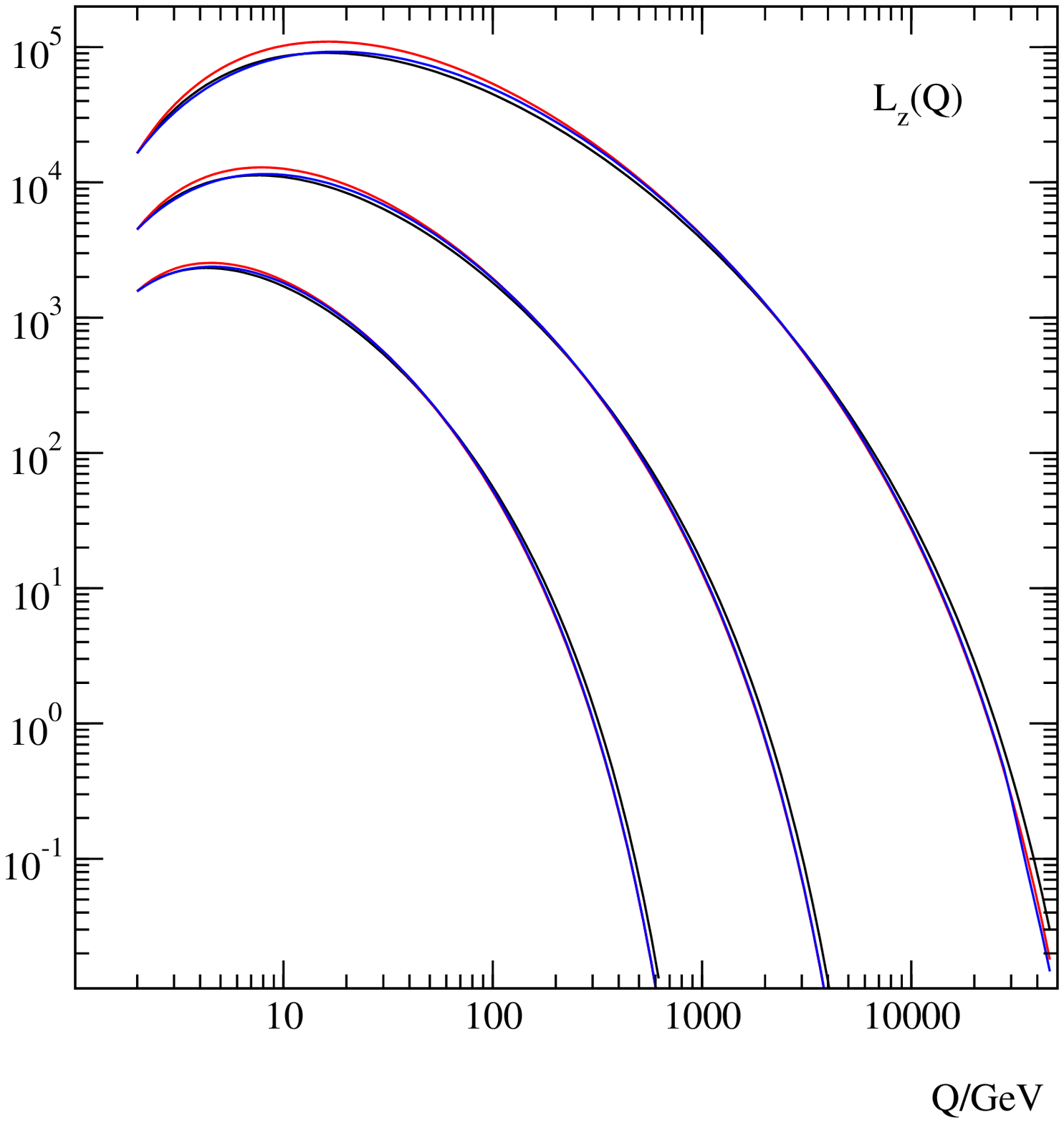}}
\vskip-1.3truecm
\bigskip
\hbox{
\vbox{\footnotefont\baselineskip6pt\narrower\noindent Figure 13: The 
gluon-gluon luminosity $L_z(Q)$ plotted against the invariant mass $Q$ in 
$\GeV$. The lower curves are for the Tevatron, the middle curves for LHC,
and the upper for a VLHC: the colour coding is the same as in fig.12.
 }}\vskip-0.1truecm}
\endinsert 

\subsec{Resumming the Dominant Infrared Singularity}

\topinsert
\vskip-1.5truecm
\vbox{\hbox{\centerline{
\hskip1truecm
\epsfxsize=10truecm
\epsfbox{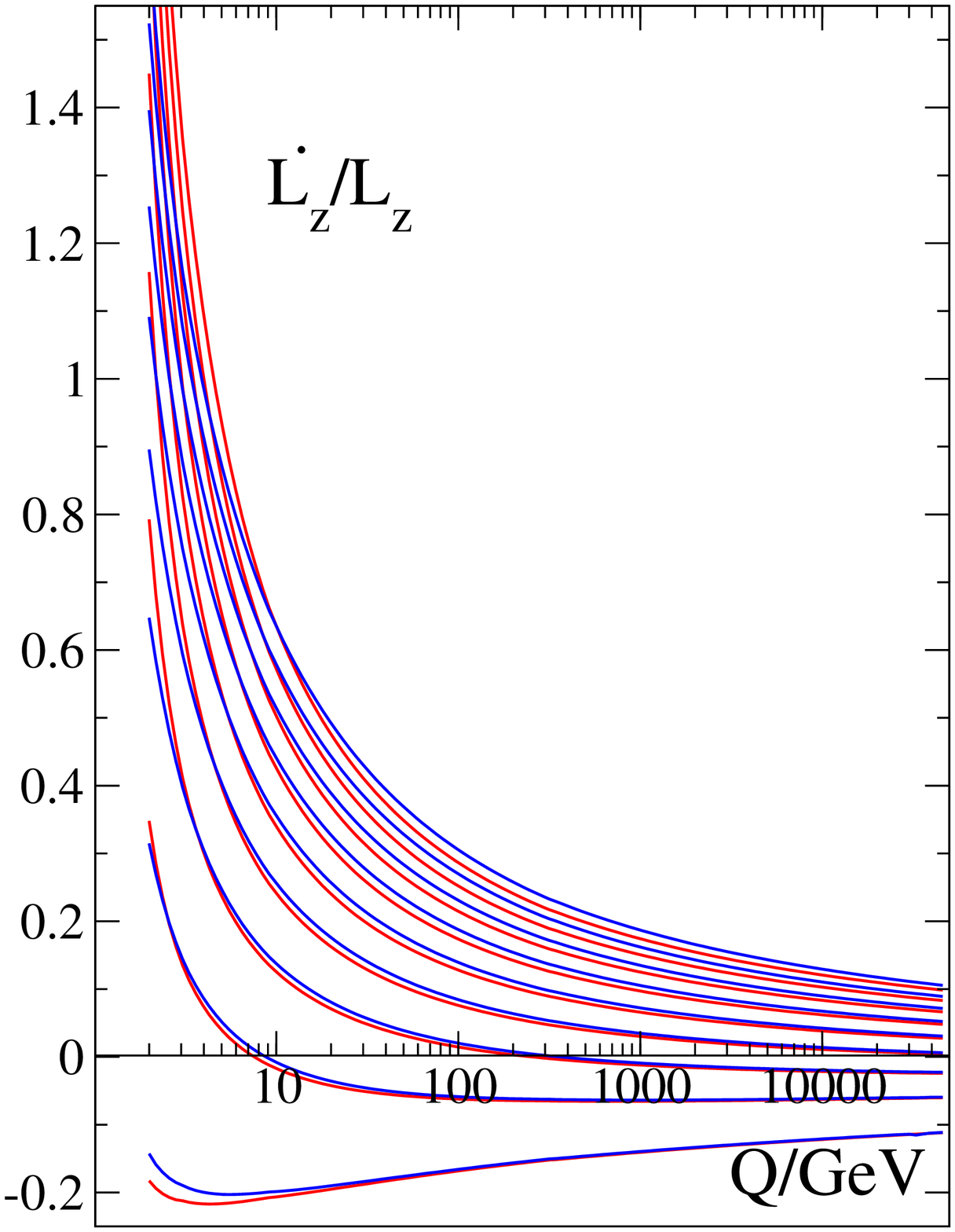}
\hskip-1truecm
\epsfxsize=10truecm
\epsfbox{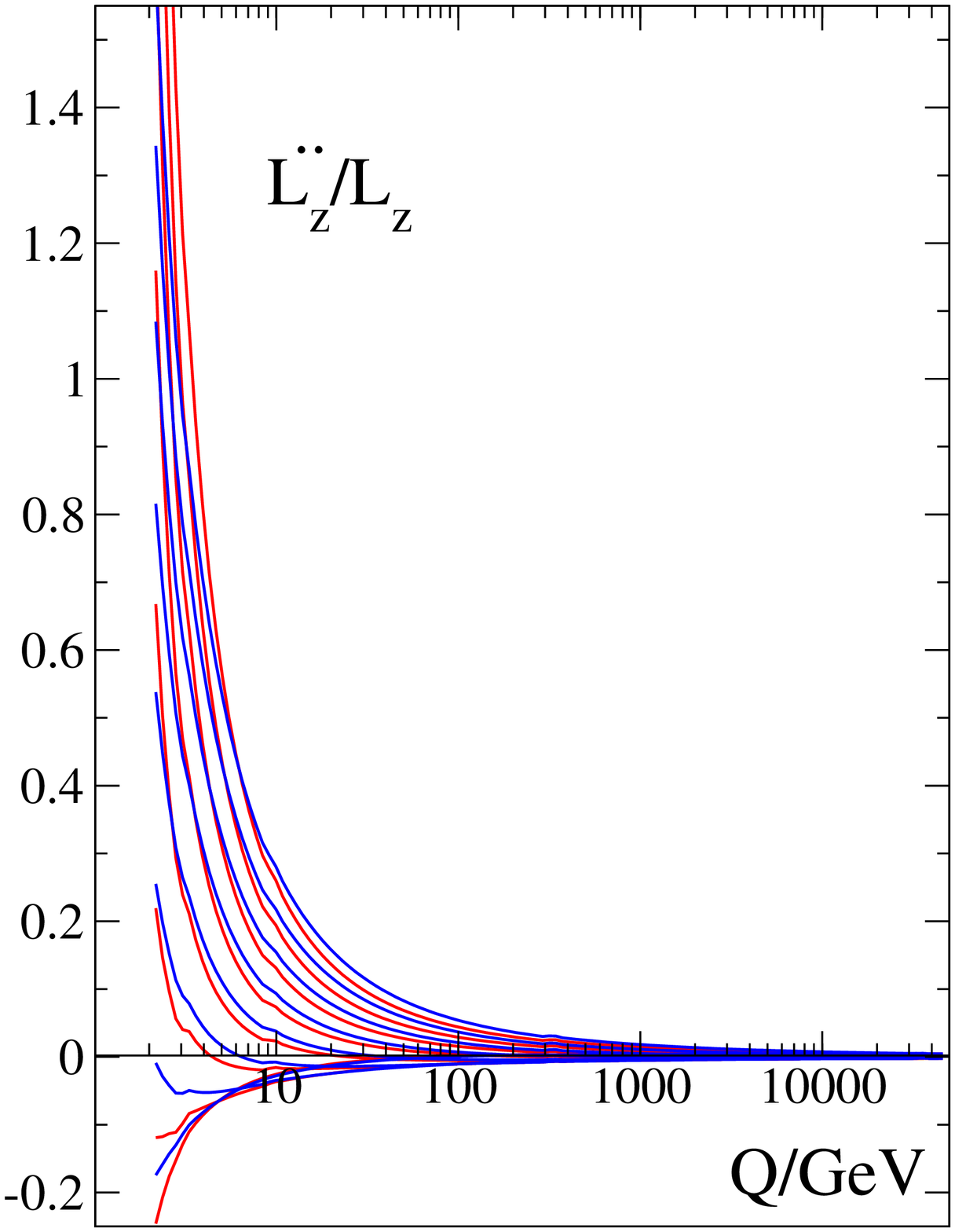}
}}
\hbox{
\vbox{\footnotefont\baselineskip6pt\narrower\noindent Figure 14: The first 
(left) and second (right) derivatives with respect to $t=\ln Q^2/\Lambda^2$ 
of the gluon-gluon luminosity $L_z(\rho ,Q)$ plotted in fig.13, normalised to 
$L_z$, plotted against $Q$ in $\GeV$, for 
$\rho=10^{-10},10^{-9},\ldots 0.01$ (from top 
to bottom). Again the blue curves are with NLO resummed evolution, the 
red with NLO GLAP evolution.
 }}
\vskip 0.0truecm}
\endinsert 
In order to compute the inclusive cross-section eq.\hadmelinvlumi\ we must
first compute the hard cross-section, and then convolute it with the 
gluon-gluon luminosity. In practice this means we must perform the inverse
Mellin transforms over $N$, $M_1$ and $M_2$. Since all the collinear and 
high energy logarithms are already included in the luminosity, the hard 
cross-section $H(N,M_1,M_2)$ is regular at $N=M_1=M_2=0$, and we may 
Taylor expand it using \xsectaylor. Just as in the photoproduction 
case eqn\powtrick, the resulting integrals over powers of 
$M_1$ and $M_2$ may then be evaluated in terms of derivatives of the 
luminosity: writing $L_z(N,t)\equiv L_z(N,t,t)$ for simplicity, then 
for all $m_1,m_2 = 0,1,2,\ldots$
\eqn\powtricksq{\int_{-i\infty}^{i\infty} \frac{dM_1}{2\pi i}
\frac{dM_2}{2\pi i}e^{t(M_1+M_2)}
\, M_1^{m_1}M_2^{m_2} L_z(N,M_1,M_2) = 
\frac{\partial^{m_1+m_2}}{\partial t_1^{m_1}\partial t_2^{m_2}}
L_z(N,t_1,t_2)\Big\vert_{t_1=t_2=t}.}
The derivatives of the luminosity are then given in turn by
\eqn\Ltderivs{\eqalign{
\frac{\partial}{\partial t_1}L_z(N,t_1,t_2)\Big\vert_{t_1=t_2=t}=
\frac{\partial}{\partial t_2}L_z(N,t_1,t_2)\Big\vert_{t_1=t_2=t}
&=\gamma(N,t)L_z(N,t),\cr
\frac{\partial^2}{\partial t_1^2}L_z(N,t_1,t_2)\Big\vert_{t_1=t_2=t}=
\frac{\partial^2}{\partial t_2^2}L_z(N,t_1,t_2)\Big\vert_{t_1=t_2=t}
&=(\gamma^2+ \dot{\gamma})L_z(N,t),\cr
\frac{\partial^2}{\partial t_1\partial t_2}L_z(N,t_1,t_2)\Big\vert_{t_1=t_2=t}
&=\gamma^2 L_z(N,t),\cr
\frac{\partial^3}{\partial t_1^3}L_z(N,t_1,t_2)\Big\vert_{t_1=t_2=t}=
\frac{\partial^3}{\partial t_2^3}L_z(N,t_1,t_2)\Big\vert_{t_1=t_2=t}
&=(\gamma^3+3\gamma \dot{\gamma}+\ddot{\gamma})L_z(N,t),\cr
\frac{\partial^3}{\partial t_1^2\partial t_2}
L_z(N,t_1,t_2)\Big\vert_{t_1=t_2=t}=
\frac{\partial^3}{\partial t_1\partial t_2^2}
L_z(N,t_1,t_2)\Big\vert_{t_1=t_2=t}
&=(\gamma^3+\gamma \dot{\gamma})L_z(N,t),
\ldots}}
The first two of these are plotted in fig.14 (to be compared with fig.8):
again even the first derivative is below unity for all except the smallest 
values of $\rho$ and $Q$, and the second derivative is considerably 
smaller than the first. The resummed derivatives are 
smaller at small $\rho$ and $Q$ than the corresponding GLAP derivatives. 
Given these facts, we expect the 
first few terms of the Taylor series to give a good approximation 
to the full cross-section at all except the highest energies and lowest 
scales, just as they did in photoproduction.

\topinsert
\vskip-6truecm
\vbox{
\epsfxsize=15truecm
\centerline{\epsfbox{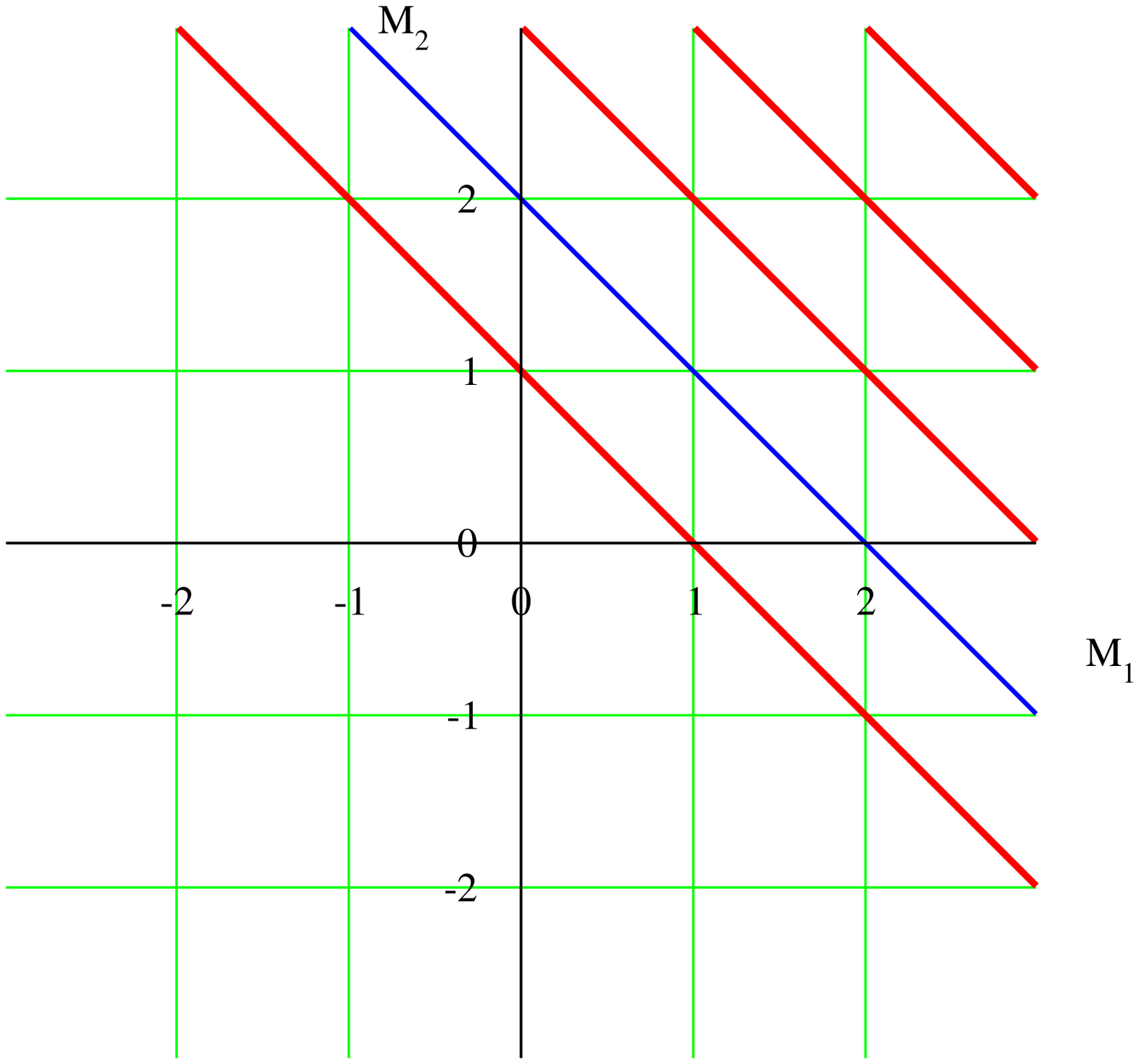}}
\vskip-2.0truecm
\bigskip
\hbox{
\vbox{\footnotefont\baselineskip6pt\narrower\noindent Figure 15: A schematic 
picture of the location of the singularities of a hadroproduction 
impact factor $H(0,M_1,M_2)$ in the $M_1$-$M_2$ plane. The particular 
singularities shown are those for heavy quark production: those in green 
(parallel to the axes) are 
simple poles, in blue double poles and in red triple poles (diagonal).
 }}\vskip0.0truecm}
\endinsert 

It remains to consider the singularities of the hard 
cross-section (see fig.15)..
Just as in the photoproduction case there will be higher twist singularities
at $M_1,M_2 = -1,-2,-3,\ldots$ which will lead to terms suppressed by inverse 
powers of the hard scale $Q^2$: these will not concern us here. The 
structure of the infrared singularities is more subtle: 
there are singularities
at $M_1,M_2 = 1,2,3,\ldots$ just as in photoproduction, but 
there are now also infrared singularities on the lines 
$M_1+M_2 = 1,2,3\ldots$, as discussed in sec.2.6.
Since these come closest to the origin, it is likely that they dominate at 
high energies, and indeed as we shall see this turns out to be the case. 

In order to integrate over the line of poles at $M_1+M_2=1$, we may employ
the same trick that we used for the $M=1$ pole in sec.3.2:
\eqn\exptrickhad{\frac{1}{(1-M_1-M_2)^n} 
= \frac{1}{n!}\int_0^\infty d\tau \tau^{n-1}e^{-\tau(1-M_1-M_2)}}
for $n=1,2,\ldots$. The nice extra feature here is that under the 
integral over $\tau$ the dependence on $M_1$ and $M_2$ has now factorised,
allowing both integrals to be performed independently:
\eqn\hadmelinvpole{\eqalign{
\Sigma^{n}_{\rm hh}(N,t)
&=\int_{-i\infty}^{i\infty} {dM_1\over 2\pi i}{dM_2\over 2\pi i} 
\frac{1}{(1-M_1-M_2)^n}e^{t (M_1+M_2)} G(N,M_1)G(N,M_2)\cr
&=\frac{1}{n!}\int_0^\infty d\tau \tau^{n-1}e^{-\tau}
\int_{-i\infty}^{i\infty} {dM_1\over 2\pi i} e^{(t+\tau)M_1} G(N,M_1) 
\int_{-i\infty}^{i\infty} {dM_2\over 2\pi i} e^{(t+\tau)M_2}  G(N,M_2)\cr 
&=\frac{1}{n!}\int_0^\infty d\tau \tau^{n-1}e^{-\tau}L_z(N,t+\tau),}}
where in the last line we have exploited the factorization to perform
both of the integrals over $M_1$ and $M_2$, writing the result in terms 
of the gluon-gluon luminosity $L_z(N,t)$. In this way we reduce a double 
integral over two complex variables with a line of singularity to a single 
real integral, which clearly converges rather rapidly when the coupling runs 
since $L_z(N,t)$ is 
very smooth (see fig.12). Again where $L_z(\rho,Q)$ increases monotonically
in $Q$ the singularity will give an enhancement of the cross-section, which 
will be most significant when the rise is steepest (i.e. small $\rho$ 
and low $Q$). 

It is easy to see that the same arguments used for the 
photoproduction and electroproduction cross-sections, 
in particular \exptrickfc\ and \exptrickrc, will also apply here, because 
of the factorization under the $\tau$ integral in \hadmelinvpole. Thus 
when the coupling runs the 
line singularity is once again pushed to infinity, and the 
only singularities of $\Sigma^{n}_{\rm hh}(N,t)$ in the $N$-plane are 
those of the anomalous dimension $\gamma(N,t)$. The running of the coupling 
thus ensures that all infrared singularities, whether in the evolution or 
in the hard cross-section, are factorised into the initial (nonperturbative)
gluon distribution.

An important consequence of this factorization is that 
the asymptotic high energy behaviour of inclusive 
hadroproduction cross-sections is determined entirely 
by the rise in the gluon distributions, just as it was in 
electroproduction and photoproduction. It follows that the high energy
powerlike rise due to the rightmost singularity in the 
anomalous dimension is universal: all inclusive cross-sections rise in
the same way.
\eqnn\kayfacpol\eqnn\khexact\eqnn\khapprox

\topinsert
\vskip-11.5truecm
\vbox{
\epsfxsize=18truecm
\centerline{\epsfbox{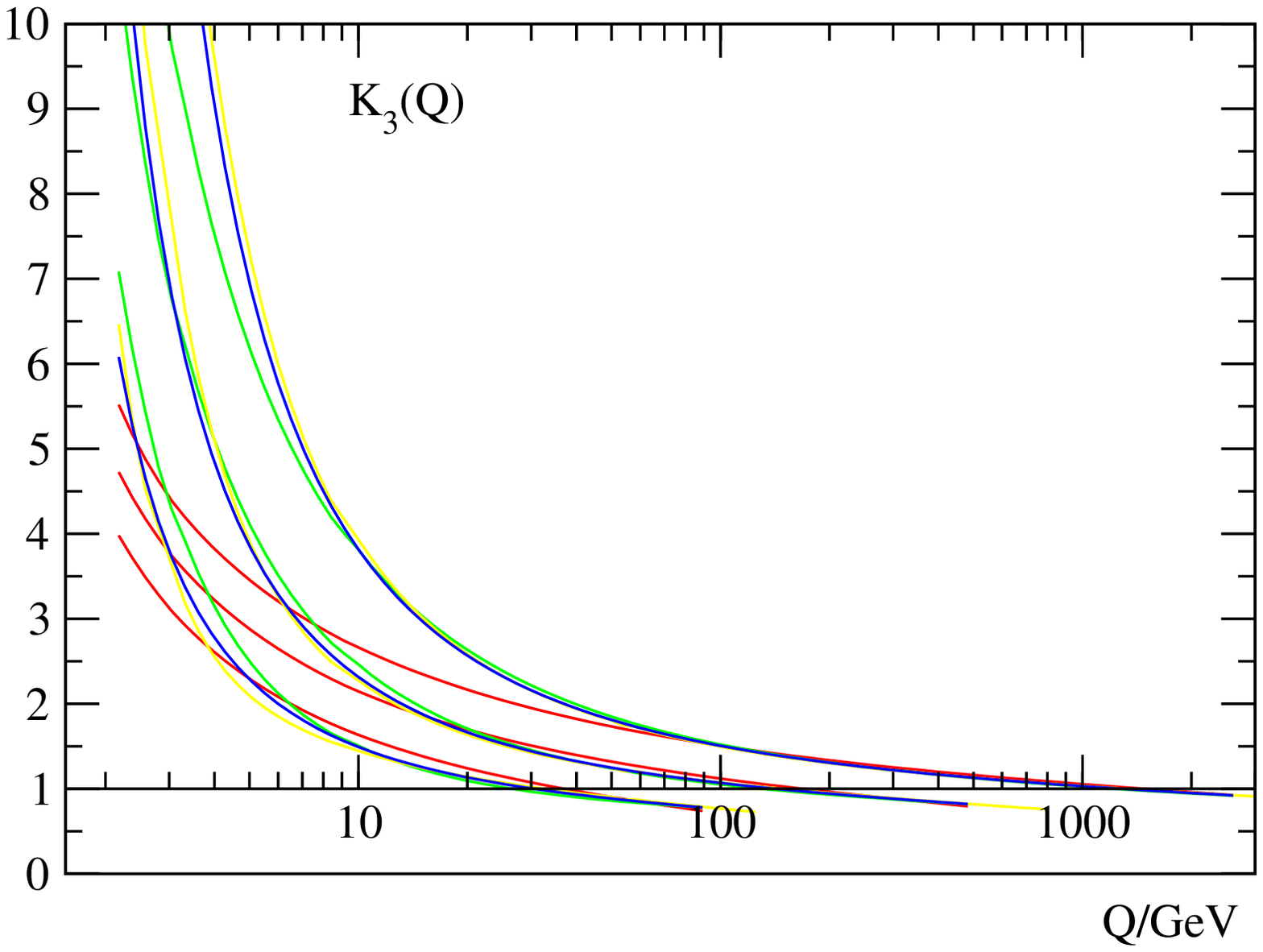}}
\vskip-2.0truecm
\bigskip
\hbox{
\vbox{\footnotefont\baselineskip6pt\narrower\noindent Figure 16: a generic 
hadroproduction $K$-factor for a triple pole $1/(1-M_1-M_2)^3$, plotted 
against invariant mass $Q$, for three different colliders: the Tevatron 
(lower curves), the LHC (middle curves) and a VLHC(upper curves). In each 
case the blue (upper) lines
correspond to the exact expression eqn.\khexact, while the red, green 
and yellow (bottom to top) lines are the NLO, NNLO and NNNLO 
approximation to it computed 
by including the second, third and fourth terms respectively of the 
series \khapprox.
 }}\hskip1truecm}
\endinsert 

To see how well all this works in practice, consider a 
triple pole, i.e. $n=3$, 
and define the $K$-factor as the ratio
$$\eqalignno{K_3(Q)&=\frac{\Sigma^3_{h h}(Q^2/S,Q)}
{\Sigma^0_{h h}(Q^2/S,Q)},&\kayfacpol}$$
where the denominator $\Sigma^0_{h h}(Q^2/S,Q)$ is simply the gluon-gluon luminosity 
distribution $L_z(\xi,t)$ when $\xi=\ln S/Q^2$. 
Then 
$$\eqalignno{K_3(Q)
&=\frac{1}{L_z(\xi,t)}\int_{-i\infty}^{i\infty}\!\frac{dN}{2\pi i}\,e^{N\xi}
\int_0^\infty\!d\tau \,\tau^2 e^{-\tau}L_z(N,t+\tau)&\khexact\cr
&= 1+\frac{3}{L_z(\xi,t)}\frac{\partial}{\partial t}L_z(\xi,t)
+\frac{6}{L_z(\xi,t)}\frac{\partial^2}{\partial t^2}L_z(\xi,t)
+\frac{10}{L_z(\xi,t)}\frac{\partial^3}{\partial t^3}L_z(\xi,t)
+\cdots,&\khapprox}$$
where in the first line we used \hadmelinvpole, and in the second we
Taylor expand $L_z(N,t+\tau)$ and integrate term by term. The diagonal 
derivatives of the luminosity are readily deduced from \Ltderivs:
\eqn\Ltdiagderivs{\eqalign{
\frac{\partial}{\partial t}L_z(N,t)
&=2\gamma(N,t)L_z(N,t),\cr
\frac{\partial^2}{\partial t^2}L_z(N,t)
&=2(2\gamma^2+ \dot{\gamma})L_z(N,t),\cr
\frac{\partial^3}{\partial t^3}L_z(N,t)=
&=2(4\gamma^3+6\gamma \dot{\gamma}+\ddot{\gamma})L_z(N,t),
\ldots}.}
Using the resummed luminosity distribution shown in fig.12 
it is now a simple matter to compute the $K$-factor, using either \khexact\ 
or the series \khapprox. The results are shown in fig.16, for $S = 2~\TeV$, 
$14~\TeV$ and $200~\TeV$. At sufficiently low invariant mass $Q$ and thus 
large $\rho$ the $K$ factor rises steeply as expected, eventually resulting
in very large corrections. In each case the exact result \khexact\ is well
approximated by only the first few terms in \khapprox. In fact above 
$Q\sim 5~\GeV$ at the Tevatron, $Q\sim 15~\GeV$ at the LHC and $Q\sim 70~\GeV$
at a VLHC the $O(M)$ (ie NLO) correction alone already gives a good 
approximation: only below these scales do the $O(M^2)$ (NNLO) corrections 
become significant. The NNLO approximation only really starts to be inadequate 
when $Q\lsim 5~\GeV$ at the VLHC, which would no doubt be outside the 
range of acceptance.

\topinsert
\vskip-11.5truecm
\vbox{
\epsfxsize=18truecm
\centerline{\epsfbox{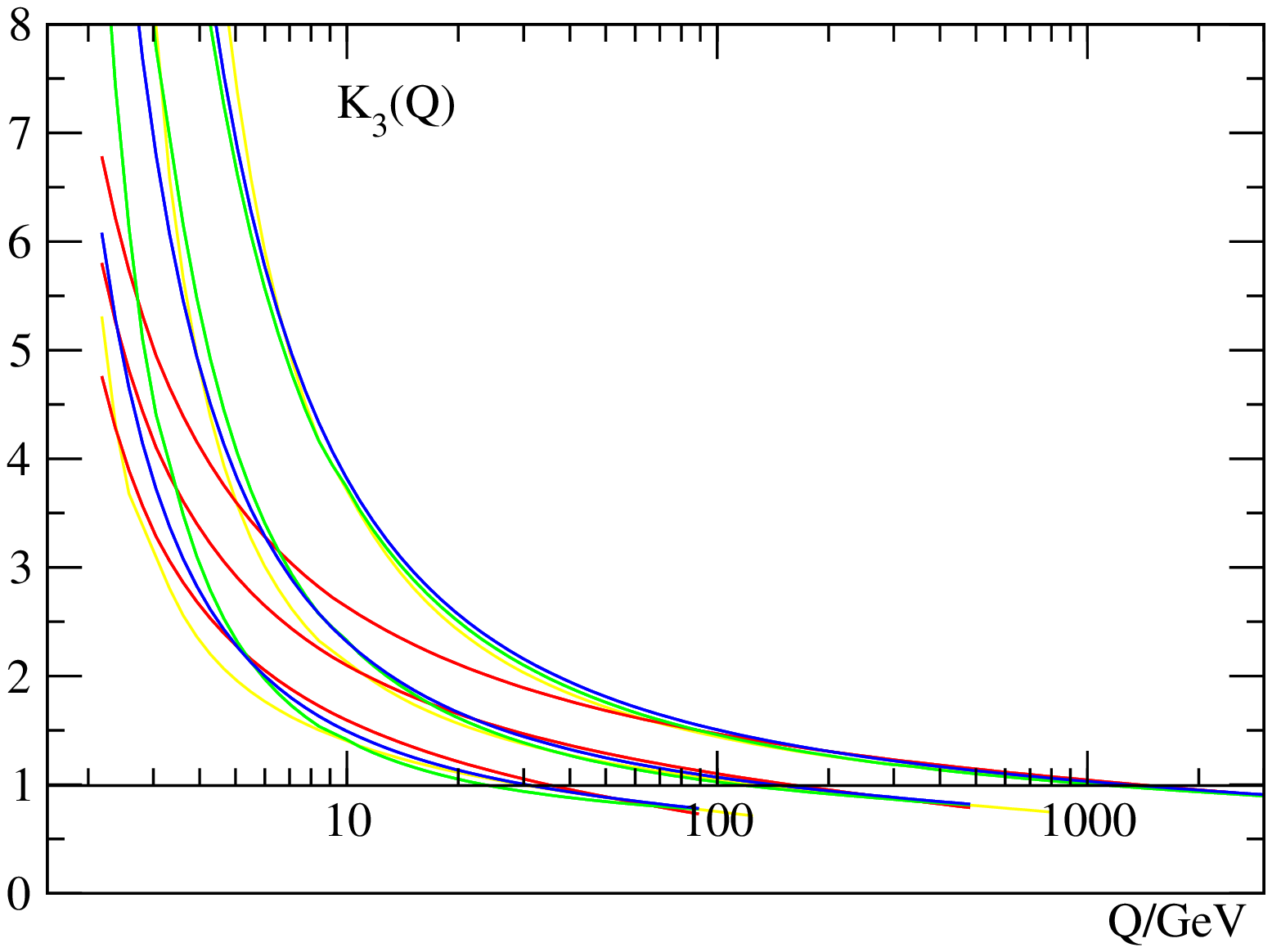}}
\vskip-2.0truecm
\bigskip
\hbox{
\vbox{\footnotefont\baselineskip6pt\narrower\noindent Figure 17: the same as 
fig.16, except that now the red, green 
and yellow lines are the NLO, NNLO and NNNLO results in fixed 
order perturbation theory (at high energy), computed 
by including the second, third and fourth terms respectively of the 
series \khapprox, with the luminosity evolved using GLAP evolution at the 
appropriate order.
 }}\hskip-1truecm}
\endinsert 
These results are interesting because, as shall show explicitly for heavy quark
production in the next section, the line of triple poles \triplepole\ is 
sufficiently singular to provide the dominant contribution to the 
cross-section. Moreover since these infrared singularities are quite generic,
we expect other inclusive cross-sections such as the inclusive jet 
cross-section to acquire $K$-factors at high energy similar to those shown 
in fig.~16. 
Calculation of the $K$-factor can then proceed in each case through 
a computation of the expansion of the hard cross-section $H(N,M_1,M_2)$ 
in $M_1$ and $M_2$, ie by perturbation about the on-shell result: the 
singularity will then generate large coefficients of the $O(M_1+M_2)$ and 
$O((M_1+M_2)^2)$ terms, and thus large NLO and (at high enough energy) 
NNLO $K$-factors.

This mechanism thus provides a simple explanation for the large $K$-factors 
commonly found in hadroproduction processes at high energy. 
Constraining the incoming
gluons to be on-shell as in the usual LO calculation keeps the 
timelike intermediate gluon in fig.6b or fig.7b away from its 
mass-shell.
Releasing this constraint either by using the off-shell formalism 
used here, or by going to higher order in $\as$ in the more 
usual on-shell formalism (with in particular contributions from 
diagrams in which one (NLO) or both (NNLO) incoming 
gluons emits another gluon),
allows the intermediate gluon to get close to its mass-shell, and thus produces
the large enhancements evident in fig.16.
\topinsert
\vskip-11.5truecm
\vbox{
\epsfxsize=18truecm
\centerline{\epsfbox{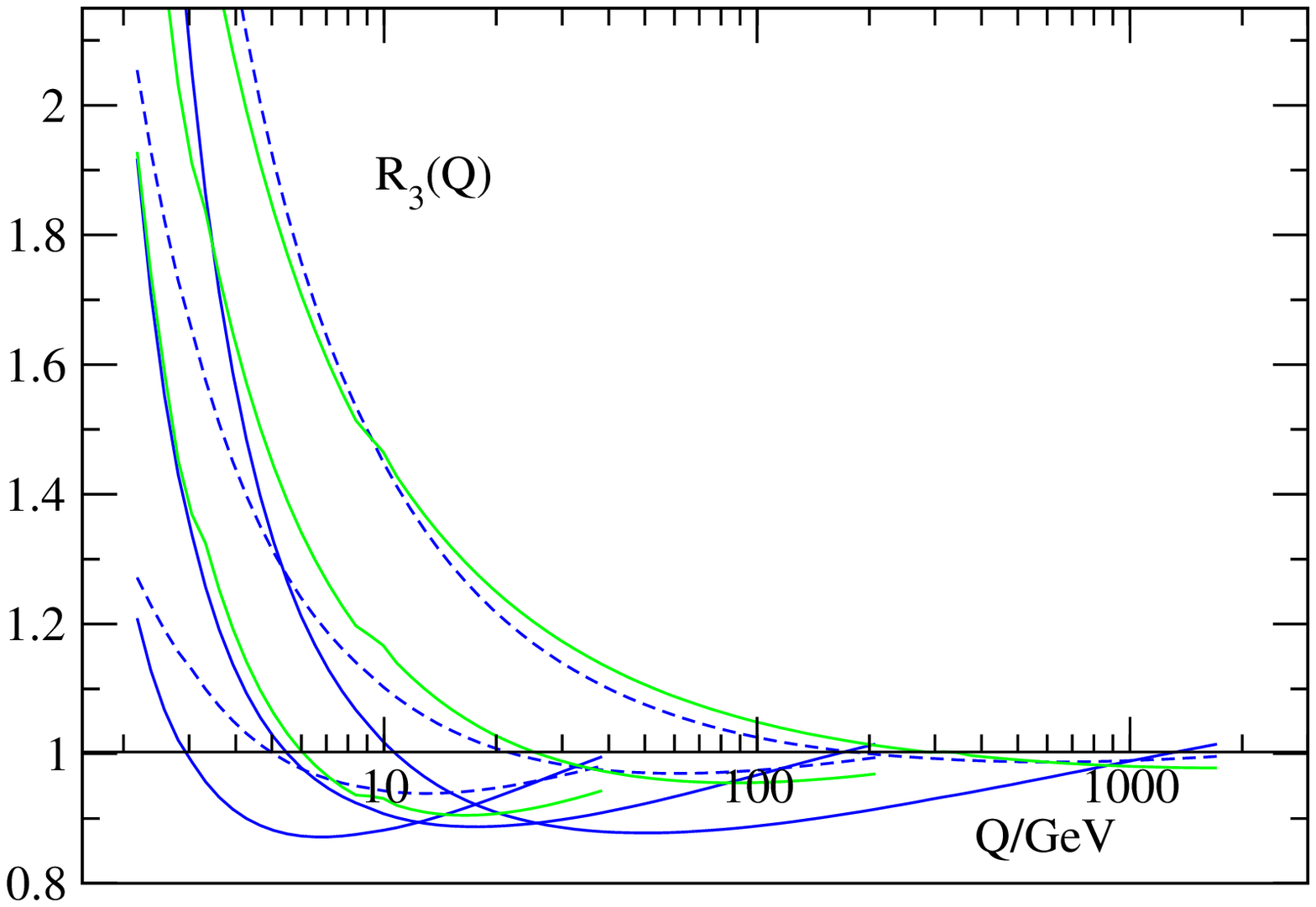}}
\vskip-2.0truecm
\bigskip
\hbox{
\vbox{\footnotefont\baselineskip6pt\narrower\noindent Figure 18: the 
resummation factor $R_3$ for a generic hadroproduction process. 
As in fig.11 the solid blue (lower) curve includes 
resummation in both evolution and hard cross-section, while the dashed 
blue curve only has resummation in the cross-section. The green (upper) 
curve compares NNLO to NLO unresummed perturbation theory.
 }}\hskip1truecm}
\endinsert 

To test this idea it is instructive to compare the resummed $K$-factors 
with those computed in 
fixed order-perturbation theory. At high energy this is easily done by using
the expansion \khapprox, applied to the luminosity evolved using GLAP 
evolution (at LO for the NLO $K$-factor, NLO for the NNLO, and NNLO for 
the NNNLO), just as we did for photoproduction in fig.10: the results are 
shown in fig.17. At the Tevatron NLO GLAP perturbation theory for the hard 
cross-section does pretty well except at very low scales (appropriate for 
charm production). At LHC NNLO is starting to be important already around the 
beauty threshold, while at a VLHC NLO is only good at scales above $60~\GeV$.
For beauty production at VLHC the NNLO correction is  
very large - around 50\% above NLO. More importantly, the NNNLO correction 
is nowhere an improvement on the NNLO result: beyond NNLO the resummed 
result is more useful.

To assess the impact of the resummation, in fig.18 we plot the 
``resummation factor''
\eqn\rthreedef{R^h_3(Q) = \frac{\Sigma^3_{\rm hh}(Q^2/S,Q)}
{\Sigma^{\rm NLO}_{\rm hh}(Q^2/S,S)},}
where, as in eqn.\rbdef, the reference cross-section
$\Sigma^{\rm NLO}_{\rm hh}$ 
is computed using the NLO hard cross-section and NLO GLAP gluon distribution.
Again we have a substantial cancellation between the suppression of 
the luminosity due to the resummation of the evolution and the 
enhancement of the cross-section due to the triple pole in the hard
cross-section. At intermediate scales ($Q\sim 8$, $20$, $60~\GeV$ for 
Tevatron, LHC and VLHC respectively) the resummation gives an overall 
suppression of around $10\%$: below these scales the suppression starts to 
turn into an enhancement due to the rapidly rising partonic cross-section.
However these enhancements are generally at scales sufficiently small that
they would escape detection in all but a very forward detector. Again we 
also show the ratio of the NNLO to NLO fixed order computation: this 
overestimates the cross-section because it underestimates the 
suppression of the gluon-gluon luminosity due to resummation.

\subsec{Hadroproduction of Heavy Quarks}

As a specific example of resummation in hadroproduction we now consider
the hadroproduction of heavy quarks, specifically $b\bar{b}$-production at the 
Tevatron, LHC and VLHC. The LO contribution to the hard cross-section then 
involves the computation of the two diagrams in fig.6 with both 
incoming gluons taken off-shell. Unlike the photoproduction cross-section 
eqn.\jay\ the hadroproduction cross-section has been 
evaluated analytically only in the high energy limit $N=0$: the result is
the impact factor \refs{\rdbrke,\ccfac}\foot{In 
refs.\refs{\cchglu,\rdbrke,\ccfac} the function $H(N,M_1,M_2)$ is denoted by 
$h_\omega(\gamma_1,\gamma_2)$.}
\eqn\aitch{\eqalign{
H(0,&M_1,M_2) = \alpha_s^2 \frac{\pi}{N_c^2-1}\Gamma(1+M_1)\Gamma(1-M_1)
\Gamma(1+M_2)\Gamma(1-M_2)\cr
&\times\Big[ 4N_c\frac{(\Gamma(3-M_1-M_2))^2}{(1-M_1-M_2)\Gamma(6-2(M_1+M_2))}
\Big(1+\Big(\frac{\Gamma(1-M_1-M_2)}{\Gamma(1-M_1)\Gamma(1-M_2)}\Big)^2\Big)\cr
&-\frac{2}{N_c}(7-5(M_1+M_2)+3M_1M_2)\frac{\Gamma(2-M_1)\Gamma(2-M_2)
\Gamma(1-M_1-M_2)}{\Gamma(4-2M_1)\Gamma(4-2M_2)}\Big].
}}
The second term in this expression is due solely to the ``abelian'' diagram
fig.6a: when one leg is on-shell (e.g. 
$M_2=0$ and $M_1=M$) this term reduces to $C(0,M)$ eqn.\jayz\ up to an overall 
constant vertex factor. The first term is due to the 
intrinsically nonabelian diagram fig.6b: it is 
this piece which contains the triple pole singularity \triplepole, which 
dominates the cross-section at high energy.

Consider the structure of the result 
eqn.\aitch\ in various regions of the $M_1$-$M_2$ 
plane fig.15. It has higher twist (simple) poles at $M_1,M_2 = -1,-2,\ldots$,
and infrared (anticollinear) poles at $M_1,M_2 = 1,2,\ldots$: for example near 
$M_1=1$
\eqn\aitchpolone{
H(0,M_1,M_2) \sim \alpha_s^2 \frac{4\pi}{N_c}\Big[
\frac{1}{1-M_1}\frac{\Gamma(M_2)\Gamma(1-M_2)(\Gamma(2-M_2))^2}{\Gamma(4-2M_2)}
+O(1)\Big],
}
thus a simple pole except when $M_2$ is an integer, at which special points
there is a double pole. However it also has lines of singularity when 
$M_1+M_2=1,2,\ldots$: writing $M_\pm\equiv M_1\pm M_2$, 
the Laurent expansion about $M_+=1$ is  
\eqn\aitchpolpm{\eqalign{
H(&0,M_1,M_2) \sim \alpha_s^2 \frac{\pi}{N_c^2-1}\Big[
\frac{N_c}{6}\frac{1-M_-^2}{(1-M_+)^3}\cr
&+\frac{1}{(1-M_+)^2}\frac{N_c}{18}\Big(3(1-M_-^2)(2\psi(1)
-\psi(\half+\half M_-)-\psi(\half-\half M_-))-11+5M_-^2\Big)\cr
&\qquad -\frac{1}{1-M_+}\Big(\smallfrac{(67+72(\ln 2)^2-132\ln 2)N_c}{54}
-\smallfrac{11\pi^3}{384N_c}+O(M_-^2)\Big)+O(1)\Big],
}}
i.e. triple, double and simple poles, except again at the special points
$M_-=\pm 1$ (and thus $(M_1,M_2)=(1,0)$ or $(0,1)$) where the triple pole 
reduces to a double pole, the double to a single.

For $M_1$ and $M_2$ close to zero the impact factor is regular, 
as it must be, with Taylor expansion
\eqn\aitchtaylor{\eqalign{
&H(0,M_1,M_2) = \alpha_s^2 \frac{\pi}{N_c^2-1}\Big[
\big(\smallfrac{4N_c}{15}-\smallfrac{7}{18N_c}\big)
+\big(\smallfrac{154N_c}{225}-\smallfrac{41}{54N_c}\big)(M_1+M_2)\cr
&+\big(\smallfrac{4924N_c}{3375}-\smallfrac{122}{81N_c}\big)(M_1^2+M_2^2)
+\big(\smallfrac{(9848-150\pi^2)N_c}{3375}
-\smallfrac{(470+21\pi^2)}{324N_c}\big)M_1M_2\cr
&+\big(\smallfrac{(150544-27000\zeta_3)N_c}{50625}
-\smallfrac{(730-189\zeta_3)}{243N_c}\big)(M_1^3+M_2^3)\cr
&+\big(\smallfrac{(150544-22500\zeta_3-1925\pi^2)N_c}{16875}
-\smallfrac{(2776+378\zeta_3+123\pi^2)}{972N_c}\big)(M_1+M_2)M_1M_2
+O(M^4)\Big]\cr
\simeq& 0.2633\alpha_s^2 \big(1+2.69(M_1+M_2)+5.78(M_1+M_2)^2-1.50 M_1M_2\cr
&\qquad\qquad\qquad\qquad
+9.41(M_1+M_2)^3-2.79(M_1+M_2)M_1M_2+O(M^4)\big).
}}
The rather large numerical coefficients in this expansion (compare them to
those of the expansion \jayztaylor), in particular of the powers of $M_1+M_2$,
are due to the dominance of the nearby triple pole singularity \aitchpolpm: 
if we first subtract the triple pole, and then Taylor expand what is left,
we find
\eqn\aitchpolx{\eqalign{
&H(0,M_1,M_2) \simeq \alpha_s^2 \frac{\pi}{N_c^2-1}\Big[
\frac{N_c}{6}\frac{1}{(1-M_1-M_2)^3}
+\big(\smallfrac{N_c}{10}-\smallfrac{7}{18N_c}\big)\cr
&\qquad+\big(\smallfrac{83N_c}{450}-\smallfrac{41}{54N_c}\big)(M_1+M_2)
+\big(\smallfrac{1549N_c}{3375}-\smallfrac{122}{81N_c}\big)(M_1^2+M_2^2)\cr
&\qquad+\big(\smallfrac{(3098-150\pi^2)N_c}{3375}
-\smallfrac{(470+21\pi^2)}{324N_c}\big)M_1M_2\cr
&\qquad+\big(\smallfrac{(66169-27000\zeta_3)N_c}{50625}
-\smallfrac{(730-189\zeta_3)}{243N_c}\big)(M_1^3+M_2^3)\cr
&+\big(\smallfrac{(66169-22500\zeta_3-1925\pi^2)N_c}{16875}
-\smallfrac{(2776+378\zeta_3+123\pi^2)}{972N_c}\big)(M_1+M_2)M_1M_2
+O(M^4)\Big]\cr
\simeq& 0.2633\alpha_s^2 \big(\smallfrac{0.74591}{(1-M_1-M_2)^3}
+0.2541+0.448(M_1+M_2)+1.31(M_1+M_2)^2\cr 
&\qquad\qquad -1.50 M_1M_2 +1.95(M_1+M_2)^3-2.79(M_1+M_2)M_1M_2+O(M^4)\big).
}}
A further reduction in the coefficients may be obtained by also 
subtracting the double pole.

It is now straightforward to compute the hadronic cross-sections using
either the Taylor expansion \aitchtaylor\ or the more precise expansion
\aitchpolx, the expressions \powtricksq\ and \Ltderivs\ for the powers 
of $M_1$ and $M_2$, the representation \hadmelinvpole\ for the
triple pole in \aitchpolx, and the gluon-gluon luminosity shown in fig.12.
The results are presented in fig.19 as a $K$-factor 
\eqn\kayhbdef{K_B(p_T)=\frac{\Sigma^B_{h h}(Q^2/S,Q)}
{\Sigma^0_{h h}(Q^2/S,Q)}\Big\vert_{Q^2=m_B^2+p_T^2},}
where $\Sigma^B_{\rm hh}$ is the fully resummed calculation, and the 
reference cross-section $\Sigma^0_{h h}$ is evaluated using 
$H(0,0,0)=\as^2\smallfrac{181\pi}{2160}$, thus
dividing out the overall normalization and the 
primary dependence on the gluon-gluon luminosity. Both
cross-sections are evaluated at $Q^2= m_B^2+p_T^2$, where 
$p_T$ is its minimum average transverse momentum of the $b\bar{b}$ pair. 
Note that we do not compute the transverse momentum 
distribution (the hard cross-section $H(N,M_1,M_2)$ is fully inclusive):
the $p_T$ dependence of \kayhbdef\ is simply a reflection of the restriction
of the phase space when we require a higher invariant mass in the final state.
We expect this to be the dominant effect here.

\topinsert
\vskip-11.5truecm
\vbox{
\epsfxsize=18truecm
\centerline{\epsfbox{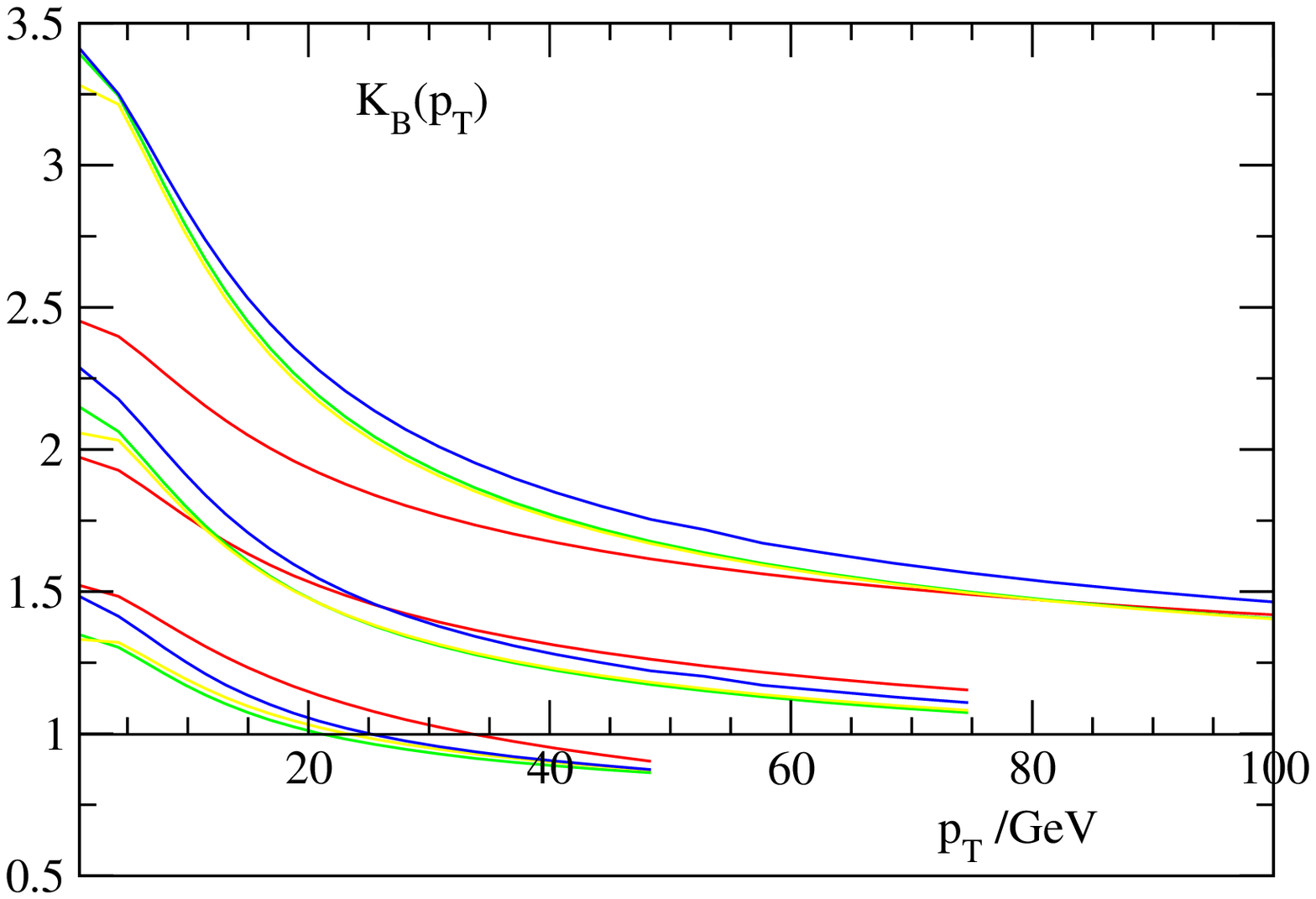}}
\vskip-2.2truecm
\bigskip
\hbox{
\vbox{\footnotefont\baselineskip6pt\narrower\noindent Figure 19: the 
$K$-factor for hadroproduction of $b\bar{b}$ pairs as a 
function of their minimum
$p_T$, at the Tevatron (lower), LHC (middle) and VLHC (upper). As in fig.17 
the blue curves are the resummed result, while the 
red green and yellow 
curves are fixed order perturbation theory at NLO, NNLO and NNNLO respectively,
computed as described in the text.
 }}\vskip0.0truecm}
\endinsert 

\topinsert
\vskip-11.5truecm
\vbox{
\epsfxsize=18truecm
\centerline{\epsfbox{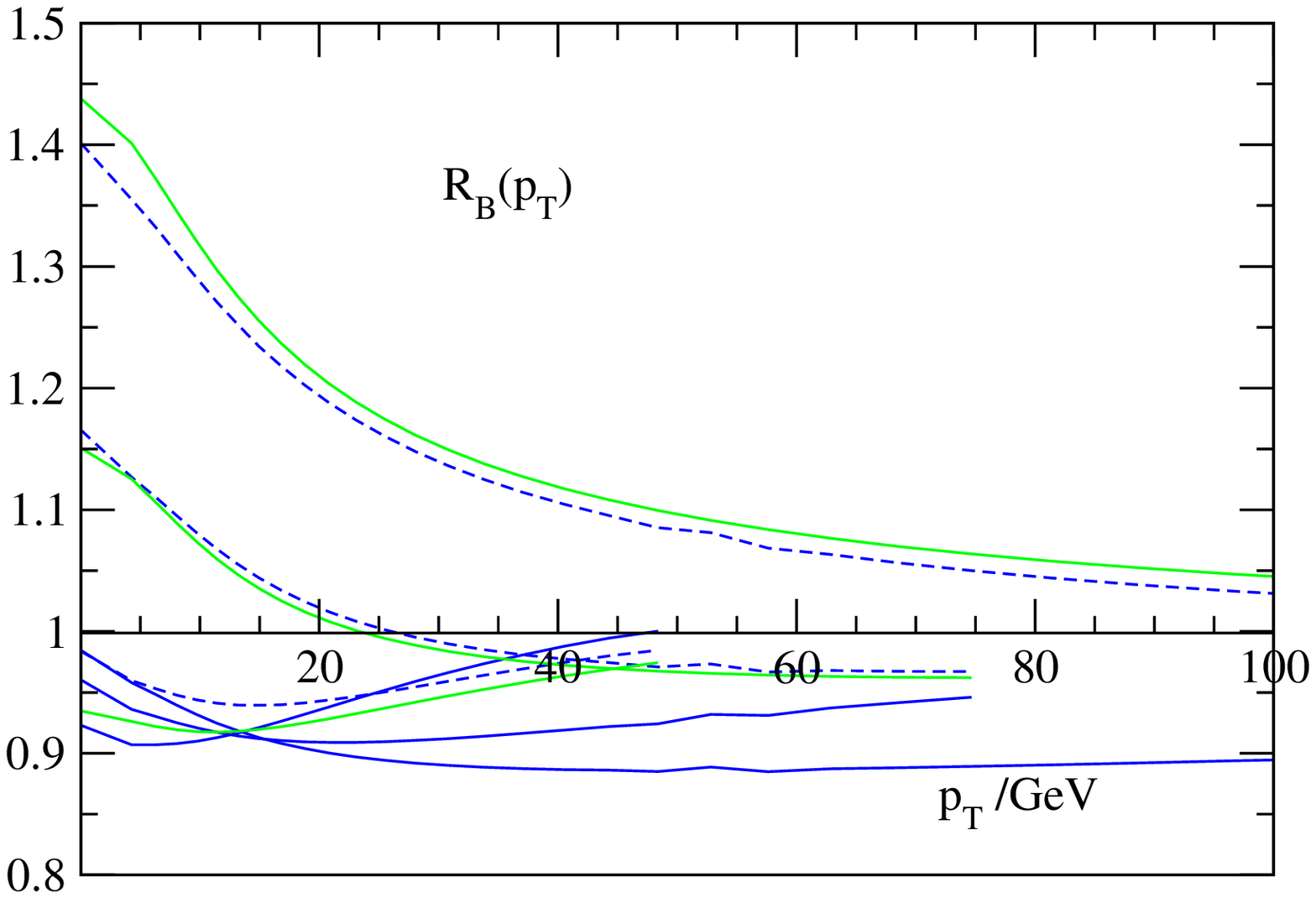}}
\vskip-2.2truecm
\bigskip
\hbox{
\vbox{\footnotefont\baselineskip6pt\narrower\noindent Figure 20:  the 
resummation factor $R_B$ for a hadroproduction of $b\bar{b}$ pairs plotted 
against their transverse momentum. As in fig.18 the solid blue (lower) 
curve includes 
resummation in both evolution and hard cross-section, while the dashed 
blue curve only has resummation in the cross-section. The green (upper) 
curve is
the same computation comparing NNLO to NLO perturbation theory.
 }}\vskip0.0truecm}
\endinsert 
As we found in the previous section the naive Taylor expansion \aitchtaylor\ 
and the pole resummed expansion \aitchpolx\ give almost identical results
throughout the entire kinematic range: the two resummed curves in fig.19 are 
indistinguishable. In fact the convergence is so rapid that 
only the $O(M)$ (NLO) and $O(M^2)$ (NNLO) terms in 
\aitchtaylor\ are actually needed: NNNLO contributions would 
only become significant for charm at a VLHC.

For comparison we also show in fig.19 results for fixed order (GLAP) 
perturbation theory at NLO, NNLO, and NNNLO, computed using the high 
energy approximation (as in fig.17) by using GLAP evolution, and keeping only 
the $O(M)$, $O(M^2)$ and $O(M^3)$ terms respectively in \aitchtaylor. At 
the Tevatron both the resummation and the NNLO GLAP corrections to the 
partonic cross-section give a slight suppression, while at the LHC there is 
a modest enhancement, at least at low $p_T$. Only at the VLHC is there a 
substantial enhancement, and even there a fixed order NNLO calculation of
the partonic cross-section would be quite sufficient.

In order to estimate the overall effect of the resummation, we 
show in fig.20 the resummation factor
\eqn\rhbdef{R_B(p_T) = \frac{\Sigma^B_{\rm hh}(Q^2/S,Q)}
{\Sigma^{\rm NLO}_{\rm hh}(Q^2/S,Q)}\Big\vert_{Q^2=m_B^2+p_T^2}.}
The reference cross-section is now the NLO fixed order cross-section 
computed with the NLO GLAP evolved gluon-gluon luminosity: unlike 
in the $K$-factor effects due to the different evolution 
of the luminosity are thus now included.
As in fig.18 we find that the overall effect of the resummation is a net
suppression by about $10\%$, which gradually goes away as $p_T$ increases.
Just above threshold this suppression is remarkably independent of the 
centre-of-mass energy $S$ of the machine, and when this 
is very high (in particular for VLHC) the suppression goes 
away very slowly. So once again the enhancement expected from the triple 
pole in the cross-section is more than compensated by a suppression of the 
gluon-gluon luminosity.

It should be noted that all these calculations are only estimates: in 
particular a proper matching to the high $\rho$ (Sudakov) region, 
inclusion of quark effects and realistic fitted parton distributions 
could all change the results substantially. In particular the 
suppression of the gluon-gluon luminosity by the resummation of the 
evolution is probably overestimated, since the starting distribution 
at $2~\GeV$ is held fixed, rather than fitted to data.  However 
the band between the 
upper and lower curves in fig.18 and fig.20 is probably a reasonable estimate 
of the current overall uncertainty due to resummation in hadroproduction 
cross-sections: for inclusive $B$ production this means roughly $-5\pm 5\%$ 
at the Tevatron, 
$5\pm 10\%$ at the LHC and $20\pm 20\%$ at a VLHC. 
In all cases it is probably comparable   
to the impact of NNLO in the partonic cross-section, and may result 
in a net suppression rather than an enhancement. At large $p_T$ the effect 
goes away as expected.

\subsec{Drell-Yan and Higher Orders}
\eqnn\khdyexact

Of course not all processes have the same structure of infrared 
singularities as heavy quark and inclusive jet production: as the number 
of particles in the final state increases, so does the range of  
possible infrared singularities. Indeed, adding one more particle to the final
state general adds one extra collinear and one extra soft singularity, thus 
increasing $n$ in eqns.\exptrickhad\ and \hadmelinvpole\ by two. 

Consider for example the gluonic contribution to Drell-Yan 
or vector boson production,
given by the diagrams in fig.3. The most singular of these diagrams is fig3c:
there is a soft singularity from the timelike gluon, a collinear 
singularity from the splitting into a $q\bar{q}$ pair, and then further 
soft and collinear singularities from the final vector boson emission. 
Unless there are accidental cancellations, we 
thus expect the impact factor for this process to have a pole of 
order five at $M_1+M_2=1$. 
The relevant $K$-factor is thus (using eqn.\hadmelinvpole\ with $n=5$)
$$\eqalignno{K_5(Q)
&=\frac{1}{L_z(\xi,t)}\int_{-i\infty}^{i\infty}\!\frac{dN}{2\pi i}\,e^{N\xi}
\int_0^\infty\!d\tau \,\tau^4 e^{-\tau}L_z(N,t+\tau)&\khdyexact
}$$
which may be evaluated either by computing the $\tau$ integral 
numerically, just as we did for $K_3$ eqn.\khexact.

\topinsert
\vskip-11.5truecm
\vbox{
\epsfxsize=18truecm
\centerline{\epsfbox{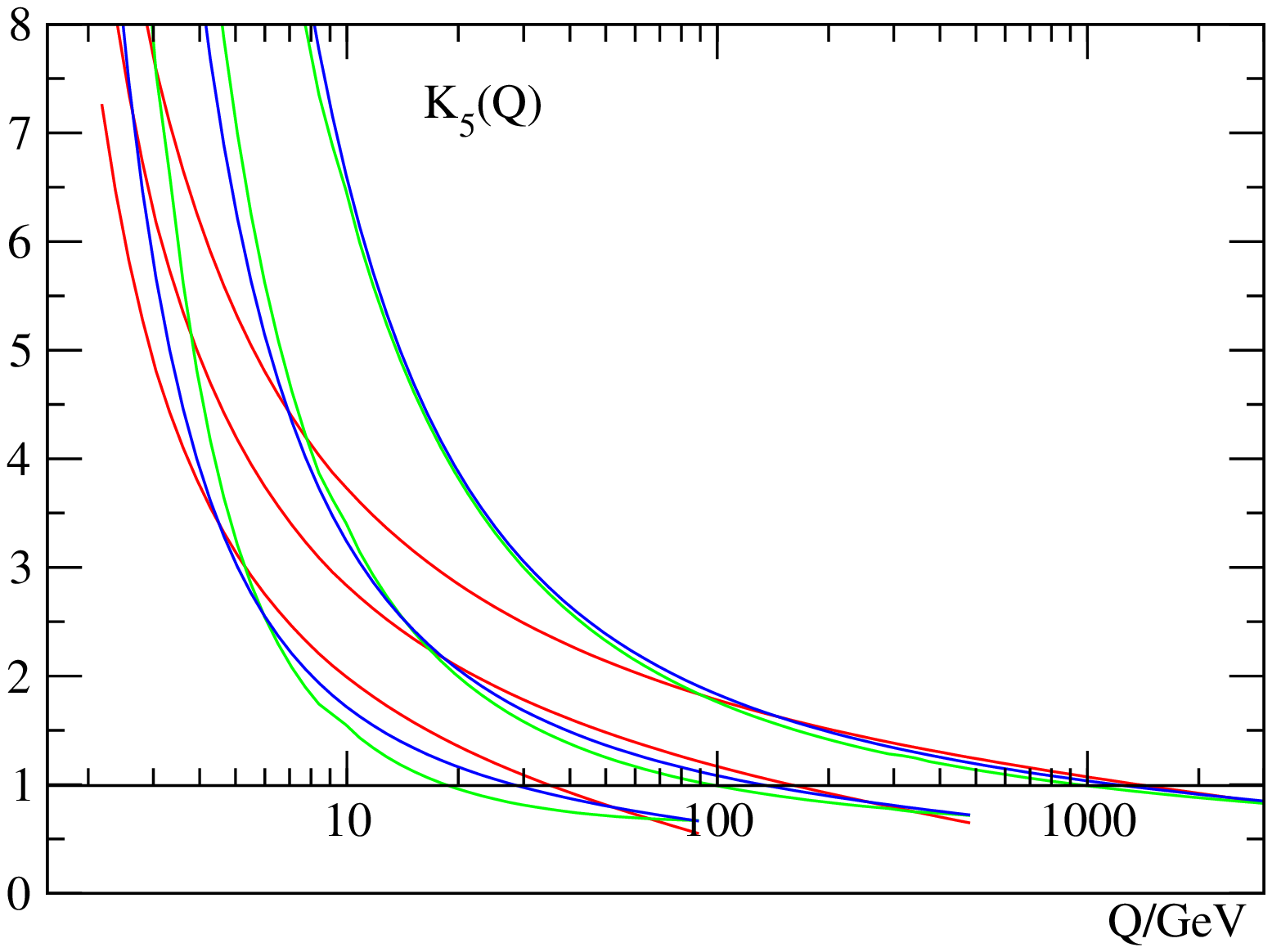}}
\vskip-2.0truecm
\bigskip
\hbox{
\vbox{\footnotefont\baselineskip6pt\narrower\noindent Figure 21: the 
$K$-factor $K_5$, eqn.\khdyexact, appropriate for the gluonic contribution 
to Drell-Yan, vector boson production and prompt photon processes. 
The curves are labelled as in fig.17: 
blue is resummed, while red and green 
are estimates of fixed order perturbation theory at NLO and NNLO.
 }}\hskip-1truecm}
\endinsert 
The results are shown in fig.21. Comparing $K_5$ to $K_3$ shown in fig.17, the 
extra soft and collinear singularities produce a further overall enhancement 
as expected. However the qualitative features of the two plots are very 
similar: in particular the resummed result starts to grow faster than the NLO
result at about the same scale in each case. For $W$ or $Z$ production at 
LHC the correction is in the region of 20-30\%: 
it only becomes large at VLHC. However for production of 
Drell-Yan pairs at around $10 \GeV$ at LHC the NLO correction is as large as 
a factor of three, and requires resummation. 

For these gluon-gluon processes NLO means $O(\alpha\as^3)$, which is NNNLO 
in the usual nomenclature of fixed order perturbation theory.
Thus unlike in the previous heavy quark and inclusive jet (ie $n=3$) 
estimates, here even the NLO curve is a new result: only 
the LO contribution to the 
gluonic contribution to Drell-Yan and vector boson production, $gg\to W+X$  
(i.e. the graphs in fig.3 but with the incoming gluons on-shell), 
has been computed exactly in fixed order perturbation theory \DYNNLO. 

We can also use fig.21 to estimate resummation 
corrections to prompt photon production, or to the 
three-jet inclusive cross-section, since the relevant 
diagrams again have the same structure as those in fig.3 
(with for jets the quarks and vector bosons replaced by gluons) 
and thus the same structure of infrared singularities. 

Since the number of infrared logarithms increases by two at each 
extra order in $\alpha_s$, and since, unlike the logarithms from 
high energy ($N=0$) and collinear ($M=0$) singularities these 
logarithms are not being explicitly resummed, one might 
worry that they might destabilise the hierarchy of terms in the 
resummed perturbation theory \cchglu.
To address this problem we
note that we can estimate the size of subleading resummation corrections to  
by computing the ratios $\alpha_s(Q) K_{n+2}(Q)/K_n(Q)$:
these are plotted in fig.22 for $n=3$ (heavy quark production and 
inclusive jets) and $n=5$ (Drell-Yan and vector boson production). 
It is clear from this plot that even at low scales and at VLHC the 
enhancements due to the two extra infrared logarithms are never 
very substantial.

\topinsert
\vskip-11.5truecm
\vbox{
\epsfxsize=18truecm
\centerline{\epsfbox{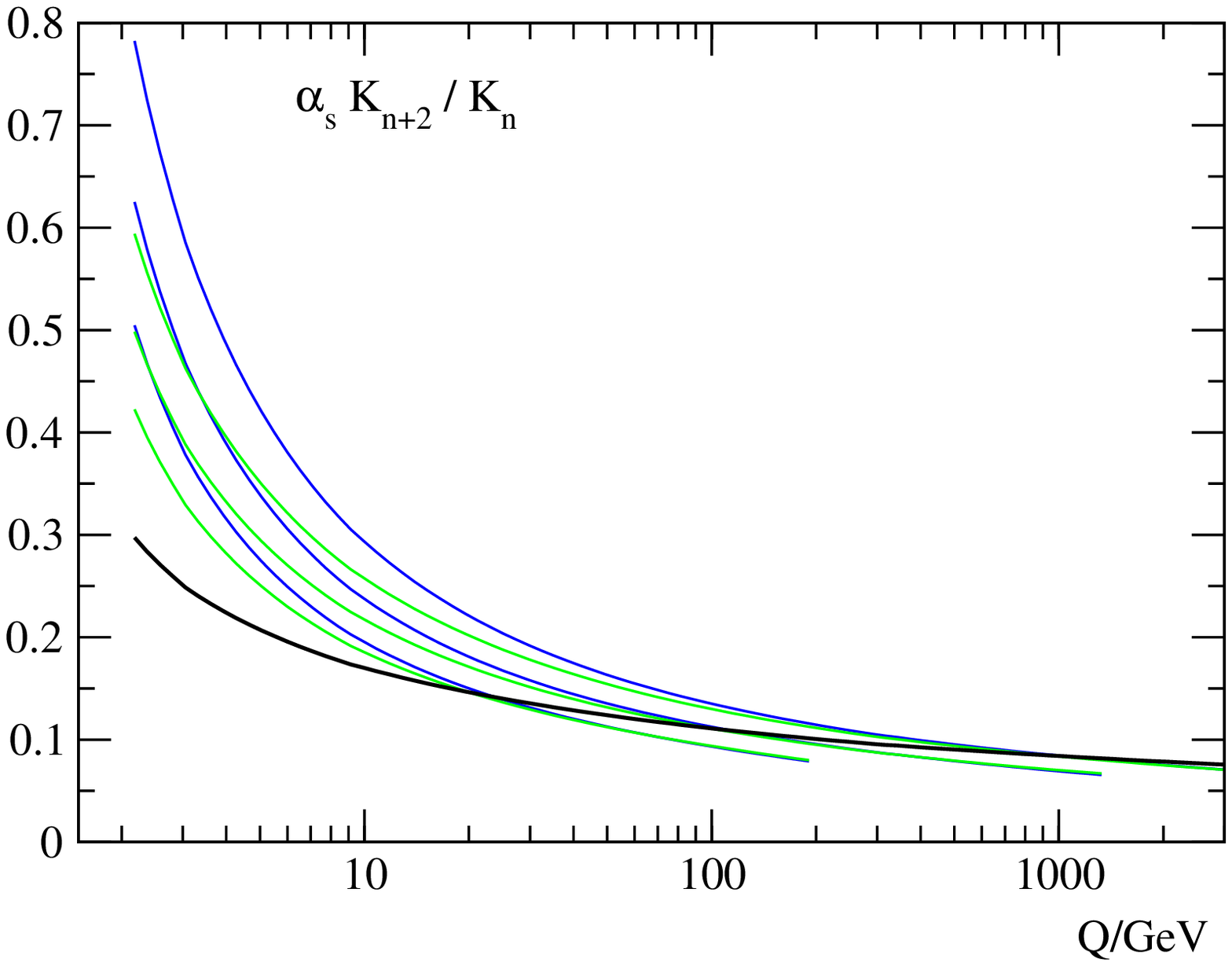}}
\vskip-1.5truecm
\bigskip
\hbox{
\vbox{\footnotefont\baselineskip6pt\narrower\noindent Figure 22: the  
ratios $\alpha_sK_{n+2}/K_n$ for  
$n=3$ (blue, upper) and $n=5$ (green, lower) showing relative enhancement of 
subleading corrections. As usual the three curves of each colour are for 
Tevatron, LHC, and VLHC.
Also shown is $\alpha_s(Q)$ (solid black) as a baseline expectation.
 }}\hskip-1truecm}
\endinsert 
Moreover as $n$ increases the enhancements are systematically reduced. The 
reason for this is not hard to find: a singularity of order $n$ results in a 
smearing of the gluon-gluon luminosity with a distribution proportional to 
$\tau^{n-1}e^{-\tau}$, eqn.\hadmelinvpole, which is peaked at $\tau = n-1$. 
Since the scale dependence of the luminosity is very smooth, thanks to 
asymptotic freedom, the main effect of the smearing at 
large $n$ is to shift the 
scale at which the luminosity is evaluated from $t$ to $t+n-1$. Thus 
at large $n$ 
\eqn\Krat{\frac{K_{n+2}(Q)}{K_n(Q)}\sim \frac{L_z(\xi,t+n+1)}{L_z(\xi,t+n-1)}
\sim 1 + \frac{2}{L_z(\xi,t+n)}\frac{\partial}{\partial t}L_z(\xi,t+n),}
which tends to one for large $n$ and large $t$ (see fig.14).
It follows that there is no reason to suspect that the infrared logarithms 
spoil our resummed perturbation theory: for high energy processes with a 
single hard scale, we have indeed resummed all large logarithms.


\newsec{Rapidity Distributions}

\subsec{Gluon-gluon Rapidity}

Besides hadroproduction total cross-sections it is also 
interesting to consider rapidity 
distributions: in the central rapidity region both partons carry roughly 
the same fraction of longitudinal momentum, but at large rapidities one
of the partons is at a much smaller value of $x$ than the other (see fig.1 
for the ranges covered at various machines) so it is perhaps here that 
one might expect the effects of resummation to be most significant.

We define
\eqn\rapdef{z\equiv x_1x_2,\qquad\qquad \eta = \half\ln (x_1/x_2),}
so that $s=zS$ is the centre-of-mass energy in the partonic collision,
and $\eta$ is the (pseudo)-rapidity: $\eta=0$ in the central region,
becoming large and positive/negative in the forward/backward regions.
In terms of $z$ and $\eta$ the fraction of longitudinal momentum 
in each of the two colliding partons is
\eqn\xonetwo{x_1=\sqrt{z}e^\eta,\qquad\qquad x_2= \sqrt{z}e^{-\eta}.}
Since $s\geq Q^2$, while $x_1,x_2\leq 1$, we must have
\eqn\raplimsa{ \rho\leq z\leq 1,\qquad\qquad 
\half\ln z\leq\eta\leq\half\ln \smallfrac{1}{z},}
\topinsert
\vskip-11.5truecm
\vbox{
\epsfxsize=18truecm
\centerline{\epsfbox{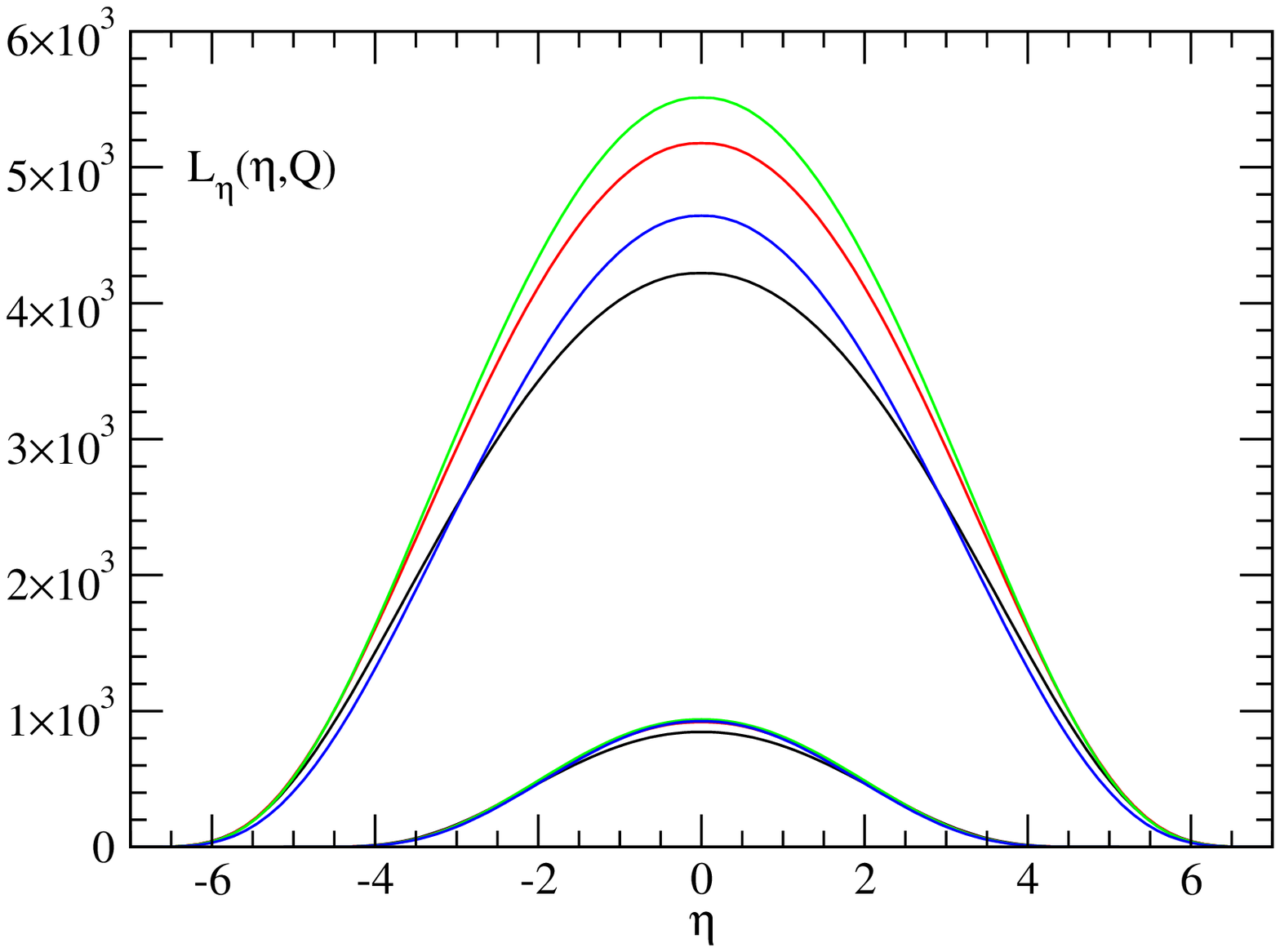}}
\vskip-1.0truecm
\bigskip
\hbox{
\vbox{\footnotefont\baselineskip6pt\narrower\noindent Figure 23: The 
gluon-gluon rapidity distribution $L_\eta(\eta,Q)$ at the LHC 
for $B$-production ($Q=10\GeV$) (upper curves) 
and $W$-production ($Q=75\GeV$) (lower curves). The blue curves are for 
gluons evolved using NLO resummation, while the black, red and green curves
are for gluons evolved using LO,NLO and NNLO GLAP.
 }}\vskip0.0truecm}
\endinsert 

In terms of $z$ and $\eta$ the factorization formula \hadrofac\ may be written 
\eqn\hadfacrapa{\Sigma_{\rm hh}(\rho,Q) 
= \int_\rho^1 
\! {dz\over z}
\int_{\half\ln z}^{\half\ln \smallfrac{1}{z}}\!\! {d\eta}
\int \! {d^2{\bf{k_1}}\over\pi{\bf{k}}_1^2}
\int \! {d^2{\bf{k_2}}\over\pi{\bf{k}}_1^2} 
\Sigma_{gg}\big({\rho\over z},
{{\bf{k_1}}\over Q},{{\bf{k_2}}\over Q}\big)
G\big(\sqrt{z}e^\eta,{\bf{k}}_1^2\big)
G\big(\sqrt{z}e^{-\eta},{\bf{k}}_2^2\big).
}
Since the hard cross-section depends only on $z$, not on $\eta$ (because 
of invariance under longitudinal boosts), we may perform the 
integral over $\eta$ first to give the gluon-gluon luminosity \lumidef.
If instead however we perform the integrals in the opposite order
\eqn\intreverse{
\int_\rho^1 
\! {dz\over z}
\int_{\half\ln z}^{\half\ln \smallfrac{1}{z}} \! {d\eta}
=\int_{-\xi/2}^{\xi/2} \! {d\eta}\int_\rho^{e^{-2|\eta|}} 
\! {dz\over z},}
where as usual $\xi = \ln 1/\rho$, the factorization \hadfacrapa\ may be 
written instead as a factorization for the differential cross-section:
\eqn\hadfacrapb{\frac{d\Sigma_{\rm hh}}{d\eta} 
= \int_\rho^{e^{-2|\eta|}} \!\! {dz\over z}
\int \! {d^2{\bf{k_1}}\over\pi{\bf{k}}_1^2}
\int \! {d^2{\bf{k_2}}\over\pi{\bf{k}}_2^2} 
\Sigma_{gg}\big({\rho\over z},
{{\bf{k_1}}\over Q},{{\bf{k_2}}\over Q}\big)
G\big(\sqrt{z}e^\eta,{\bf{k}}_1^2\big)
G\big(\sqrt{z}e^{-\eta},{\bf{k}}_2^2\big).
}

At high energy the explicit dependence of the hard cross-section 
$\Sigma_{gg}\big({\rho\over z},
{{\bf{k_1}}\over Q},{{\bf{k_2}}\over Q}\big)$ on $z$ is relatively
weak: the dominant contribution to the $z$ dependence of the collinear 
cross-section comes about through the dependence of the 
off-shell cross-section on the transverse momenta. This makes the 
computation of rapidity distributions at high energy  
particularly simple, since the rapidities of the final state particles are
directly related to the rapidities of the colliding partons.
It is then useful to define gluon-gluon rapidity distribution
\eqn\ggrap{L_\eta(\eta,k_1^2,k_2^2,\rho)=
\int_\rho^{e^{-2|\eta|}} \! {dz\over z} 
G\big(\sqrt{z}e^\eta,k_1^2\big)
G\big(\sqrt{z}e^{-\eta},k_2^2\big).}
Clearly the normalization of this distribution is not independent
of that of the gluon-gluon luminosity \lumidef: in fact
\eqn\intggrap{\int_{-\xi/2}^{\xi/2} \! {d\eta}\,
L_\eta(\eta,k_1^2,k_2^2,\rho) = \int_\rho^1 
\! {dz\over z}L_z(z,k_1^2,k_2^2).}
Again the gluon-gluon rapidity distribution may be readily computed from
$G(x,Q)$ displayed in fig.4: the result for $L_\eta(\eta,Q)\equiv 
L_\eta(\eta,Q^2,Q^2,\rho)$ 
is shown in fig.23 for both 
$B$-production and $W$-production at LHC. The falloff as $\eta\to\pm\xi/2$ is
very rapid, essentially because the gluon is very small at large $x$ (remember
that in all these calculation we are suppressing the quark contribution,
so the valence region is underpopulated).

In terms of the gluon-gluon rapidity at high energy the differential 
cross-section is then given by
\eqn\diffxsec{\eqalign{\frac{d\Sigma_{\rm hh}}{d\eta}&\simeq
\int \! {d^2{\bf{k_1}}\over\pi\bf{k_1^2}}
\int \! {d^2{\bf{k_2}}\over\pi\bf{k_2^2}} 
\Sigma_{gg}\big(1,{{\bf{k_1}}\over Q},{{\bf{k_2}}\over Q}\big)
L_\eta(\eta,k_1^2,k_2^2,\rho)\cr
&=\int_{-i\infty}^{i\infty}\!\frac{dM_1}{2\pi i}\frac{dM_2}{2\pi i}
\,e^{t(M_1+M_2)}
H(0,M_1,M_2)L_\eta(\eta,M_1,M_2,\rho),}}
where the (double) Mellin transform is defined in the usual way:
\eqn\ggrapmel{
L_\eta(\eta,M_1,M_2,\rho)\equiv \int_0^\infty \! {dk_1^2\over k_1^2}
\,\Big({k_1^2\over \Lambda^2}\Big)^{-M_1}
\int_0^\infty \! {dk_2^2\over k_2^2}\, 
\Big({k_2^2\over \Lambda^2}\Big)^{-M_2}
L_\eta(\eta,k_1,k_2,\rho).}
Note that unlike \hadmelinvlumi\ the expression \diffxsec\ only holds in 
the approximation where we 
ignore the $N$ dependence in the hard cross-section $H(N,M_1,M_2)$.

\subsec{$B$-production and $W$-production at LHC}

For a particular impact factor $H(0,M_1,M_2)$ the 
corresponding differential cross-section \diffxsec\ may be readily
evaluated using eqns.\Ltderivs\ for powers of $M_1$ and $M_2$, and 
the exponentiation trick eqn.\exptrickhad\ for singularities. We consider 
two examples: $B$-production at the LHC, for which we use the impact factor 
\aitch\ expanded as \aitchtaylor\ or \aitchpolx, and the gluon-gluon 
contribution to $W$-production at LHC, which we assume is dominated at 
high energy by the infrared singularities in final state emission 
(fig.3c), and may thus be modelled by an $n=5$ pole:
\eqn\diffxsec{\eqalign{
\frac{d\Sigma^W_{\rm hh}}{d\eta}&\simeq
\alpha\as^2\int_{-i\infty}^{i\infty}\!\frac{dM_1}{2\pi i}\frac{dM_2}
{2\pi i}\,e^{t(M_1+M_2)}
\frac{r}{(1-M_2-M_2)^5}L_\eta(\eta,M_1,M_2,\rho)\Big\vert_{Q=m_W},\cr
&=\alpha\as^2 r\int_0^\infty\!d\tau \,\tau^4 e^{-\tau}L_\eta(\eta,t+\tau)
\Big\vert_{t=\ln m_W^2/\Lambda^2},}}
where $r$ is a normalization factor (just a number), and in the second 
line we have used eqn.\exptrickhad\ to evaluate the integrals 
over $M_1$ and $M_2$ just as we did in eqn.\khdyexact.  For 
$B$-production we use a similar expression, but here of course we take $n=3$.
Note however that now the 
integration over $z$ in eqn.\ggrap\ can take us into the region of large $x_1$
or $x_2$ at high rapidities, so here we must take care to match the 
cross-section smoothly in this region to the fixed order calculation 
to avoid spurious contributions.

\topinsert
\vskip-11.5truecm
\vbox{
\epsfxsize=18truecm
\centerline{\epsfbox{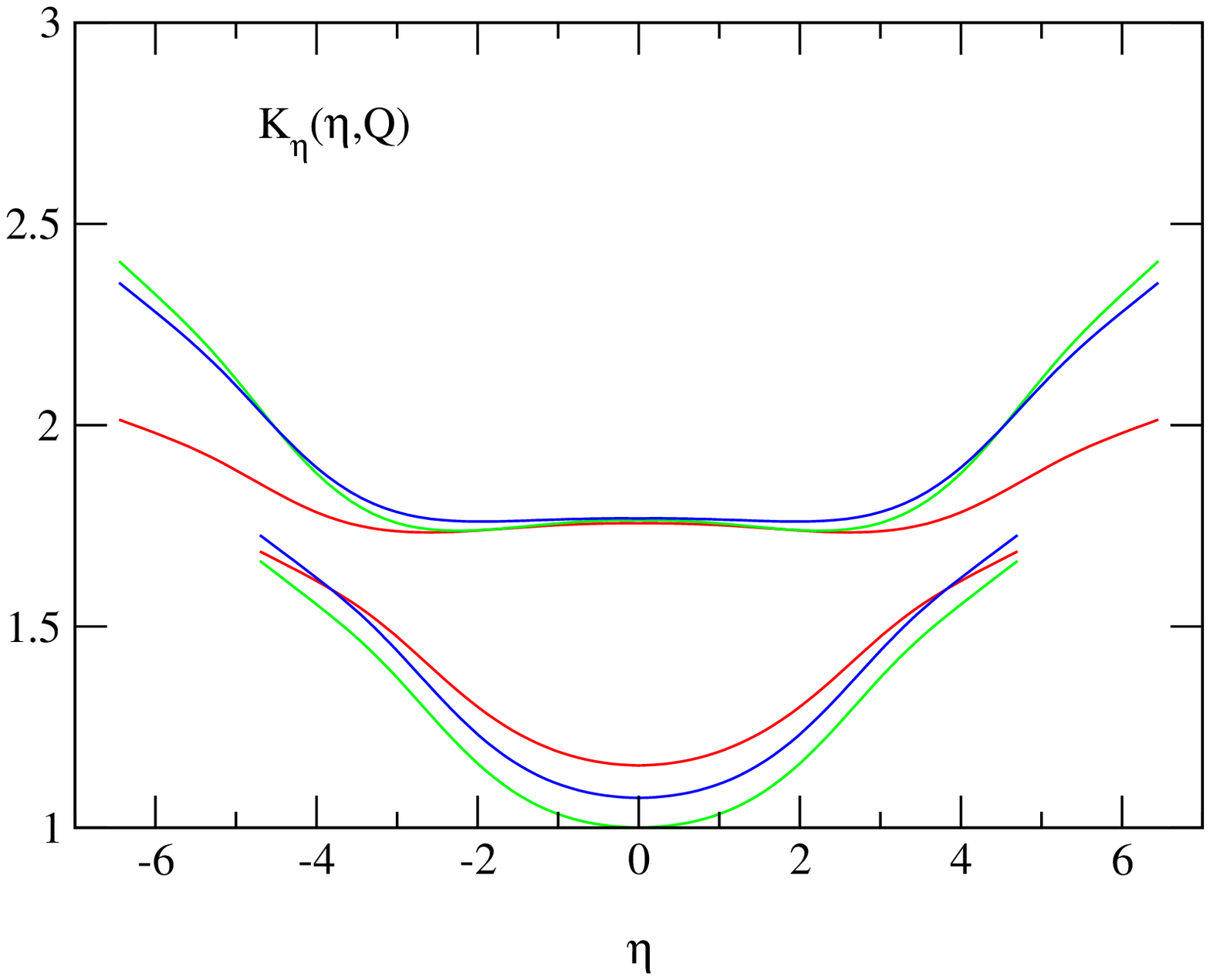}}
\vskip-0.5truecm
\bigskip
\hbox{
\vbox{\footnotefont\baselineskip6pt\narrower\noindent Figure 24: the 
$K$-factors $K_\eta(\eta,m_B)$ (upper curves) $K_\eta(\eta,m_W)$ 
(lower curves) for the rapidity distribution of $B$-production and 
(the gluon-gluon component of) $W$-production at the LHC. As in fig.17 
and fig.19 the blue curves are the resummed result, 
while the red and green curves are NLO and NNLO 
perturbation theory, estimated 
in the usual way.
 }}\vskip0.0truecm}
\endinsert 
Again we express the results of these calculations as $K$-factors: here
\eqn\kayrapdef{K_\eta(\eta,m_W)=\frac{d\Sigma^W_{\rm hh}}{d\eta}\Big/
\frac{d\Sigma^0_{\rm hh}}{d\eta}\Big\vert_{Q=m_W},}
and a similar expression for $K^B_\eta(\eta,m_B)$. The denominator is in 
each case the result obtained by setting $M_1=M_2=0$ in the impact factor, 
and thus divides out the unknown normalization $r$ and the 
primary dependence on the gluon-gluon rapidity.
The corresponding $K$-factors in perturbation theory may be estimated by 
expansion in powers of $M_1$ and $M_2$, and using the appropriate GLAP evolved
gluon-gluon rapidity distribution. 
\topinsert
\vskip-11.5truecm
\vbox{
\epsfxsize=18truecm
\centerline{\epsfbox{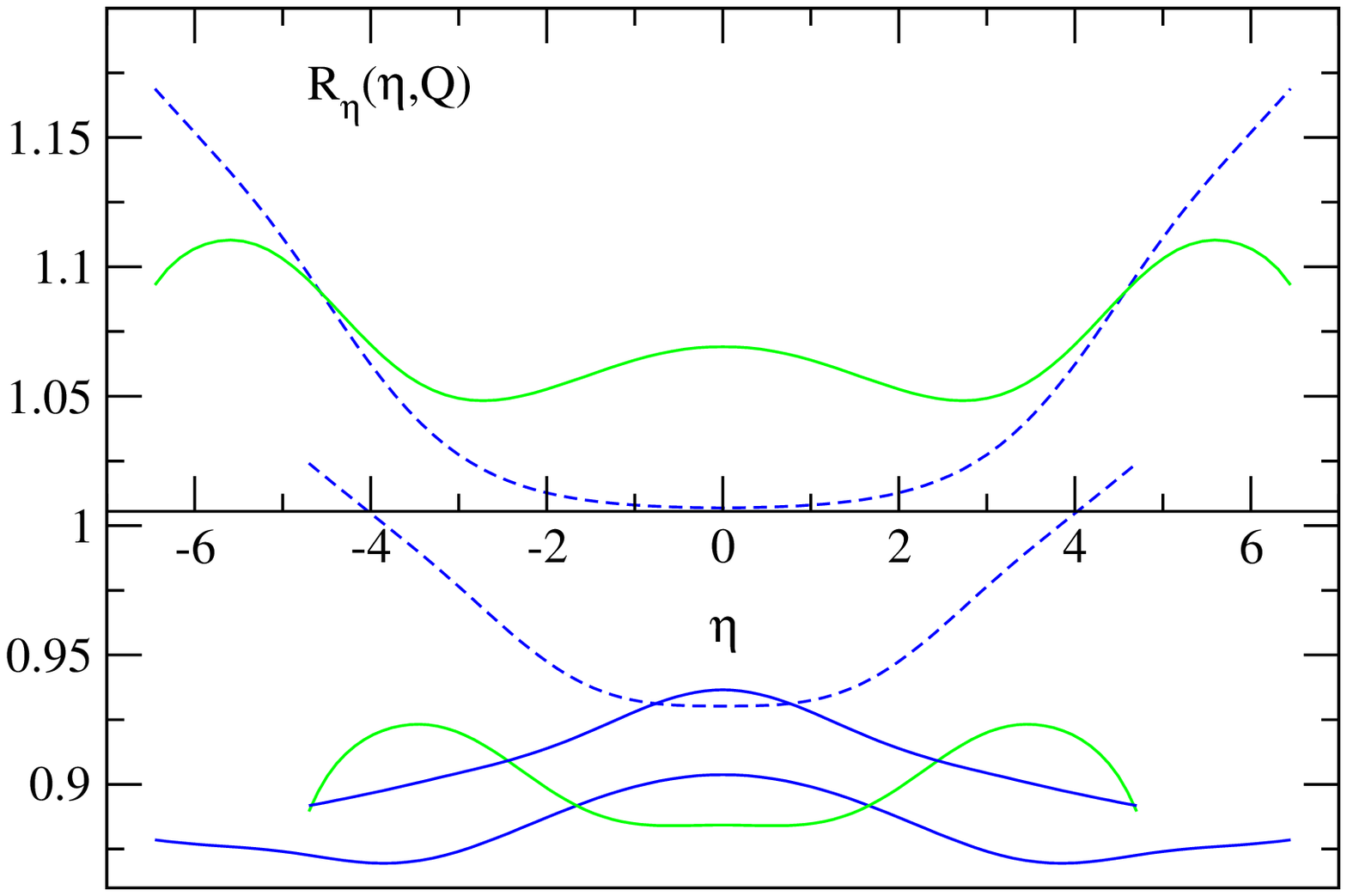}}
\vskip-2truecm
\bigskip
\hbox{
\vbox{\footnotefont\baselineskip6pt\narrower\noindent Figure 25: the 
resummation factor $R_\eta$ for $B$-production (outer curves) and 
$W$-production (inner curves) at the LHC plotted 
against the rapidity $\eta$. The solid blue (lower) curves correspond 
to fully resummed calculations, while the dashed curves are 
the same calculations but with the effect of the evolution on the 
gluon-gluon rapidity distribution factored out. The true size of the 
resummation effect probably lies somewhere between these two extremes. 
The green (upper) curves are the result of a NNLO fixed order calculation.
 }}\vskip0.5truecm}
\endinsert

The results of these calculations are shown in fig.24. For $B$-production
the $K$-factor is already substantial in the central rapidity region, and 
rises further at large rapidities where one of $x_1$ or $x_2$ becomes very 
small. This rise is rather steeper for the resummed and NNLO calculations 
than it is at NLO, as expected. However this prediction 
needs to be interpreted with care, since at large rapidities 
one of the gluons is moving towards the high $x$ region, which is 
poorly modelled in this calculation since there are no quark contributions.
For the gluon-gluon contribution to $W$ production the $K$-factors 
in the central region are all rather small, but increase quite quickly 
with rapidity. However here the resummed calculation is close to the NLO and 
NNLO calculations throughout the whole range. Note again that here even the
NLO curve is a new result. The change in the shape of the rapidity
distribution due to what is formally a NNNLO (i.e. $O(\alpha\as^3)$) 
contribution is thus quite striking.

To see more clearly the effect of the resummation alone, we also compute the 
``resummation factor''
\eqn\kayrapdef{R_\eta(\eta,m_W)=\frac{d\Sigma^W_{\rm hh}}{d\eta}\Big/
\frac{d\Sigma^{\rm NLO}_{\rm hh}}{d\eta}\Big\vert_{Q=m_W},}
again with a similarly expression for $B$-production, 
where now the reference differential cross-section is the NLO fixed 
order cross-section computed with a NLO GLAP evolved rapidity 
distribution, so that the effect of the resummation in both evolution and 
partonic cross-section are combined. The results are shown in fig.25. 
Resummation reduces the $B$ cross-section by around 10\% across the whole 
rapidity region, this effect being due almost entirely to evolution since 
without it there is an enhancement rising to around 15\% at large rapidity. 
The effect of resummation on $W$-production is rather less pronounced. 
Note that 
in both cases the enhancement due to the infrared singularities in the 
hard cross-section is largely cancelled by the suppression of the luminosity,
leaving a relatively flat distribution.

\newsec{Conclusions}

We have shown that the problem of integration of infrared 
singularities in photoproduction, electroproduction and hadroproduction 
cross-sections due to final state gluons becoming 
soft or collinear \refs{\cchglu,\rdbrke} may be solved through an 
exponentiation trick eqn.\exptrick\ and eqn.\exptrickhad\ 
respectively. This enables us to show that despite the dramatic enhancements 
found when when the coupling is fixed, when the coupling runs the effects
are much more modest. 
In particular we have shown that when the coupling runs the 
growth of the inclusive 
cross-sections at asymptotically high energy is given universally by the growth
of the resummed integrated gluon distribution, with no process dependent power
enhancements: all the infrared singularities are correctly 
factorised into the initial nonperturbative distribution, as they should be.

We find furthermore that except at very low scales 
the inclusive cross-sections are well approximated 
by keeping only the first few terms in the Taylor 
expansion of the partonic cross-section in powers of $M$, irrespective of the
presence of the nearby infrared singularities. Since this 
expansion in powers of $M$ is closely related to the usual perturbative 
expansion of the hard cross-section in powers of $\as/N$, this enables 
us to understand the behaviour of the expansion to fixed orders 
in $\as$. In 
particular we have shown that the large $K$-factors commonly found 
in hadroproduction processes at NLO and NNLO are at high energy due mainly 
to the infrared final state singularities (compare fig.16, fig.17, fig.19, 
fig.21 and fig.24), and moreover that although in some
kinematic regions (high energy, low invariant mass and large rapidity) 
the NNLO correction to the hard cross-section may be important, the 
series converges sufficiently rapidly that NNNLO corrections are in practice 
usually small. This is reassuring. 

We also find that although in the resummed perturbation theory the number of 
infrared logarithms increases by two for every extra power of $\as$, the 
effect of the extra logarithms is sufficiently benign that the hierarchy of 
the resummed perturbative expansion is not spoiled, and thus no further 
resummation is necessary. This again is due to asymptotic freedom, in 
the sense that it is only true when the coupling runs.

A useful way to characterise this interplay between collinear and 
high energy resummation is to ask in which regions of the kinematic
plane fig.1 logarithms of $Q^2$ or logarithms of $x$ are more important.
If the former are dominant, we may expand the partonic cross-section
$H(N,M_1,M_2)$ in powers of $M_1$ and $M_2$, keeping the full dependence 
on $N$, while if the small $x$ logarithms dominate, we may expand in 
powers of $N$ but keeping the full dependence on $M_1$ and $M_2$. The 
relevant regions are thus characterised by the relative importance of 
factors of $M$ and factors of $N$, or more specifically of the two
integrals
\eqn\logrel{\eqalign{
\dot L_z(\xi,t)&\equiv \int_{-i\infty}^{i\infty}\! {dN\over 2\pi i}\, e^{\xi N}
\int_{-i\infty}^{i\infty} {dM_1\over 2\pi i}\!
{dM_2\over 2\pi i}\, e^{t (M_1+M_2)}\,(M_1+M_2)\,L_z(N,M_1,M_2)
=\frac{\partial}{\partial t}L_z(\rho,t),\cr
L_z'(\xi,t)&\equiv \int_{-i\infty}^{i\infty}\! {dN\over 2\pi i}\, e^{\xi N}
\int_{-i\infty}^{i\infty} \!{dM_1\over 2\pi i}
{dM_2\over 2\pi i}\, e^{t (M_1+M_2)}\,N \,L_z(N,M_1,M_2)
=\frac{\partial}{\partial \xi}L_z(\rho,t).}}
These integrals may be computed for the gluon-gluon luminosity shown in 
fig.12: the results are displayed in fig.1. 
Clearly when $L_z'\gg\dot L_z$, high energy logarithms are relatively 
unimportant and the usual on-shell perturbative approximation to the hard 
cross-section is applicable. However when $L_z'\ll\dot L_z$, the high 
energy logarithms are the most important, and we should use the off-shell 
perturbative expansion of the hard cross-section, perhaps with $N=0$. 
It can be seen from fig.1 that this is only true at 
very low scales, close to the initial boundary condition. More interesting
is the intermediate region $L_z'\sim\dot L_z$, in which resummation of both
types of logarithm is necessary, and in which we may usefully expand both
in $N$ and in $M$. This region is important 
not only at high rapidity, but also in the central region when the 
invariant mass of the produced particles is not too high. 

The reason that
the region dominated by logarithms of $Q^2$ is so much larger than that 
dominated by logarithms of $x$ is very simple: when the coupling runs with 
$Q^2$ the variation of the luminosity with $t=\ln Q^2$ is 
much weaker than its variation with $\xi=\ln S/Q^2$, so factors of $M$ are
less significant than factors of $N$. This observation underpins the success
of unresummed perturbation theory (which amounts to expansion 
in powers of $M$) in computing hard cross-sections in kinematic 
regions where one might naively have expected it 
to fail due to unresummed logarithms of $x$.

We have shown a variety of estimates of the size of the high energy resummation
effects compared to standard NLO perturbation theory. Our basic conclusion is
that the effect of resummation in the partonic cross-section is an 
enhancement similar in size to that of a perturbative NNLO correction, 
while in the full cross-section the resummation of the gluon distribution 
produces an effect of similar magnitude but opposite sign. The net result
is thus rather less than might be expected from NNLO considerations alone,
since there is substantial cancellation. In fact we find that for 
a wide range of processes, each over a wide kinematic range, the net
effect of resummation seems to be a suppression of between $5$ and $10\%$ 
(compare the blue curves in fig.11, fig.18, fig.20 and fig.25). 
This seems to suggest
that the hard scale $Q$ may not be the optimal factorisation scale for 
high energy processes.

However it must also be remembered that the suppression of the gluon 
due to resummed evolution is probably being overestimated in these 
calculations, since the initial distribution at $2\GeV$ 
is kept fixed, rather than 
fitted to data (see for example the resummed fits in 
refs.\refs{\abfsxphen,\bfalf,\thornefits}). The true band of 
current uncertainty thus probably lies between these two extremes, i.e. 
between the solid and dashed blue curves in the figures. 
Thus for example in the total cross-section for 
hadronic $B$-production (fig.20) the  
resummation corrections are as small as $-5\pm 5\%$ at the Tevatron, rising
to $5\pm 10\%$ at the LHC and $20\pm 20\%$ at a VLHC. These figures are still 
rather lower than suggested by the leading order 
calculations in ref.\colell\ and substantially smaller than
the fixed coupling estimates in \refs{\cchglu,\rdbrke}. 
The effects on rapidity distributions 
may be rather larger, particularly at large rapidities: for $B$ production
at LHC resummation effects might be as large as $\pm 15\%$ at rapidities of
$5$ or so. Similar estimates should hold for inclusive jet cross-sections.

For the gluonic contribution to Drell-Yan at LHC the effects are 
a little larger, since the infrared singularity is stronger. However 
for vector boson production the corrections are still modest 
since the hard scale is relatively large, though they are 
even so comparable to other sources of uncertainty. 
The predicted enhancement at large rapidities
shown in fig.24 is particularly striking.

Refining these estimates into precise predictions is now straightforward: 
there are no longer any theoretical obstacles to computing cross-sections for 
hadroproduction processes correct to NLO which simultaneously resum all
leading and next-to-leading logarithms of $S$ and $Q^2$. However there 
is still a lot of work to be done. 

Firstly it would be useful to have complete 
calculations of off-shell partonic cross-sections for a wider variety of 
hadronic processes: at the moment all we have are cross-sections for 
heavy quark production \refs{\rdbrke,\ccfac} and Higgs production 
in the $m_t\to\infty$ limit \refs{\higgs}. The key partonic calculation for 
the future is clearly Drell-Yan and vector-boson production, both to 
confirm the conjectured structure of infrared singularities and 
provide a firm prediction for these benchmark processes. As 
explained above, it is important to match 
these cross-sections to the standard perturbative (on-shell) cross-sections,
preferably by keeping the full $N$ dependence when going off-shell. However 
for most purposes it is probably sufficient to treat the off-shellness 
perturbatively, keeping only the first few powers of $M$. The off-shell 
calculations are then complementary to the more usual fixed order (on-shell)
calculations, offering a short cut to new results. 

Secondly, for more precise calculations we clearly need to include 
quark effects, particularly in the high rapidity region where one 
of the partons is in the valence region. Including quarks in the 
resummed singlet evolution is no longer a problem \abfnf, and including 
the contribution of initial state quarks to the resummed partonic 
cross-sections is also well understood \refs{\CH,\rdbrke,\abfsxphen}. 

Finally, for accurate 
resummed predictions it is necessary to produce resummed parton densities,
fitted to data using resummed theoretical predictions. Previous experience
\refs{\abfsxphen,\bfalf} suggests that much of the effect of resummation in 
the evolution might then be absorbed into the parton distributions, so 
without this ingredient resummation effects in the parton distribution 
functions are probably being overestimated. In order 
to obtain unbiased resummed parton distributions with sensible 
experimental uncertainty distributions, it will be necessary to use 
a statistical approach such as that currently being developed by the 
NNPDF collaboration \nnpdf.


\vfill\eject
\centerline{\bf Acknowledgements} 
\bigskip
I would like to thank M.~Ciafaloni for 
emphasising to me long ago the importance of going beyond the saddle 
point approximation when the coupling runs, R.K.~Ellis for 
encouraging me to persevere 
with the hadronic singularity problem, L.~Magnea and 
G.~Sterman for encouraging me 
to write it up, R.K.~Ellis and S.~Forte for comments 
on the completed manuscript, and finally an anonymous referee for 
several constructive remarks on the nature of the infrared singularities. 
This work was done in the context of the  
Scottish Universities Physics Alliance.

\bigskip
\footatend\immediate\closeout\rfile\writestoppt
\baselineskip=14pt\centerline{{\bf References}}\bigskip{\frenchspacing%
\parindent=20pt\escapechar=` \input refs.tmp\vfill\eject}\nonfrenchspacing
\vfill\eject
\bye